\documentclass[useAMS,usegraphicx,usenatbib]{mn2e}
\usepackage{times}   %for Times Roman font (used by MNRAS)
\usepackage{mathptm} %Type1 fonts for mathematical symbols
\usepackage{color}

\title[Detectability of macronovae]
{Detectability of compact binary merger macronovae}
\author[Rosswog et al.]
{\Large \bf S. Rosswog\thanks{E-mail: stephan.rosswog@astro.su.se}$^{1}$, U. Feindt$^2$, O. Korobkin$^3$,  M.-R. Wu$^4$, J. Sollerman$^{1}$, A. Goobar$^{2}$,  G. Martinez-Pinedo$^5$\\
$^1$ The Oskar Klein Centre, Department of Astronomy, AlbaNova, Stockholm University, SE-106 91 Stockholm, Sweden\\
$^2$ The Oskar Klein Centre, Department of Physics, AlbaNova, Stockholm University, SE-106 91 Stockholm, Sweden\\
$^3$ Los Alamos National Laboratory, Los Alamos, NM 87545, USA\\
$^4$ Niels Bohr International Academy, Niels Bohr Institute, Blegdamsvej 17, 2100 Copenhagen, Denmark\\
$^5$ GSI Helmholtzzentrum f\"{u}r Schwerionenforschung, Planckstr.~1, 64291 Darmstadt, Germany  \& \\
$\;$ Institut f\"{u}r Kernphysik (Theoriezentrum), Technische Universit\"{a}t Darmstadt, 64289 Darmstadt, Germany
}

\def\msun{M$_{\odot}$}
\def\Msun{M$_{\odot}$ }
\def\be{\begin{equation}}
\def\ee{\end{equation}}
\def\bi{\begin{itemize}}

\def\ei{\end{itemize}}
\def\ben{\begin{enumerate}}
\def\een{\end{enumerate}}
\def\bea{\begin{eqnarray}}
\def\eea{\end{eqnarray}}
\def\bt{\begin{tabbing}}
\def\et{\end{tabbing}}

\def\edo{\end{document}}

\def\M4{M$_4$}
\def\M6{M$_6$}
\def\lcm6{LCM$_6$}
\def\qcm6{QCM$_6$}

\def\wh3{W$_{\rm H,3}$}
\def\wh4{W$_{\rm H,4}$}
\def\wh5{W$_{\rm H,5}$}
\def\wh6{W$_{\rm H,6}$}
\def\wh7{W$_{\rm H,7}$}

\begin{document}

\date{Draft version}

\pagerange{\pageref{firstpage}--\pageref{lastpage}} \pubyear{2016}

\maketitle

\label{firstpage}

\begin{abstract}
We study the 
optical and near-infrared luminosities and 
detectability of radioactively powered electromagnetic 
transients ('macronovae') occuring in the aftermath of  binary neutron 
star and neutron star black hole mergers. We explore the transients 
that result from the dynamic ejecta and those from different types of
wind outflows. Based on full nuclear network simulations we calculate 
the resulting light curves in different wavelength bands. We scrutinize 
the robustness of the results by comparing  a) two different nuclear 
reaction networks and b) two macronova models.  
We explore in particular how sensitive the results are to  
the production of $\alpha$-decaying trans-lead nuclei. We compare
two frequently used mass models: the Finite-Range Droplet Model (FRDM) and the nuclear mass model of Duflo and Zuker (DZ31). \\
We find that the abundance of $\alpha$-decaying trans-lead nuclei
 has a significant impact 
on the observability of the resulting macronovae. For example, the
DZ31 model yields considerably larger abundances resulting in larger
heating rates and thermalization efficiencies and therefore predicts
substantially brighter macronova transients. We find that 
the dynamic ejecta from NSNS models can reach  peak  K-band 
magnitudes in excess of $-15$ while those from NSBH  
cases can reach beyond $-16$. Similar values can be reached 
by some of our wind models. Several of our models (both wind 
and dynamic ejecta)  yield properties that are similar
to the transient that was observed in the aftermath of the short GRB 
130603B. We further explore the expected macronova detection 
frequencies for current and future instruments such as  VISTA, ZTF 
and LSST.\\
\end{abstract}

\section{Introduction}
% start with GW-event
The recent, direct detection of gravitational waves \citep{abbott16a} has 
finally opened the door to the long-awaited era of gravitational wave (GW) 
astronomy. With a sky localization uncertainty of  $\sim 600$ square degrees 
for the first event, however, the astronomical environment (e.g. type of 
galaxy or ambient medium density) in which the black hole (BH) merger took 
place is essentially unknown. 
To extract this information, measure the event's redshift and to constrain
the evolutionary channels that lead to the merger in the first place one needs
a coincident  electromagnetic (EM) signal.\\ 
While today the majority of LIGO-detectable sources 
is believed to be binary black holes, advanced detectors should also be able 
to observe compact binary systems \citep{belczynski16a} that contain at least one neutron star 
(NS; either NSNS or NSBH; hereafter collectively referred to as compact 
binary mergers, CBMs).
% the idea of r-process in NSM and some "milestones"
In such mergers neutron star matter is decompressed and ejected into space. 
Within the ejecta rapid neutron capture (r-process) synthesizes a range of 
heavy elements \citep{lattimer74,eichler89,rosswog99,freiburghaus99b} up to 
and beyond the platinum peak near A=195. The subsequent radioactive decay of 
the freshly synthesized r-process elements in the expanding ejecta 
causes EM transients known as  "macronovae" (MN)\footnote{The same 
phenomenon is often referred to as "kilonova". We prefer "macronova" (MN), 
since the phenomenon is {\em not} a thousand times brighter than
a nova, as originally thought. See~\cite{metzger16a} for discussion of
the naming conventions.} \citep{li98,kulkarni05,rosswog05a,metzger10b,roberts11}.
They are promising EM counterparts of GWs \citep{metzger12a}, since they are 
--contrary to gamma-ray bursts-- close to isotropic and viewing angle effects 
are only of order unity \citep{grossman14a}.  By now, there are
three cases where near-infrared excesses in the aftermath of gamma-ray 
bursts have been interpreted as being due to macronovae \citep{tanvir13a,berger13b,jin15,yang15}.\\
%%
%%Mass ejection channels
%%
%----------------------------------------------------------------
\begin{figure*} %  figure placement: here, top, bottom, or page
 \hspace*{-0.cm}\includegraphics[width=13cm,angle=0]{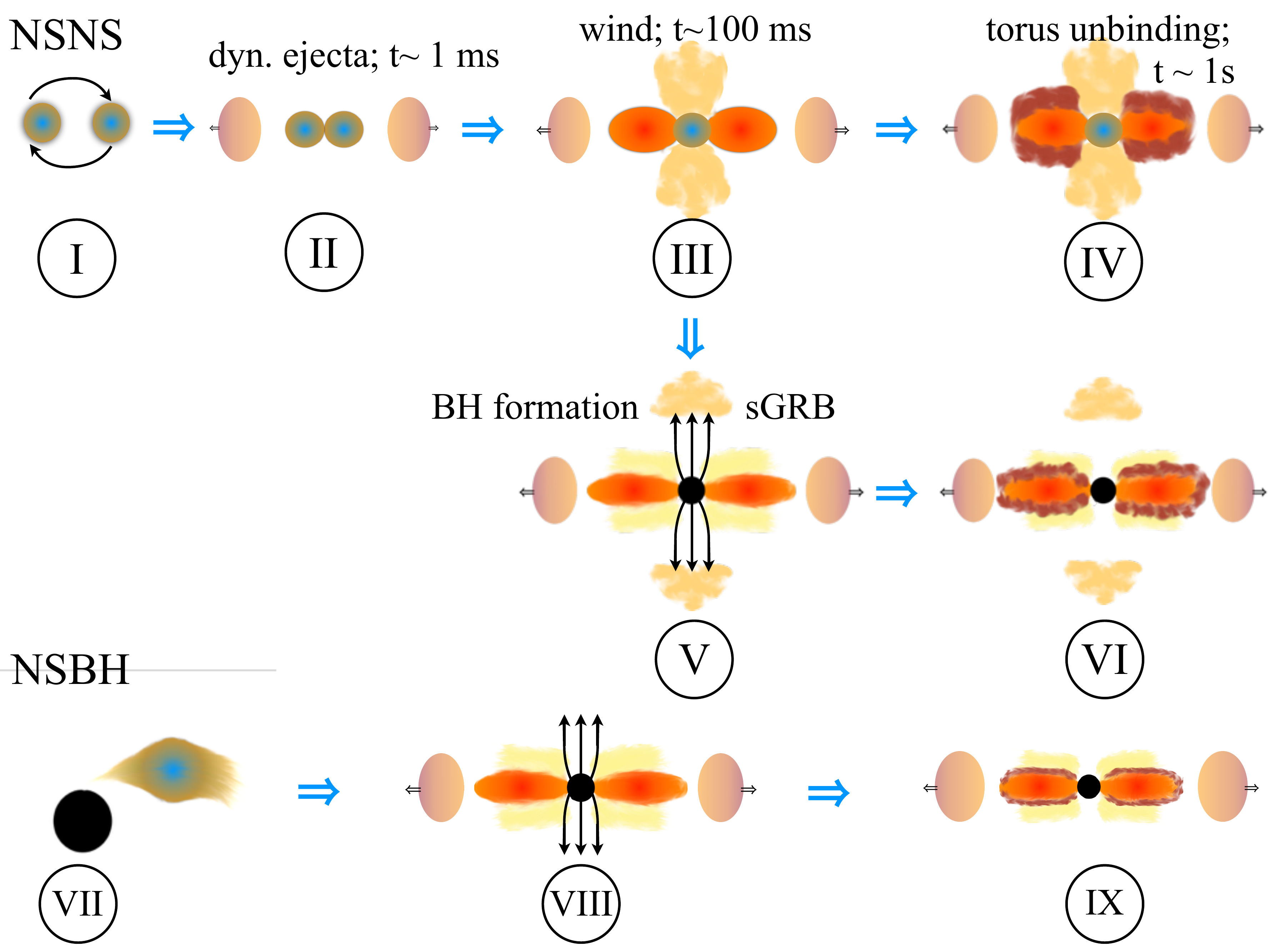}
   \caption{Overview of the different mass loss channels in both the NSNS and NSBH case.}
   \label{fig:mass_loss}
\end{figure*}
%----------------------------------------------------------------
%----------------------------------------------------------------
\begin{figure*} %  figure placement: here, top, bottom, or page
 \hspace*{0cm}\includegraphics[width=17cm,angle=0]{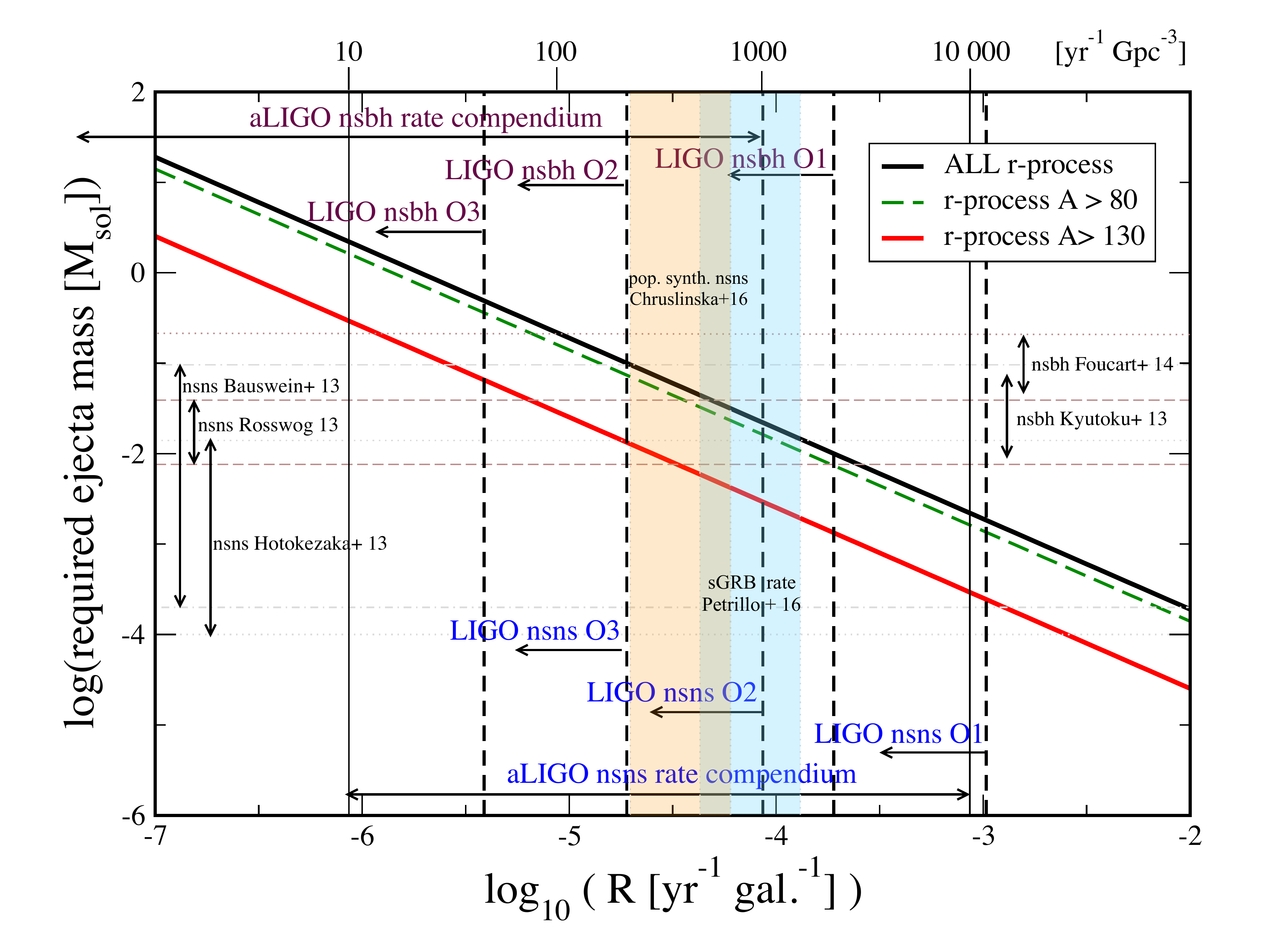}
   \caption{Summary of various rate constraints. The lines from the upper left to lower right indicate
   the typical ejecta mass required to explain all r-process/all r-process with $A>80$/all r-process with $A>130$
   for a given event rate (lower panel per year and Milky Way-type galaxy, upper panel per year and Gpc$^3$). Also marked
   is the compiled rate range from Abadie et al. (2010) for both double neutron stars and neutron star black hole systems and (expected)
   LIGO upper limits for O1 to O3 (Abbott et al. 2016b). The dynamic ejecta results from some hydrodynamic simulations 
   are also indicated.}
   \label{fig:rate_constraints}
\end{figure*}
%----------------------------------------------------------------

CBMs eject matter via several channels, see Fig.~\ref{fig:mass_loss} and e.g. \cite{fernandez16} and 
\cite{rosswog15a} for recent reviews.
The best-studied channel (stage "II" in Fig.~\ref{fig:mass_loss}) is the so-called {\em dynamic ejecta} \citep{rosswog99,ruffert01,oechslin07a,Bauswein13a,hotokezaka13a,rosswog13b,lehner16a}
that become unbound immediately at merger. They are launched by either gravitational torques ("tidal component") or, in the NSNS case, 
by hydrodynamic interaction at  the interface between the stars ("interaction component", see \citet{oechslin07a} and \cite{korobkin12a}\footnote{See Fig.~2 in the latter publication for an illustration.}). 
While the tidal component carries away matter with the original, very low electron fraction ($Y_e\approx0.04$) which is set by cold $\beta$-equilibrium in the original neutron 
star, the interaction component is heated and can increase its $Y_e$ via positron captures 
$e^+ + n \rightarrow p + \bar{\nu}_{e}$.
Both sub-components, however, share the property that they are hardly 
impacted by neutrino irradiation \citep[see e.g.][]{radice16a},
simply because they have already reached large distances from the 
remnant before the neutrino luminosity starts rising in earnest 
(after accretion torus formation at $t >10$ ms after contact). 
Nevertheless, for material that is not immediately ejected, weak
interactions can change the electron fraction substantially, see 
e.g. \cite{wanajo14} who find a particularly large fraction of
high $Y_e$ material in the ejecta.\\ 
The dynamically ejected matter is complemented by baryonic {\em winds} that are driven 
by neutrino-energy deposition, magnetic fields, viscous evolution and/or nuclear recombination energy 
\citep{beloborodov08,metzger08a,dessart09,fernandez13b,perego14b,siegel14a,just15,ciolfi15,martin15}. 
{\em If} a completely or temporarily stable neutron star survives in the centre of a 
merger remnant (see stage "III" in Fig.~\ref{fig:mass_loss}) the neutrino- and the magnetically driven 
winds are substantially enhanced in comparison to the case where a BH forms immediately. 
Simulations indicate that a prompt collapse of the massive neutron star to a black hole
can be avoided in many cases \citep{baumgarte00,kaplan14,kastaun15,takami14,gondek16}
even if the central mass substantially exceeds  the  Tolman-Oppenheimer-Volkoff (TOV) limit.
The calculations of \cite{shibata06c} and \cite{hotokezaka11,hotokezaka13b} indicate
that  prompt collapse is avoided, unless the initial  mass of the binary exceeds 
the TOV mass limit by more than $\sim$ 35\%. The two precisely determined neutron star masses 
near 2.0 \Msun (J1614-2230,~ \cite{demorest10} and PSR J0348+0342,~\cite{antoniadis13}), place 
this threshold value to at least $2.7$ \msun. Thus,  a large fraction of the mergers may go 
through such a metastable phase with a strong neutrino-driven wind phase and binaries at the low 
mass end may even produce stable, very massive neutron stars.\\
Recent studies \citep{murguia14,murguia16} indicate, however,  that a BH needs to form on a time scale of
less than $\sim 100$ ms in order to  launch a short GRB (stage "V" in Fig.~\ref{fig:mass_loss}), otherwise 
outflows will be baryon-overloaded and will not reach relativistic speeds. Seconds after the burst the resulting 
BH-disk system can release a good fraction of its disk (stage "VI"). Recent work \citep{fernandez13b,just15}
suggests that neutrinos play a sub-dominant role for the mass loss once a BH has formed.\\
NSBH mergers (stage "VII") can release substantially more matter in dynamic ejecta than NSNS binaries
\citep{rosswog05a,duez10b,deaton13a,foucart13,kyutoku13,foucart14,bauswein14b}. If a sufficiently massive torus
forms around the BH, it is expected that a jet can be launched into a relatively unpolluted surrounding (stage "VIII").
But large BH-masses in combination with low spins may lead to very small or no disks \citep{foucart12,rosswog15a}.
On longer times scales a fair fraction of the initial disk mass is expected to become unbound (stage "IX"), just as in the
NSNS case.\\
Recently, a fast neutron component has been claimed to produce a
macronova precursor signal \citep{metzger15a}. This model is based on
SPH simulations in the conformal flatness approximation where
a few particles are ejected early on with a high velocity from the
shear interface between the two merging neutron stars. While such a
precursor is a possibility, we do not see it in our simulations and
neither do, e.g., \cite{lehner16a} in their recent GR simulations. Therefore,
we will not consider this possibility in the following discussion.
\\
%%%%%%%%%%%%%%%%%%%%%%%%%%%%
Our knowledge of the rates of compact binary mergers is  still plagued  by large uncertainties. Simple constraints can be derived by assuming that they are related to the production of r-process
elements. If we take the solar-system r-process abundance pattern \citep{arnould07} as representative
and define a quantity $\sigma(A)= \sum_{A_i > A} X^r_{A_i}$, where $X^{r}_{A_i}$ is the r-process mass fraction with nucleon number $A_i$ and use a baryonic mass of $M_{\rm b, MW}= 6 \times 10^{11}$ \Msun \citep{mcmillan11a}, we find that the Milky Way contains $\approx 19000$ \Msun of r-process material
in total. About $M_{\rm r, >130}= \sigma(130)M_{\rm b, MW}=$ 2530 \Msun of this matter has $A>130$ and about 500 \Msun are beyond the
"platinum peak" ($A\ge195$).  With an age of the Galaxy of $\tau_{\rm MW} \approx 10^{10}$ yrs, this
yields average production rates of $\dot{M}_{\rm r, all}= 1.9 \times 10^{-6}$ \Msun yr$^{-1}$, 
$\dot{M}_{\rm r, A>130}= 2.5 \times 10^{-7}$ \Msun yr$^{-1}$ and
$\dot{M}_{\rm r, A>195}= 5 \times 10^{-8}$ \Msun yr$^{-1}$.  The product of  average ejecta mass and event rate is known but the individual factors are not. 
This is shown as lines from the upper left to the lower right  in Fig.~\ref{fig:rate_constraints} (e.g. r-process with $A>130$ in red).  As an example, if an event that produces all r-process $A>130$ occurs at a  rate of $10^{-5}$ yr$^{-1}$ it has to eject $2.5 \times 10^{-2}$ \Msun each time. 
For comparison, some representative simulation results for both NSNS \citep{hotokezaka13a, bauswein13b,rosswog13b} and NSBH \citep{kyutoku13,foucart14} are also indicated. We also translated the rates from yr$^{-1}$ MWEG$^{-1}$ (bottom axis) to  yr$^{-1}$ Gpc$^{-3}$ (axis on top) via a density of $1.16 \times 10^{-2}$ MWEG  Mpc$^{-3}$ \citep{abadie10}. Here MWEG abbreviates "Milky Way  equivalent galaxy". Also indicated  are the  expected LIGO upper limits of science runs O1-O3 \citep{abbott16h}. We have further marked the NSNS merger rates from the population synthesis studies of \cite{chruslinska16} and the range of sGRB rates   estimated by \cite{petrillo13}.
\\
%%%%%%%%%%%%%%%%%%%%%%%%%%%
The paper is organized as follows. In Sec. 2 we describe the methodological 
elements that enter our study. We discuss in particular our hydrodynamic 
simulations, the used nuclear reaction networks and our two macronova models. 
Section 3 presents our results on nucleosynthesis and discusses the impact 
of different nuclear mass formulae on the resulting abundances, the thermalization 
efficiency and the radioactive heating rates. We further present optical and 
near-infrared lightcurves and we discuss the detection feasibility, the 
follow-up of LIGO triggers and the use of GRB-triggers to search for 
macronovae. Sec. 4 summarizes the main findings of this study.

\section{Methodology}
In this study we explore macronova transients based on nuclear network calculations
along thermodynamic trajectories. We perform a set of hydrodynamic simulations to
obtain trajectories for dynamic NSNS ejecta, for other cases we use a parametrized 
treatment with numerical values based on existing hydrodynamic studies.

\subsection{NSNS merger simulations}
\begin{figure*} 

\vspace*{-3cm}

\center{\includegraphics[width=12cm,angle=-90]{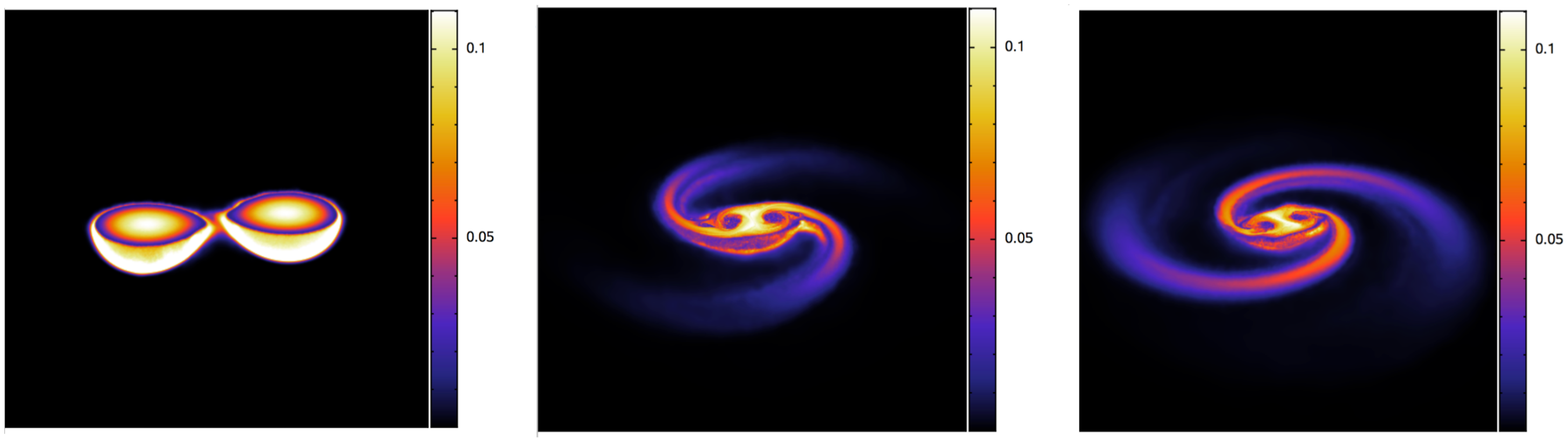}}
 
 \vspace*{-3.4cm}
 
   \caption{Electron fraction in a 1.3-1.3 \Msun merger (model N2; only matter below orbital plane shown) at t= 7.06, 11.6 and 12.4 ms.}
   \label{fig:runN2}
\end{figure*}

The  simulations performed for this study make use of the Smooth Particle Hydrodynamics method (SPH)
\citep{monaghan05,rosswog09b,springel10a,price12a,rosswog15c} coupled with a temperature-dependent
nuclear equation of state \citep{shen98a} and an opacity-dependent multi-flavour neutrino leakage scheme \citep{rosswog03a}.
This treatment includes in particular electron and positron captures and therefore the nuclear matter can change 
its electron fraction $Y_e$ in the course of the merger.
A concise summary of the implemented physics and of our numerical techniques can be found in \cite{rosswog13a}. The
only noticeable change is that we are using the Wendland C6 kernel \citep{wendland95,schaback06} together with 200 
neighbours per particle. We have recently scrutinized various ingredients in the SPH method \citep{rosswog15b}
and found that this kernel has excellent numerical properties and in particular drastically reduces the noise in
an SPH simulation in comparison to the cubic spline kernel that is commonly used.
The binary systems that are explored are summarized in Table~\ref{tab:dynamic_ejecta}, each of them  is modeled
with $10^6$ SPH particles and the stars have negligible initial spin as expected in nature due to the very short
tidal interaction time and the low viscosity of neutron star matter \citep{bildsten92,kochanek92}. Our typical
simulation time is $\sim 30$~ms.
As an example, a series of snapshots from a NSNS simulation (1.3-1.3 \msun; run N2; color-coded is the electron fraction
$Y_e$)  is shown in Fig.~\ref{fig:runN2}.

%----------------------------------------------------------------
%----------------------------------------------------------------

\begin{table*}
 \centering
 \begin{minipage}{140mm}
  \caption{Overview over the simulation parameters. Also shown are the resulting mass fractions of lanthanide ($X_{\rm lan}$) and actinide ($X_{\rm act}$)  elements. All NSBH cases start with an initial entropy of 2 $k_B$/nuc and $Y_e$= 0.06.}
\centerline{\bf Dynamic ejecta NSNS mergers}
\vspace*{0.3cm}
\begin{tabular}{@{}rccccccccl@{}}
\hline
   Run   &  $m_1$ [\msun] & $m_2$ [\msun]  & $t_{\rm end}$ [ms] & $m_{\rm ej}$ [$10^{-2}$\msun]&  $\langle v_{\rm ej,\infty}\rangle$ [c] & $m_{\rm ej, max} $     [$10^{-2}$ \msun]  & $X_{\rm lan} \; [10^{-2}]$ & $X_{\rm act} \; [10^{-2}]$ \\
   \hline \\
%\hline                            
N1  & 1.2 & 1.2  & 32.1 &  0.79& 0.12 & 3.17 & 17.22 & 5.06\\
N2  & 1.3 & 1.3  & 31.1 &  1.26& 0.11 & 2.70 & 17.01& 5.69 \\
N3  & 1.4 & 1.4  & 38.3 & 0.84 & 0.11 & 2.25 & 17.93 & 6.03\\
N4  & 1.2 & 1.4  & 30.6 & 1.59 & 0.11 & 2.92 & 17.82 & 5.39\\
N5  & 1.4 &  1.8 & 25.3 & 3.40 & 0.12 & 4.76 & 16.90 & 7.57 \\
\end{tabular}
\\
\\
\\
%%%%%%%%%%%%%%%%%%%%%%%%%%%%%%%%%%%
%                                                nsbh                                                  %
%%%%%%%%%%%%%%%%%%%%%%%%%%%%%%%%%%%
\centerline{\bf \hspace*{2.5cm} Dynamic ejecta NSBH mergers\\}
\vspace*{0.0cm}
\\
\begin{tabular}{@{}rcccccccccccl@{}}
\hline
   Run   &  $m_{\rm ns}$ [\msun] & $m_{\rm bh}$ [\msun] & $\chi$ & $m_{\rm ej}$ [$10^{-2}$\msun]& $\langle v_{\rm ej, \infty} \rangle$ [c] & $X_{\rm lan} \; [10^{-2}]$ & $X_{\rm act} \; [10^{-2}]$ & comment \\
   \hline
   \\
    B1 &  1.4 & 7.0 &   0.7 & 4.0 & $0.20 $  & 19.87 & 4.10 &Foucart et al. (2014), run M14-7-S7\\
    B2 &  1.4 & 7.0 &   0.9 & 7.0 & $0.18 $  & 19.27 & 4.83 & Foucart et al. (2014), run M14-7-S9\\
    B3 &  1.2 & 7.0 &   0.9 & 16.0 & $0.25 $ & 19.48 & 4.64 & Foucart et al. (2014), run M14-7-S9\\    
    \label{tab:dynamic_ejecta}
\end{tabular}
\end{minipage}
\end{table*}

%%%%%%%%%%%%%%%%%%%%%%%%%%%%%%%%%%%
%                                                winds                                                  %
%%%%%%%%%%%%%%%%%%%%%%%%%%%%%%%%%%%
\begin{table*}
 \centering
 \begin{minipage}{140mm}
\vspace*{0.3cm}
{\bf \hspace*{2.5cm} Parametrized winds\\}
\vspace*{0.0cm}
\\
\begin{tabular}{@{}cccccccll@{}}
\hline
     run          & $m_W$  [\msun]   & $Y_{\rm e}$ &   $v_{\rm w,\infty}$ [c] &  $E_{\rm kin}$ [erg] & $X_{\rm lan}$ & $X_{\rm act}$& comment\\
\hline \\
wind $\; $ 1  & 0.01   & 0.30 &  0.05 & $2.2 \times 10^{49}$ & $1.61 \times 10^{-7}$ & \quad \quad$<10^{-15}$& insp. by Perego et al. (2014)\\
wind $\; $ 2  & 0.01   & 0.25 &  0.05 & $2.2 \times 10^{49}$ & $6.30 \times 10^{-5}$ & \quad \quad $<10^{-15}$& insp. by Perego et al. (2014)\\
wind $\; $ 3  & 0.01   & 0.35 &  0.05 & $2.2 \times 10^{49}$ & \quad \quad $< 10^{-15}$  & \quad \quad $<10^{-15}$& insp. by Perego et al. (2014)\\
%-----------------------------
wind $\; $ 4  & 0.05   & 0.25 &  0.05 & $1.1 \times 10^{50}$ & $2.41 \times 10^{-4}$ & \quad \quad $<10^{-15}$& unb. disk material\\
wind $\; $ 5  & 0.05   & 0.30 &  0.05 & $1.1 \times 10^{50}$ & $2.45 \times 10^{-7}$ & \quad \quad $<10^{-15}$& unb. disk material; low-viscosity\\
wind $\; $ 6  & 0.05   & 0.35 &  0.05 & $1.1 \times 10^{50}$ & \quad \quad $< 10^{-15}$ & \quad \quad$<10^{-15}$& unb. disk material; low-viscosity\\
wind $\; $ 7  & 0.05   & 0.25 &  0.10 & $4.5 \times 10^{50}$ & $4.28 \times 10^{-5}$ & \quad \quad$<10^{-15}$\\
wind $\; $ 8  & 0.05   & 0.30 &  0.01 & $4.5 \times 10^{48}$ & $1.57 \times 10^{-4}$ & \quad \quad$<10^{-15}$\\
%----------------------------
wind $\; $ 9  & 0.10   & 0.25 &  0.10 & $9.0 \times 10^{50}$  & $7.70 \times 10^{-5}$& \quad \quad $<10^{-15}$\\
%---------------try some low mass - high velocity combinations----------
wind 10  & 0.01   & 0.25 &  0.10 & $9.0 \times 10^{49}$  & $3.49 \times 10^{-5}$ & \quad \quad$<10^{-15}$\\
wind 11  & 0.01   & 0.25 &  0.25 & $5.9 \times 10^{50}$  & $2.13 \times 10^{-2}$ & $6.34 \times 10^{-6}$\\
wind 12  & 0.01   & 0.25 &  0.50 & $2.8 \times 10^{51}$  & $7.50 \times 10^{-2}$& $1.66 \times 10^{-3}$\\
%------------------and some more parameter space------------------------
wind 13  & 0.10   & 0.35 &  0.01 & $8.9 \times 10^{48}$ & \quad \quad  $< 10^{-15}$& \quad \quad$<10^{-15}$\\
wind 14  & 0.10   & 0.30 &  0.05 & $2.3 \times 10^{50}$ & $1.35 \times 10^{-7}$& \quad \quad$<10^{-15}$\\
wind 15  & 0.20   & 0.35 &  0.01 & $1.8 \times 10^{49}$ & \quad \quad $< 10^{-15}$& \quad \quad $<10^{-15}$\\
wind 16  & 0.20   & 0.30 &  0.05 & $4.5 \times 10^{50}$ & $8.39 \times 10^{-8}$& \quad \quad$<10^{-15}$\\
wind 17  & 0.20   & 0.25 &  0.10 & $1.8 \times 10^{51}$ & $1.27 \times 10^{-4}$& \quad \quad$<10^{-15}$\\
wind 18  & 0.01   & 0.35 &  0.01 & $8.9 \times 10^{47}$ & \quad \quad $< 10^{-15}$& \quad \quad$<10^{-15}$\\
wind 19  & 0.05   & 0.25 &  0.25 & $2.9 \times 10^{51}$ & $3.91 \times 10^{-4}$& \quad \quad$<10^{-15}$\\
wind 20  & 0.10   & 0.25 &  0.25 & $5.8 \times 10^{51}$ & $4.95 \times 10^{-5}$& \quad \quad $<10^{-15}$\\
wind 21  & 0.20   & 0.25 &  0.25 & $1.1 \times 10^{52}$ & $3.20 \times 10^{-5}$& \quad \quad $<10^{-15}$\\
\end{tabular}
\caption{Parameters of the wind models. In all cases a specific entropy of 15 $k_{\rm B}$/nucleon is used, $X_{\rm lan}$ is the mass fraction in lanthanides.} 
\label{tab:winds}
\end{minipage}
\end{table*}

%----------------------------------------------------------------
\begin{figure} %  figure placement: here, top, bottom, or page
 \hspace*{-0.5cm}\includegraphics[width=10cm,angle=0]{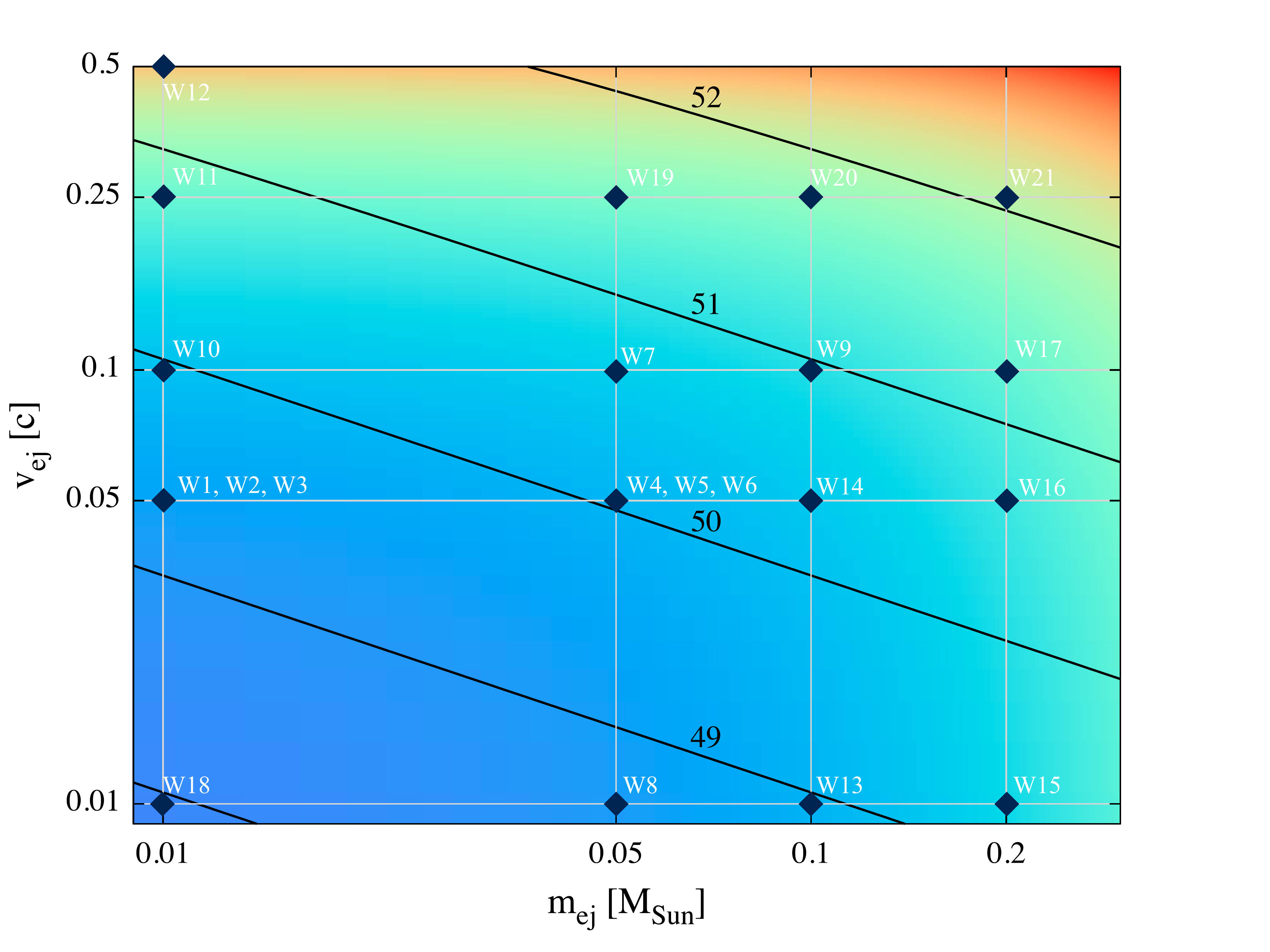}
   \caption{Explored parameter space for the parametric wind studies (different winds are annotated, e.g., as "W1"
   for wind 1). For the corresponding parameters are summarized in Tab.~\ref{tab:winds}. The black lines indicate
   the kinetic energy in the winds with the line labeled "50", for example, indicating $10^{50}$ erg.}
   \label{fig:wind_param_space}
\end{figure}
%----------------------------------------------------------------

 \subsection{Ejecta}
We identify unbound matter in our hydrodynamic simulations by the criterion 
$v_a^2/2 + \phi_a > 0$ at the end of our simulation, where $a$ labels the SPH particle and $v$ and $\phi$ are velocity and gravitational potential.
To double-check this criterion we compare with a criterion based on the 
outward radial velocity being larger than the local escape velocity. The ejecta masses based 
on both criteria agree with each other to within $\sim 2$\%. To have a robust 
upper limit, we also examine $v_a^2/2 + \phi_a  + u_a > 0$, 
where $u_a$ is the specific internal energy of a particle $a$. The resulting mass is noted as $m_{\rm ej, max} $ 
in Table~\ref{tab:dynamic_ejecta}, and it is used to set the upper limit that can be plausibly expected for the 
electromagnetic signal from dynamic ejecta.\\
Studies that focus on the long-term evolution of accretion disks around BHs  \citep{wanajo12,fernandez13a,fernandez13b,just15,wu16}
find that a substantial fraction ($\sim 20\%$) of the initial torus can become unbound. According to \cite{giacomazzo13} the
torus masses can, depending on the initial mass ratio, be very large and reach multiples of 0.1 \msun.
Thus the unbound mass from accretion tori can be large and actually rival the dynamic ejecta
masses of even asymmetric mass ratio NSNS/NSBH mergers. The exploratory numerical studies, however, 
suggest that the velocities are substantially lower than in the dynamic ejecta case.
\cite{just15}, for example, find that the  average velocities of the torus component never exceed 0.06 c, while 
the dynamic ejecta in the NSBH cases can be larger than 0.2c, see Table~\ref{tab:dynamic_ejecta}.\\
We subsume all the ejecta types other than dynamic ejecta under the broad category "winds". 
Despite their different origin, they all have in common that they are exposed for longer time 
to the neutrino irradiation, therefore this material has a larger $Y_e$ and consequently a different nucleosynthesis.
In particular, such matter has a lower content in those elements that are the major opacity sources  \citep{kasen13a,tanaka13a,fontes15}.
With potentially different nuclear heating rates  
and lower opacities, these winds are promising sources for EM transients, that could potentially outshine 
the signals from the dynamic ejecta, if they are not obscured by them. A first study by \cite{fernandez15} finds,
however, that due to their different velocities, the wind expansion at large radii is not influenced 
by the dynamic ejecta. \\
Our parametric wind model contains four parameters: the wind mass $m_w$, the initial entropy $s_0$, the electron 
fraction $Y_e$ and the (terminal) wind velocity $v_w$. To create initial conditions for the network and subsequent
macronova calculations, we also need a starting temperature $T_0$ and an initial radius $R_0$. The starting temperature
is chosen as $T_0= 9 \times 10^9$ K, so that we are safely above the threshold temperature for nuclear statistical
equilibrium (NSE, $T_{NSE}\approx 5 \times 10^9$ K). Thus, the initial abundance distribution is set by NSE
and, below $8 \times 10^9$ K, the nuclear reaction network takes over. The initial radius $R_0$ is found by the 
requirement that the average density $\rho_0= 3m_W/(4\pi R_0^3)$ together with $T_0$ reproduces the desired initial
entropy $s_0$. To keep the parameter space under control and motivated by the strongly peaked entropy distribution
of \citet[][their Fig.~7]{perego14b}, \cite{just15} and \cite{fernandez15}, we fix the initial entropy to values of 15 k$_B$ per
baryon. Typical values for $R_0$ are $\sim 650$ km, in reasonable agreement with the numerical studies of
neutrino-driven winds \citep{dessart09,perego14b}. Since we are interested here in exploring the lower opacity
case, we restrict our study to electron fractions above the threshold value for heavy r-process, $Y_{e, \rm thresh} = 0.25$ 
\cite[see][ their Fig.~8]{korobkin12a}. 
From the results shown in \cite{perego14b}, we 
expect terminal wind velocities around 0.05 c. These numbers motivate our ``wind 1'' simulation.
Once the parameters have been set, the density evolves according to
$\rho(t)= \rho_0 (1 + v_w t/R_0)^{-3}$ and the temperature evolution is calculated using
the HELMHOLTZ equation of state \citep{timmes00a} according to the entropy change from nuclear 
reactions \citep{freiburghaus99b}. The explored wind parameters are given in Tab.~\ref{tab:winds} and visualized in Fig.~\ref{fig:wind_param_space}.

\subsection{Nucleosynthesis calculations}
For each simulation we perform a nucleosynthesis calculation
to obtain the final abundance pattern and  the nuclear
heating rate $\dot{\epsilon}_{\rm nuc}$ that is needed for the macronova
models. 
We calculate an average out of 1000 randomly chosen hydrodynamic ejecta trajectories
for each of our merger simulations (N1 - N5).
Since there is little variation between individual trajectories this is a fair representation
of the overall ejecta dynamics. These average trajectories are used in the network 
calculations to obtain the nuclear energy generation rate $\dot{\epsilon}_{\rm nuc}$ that
is used in the macronova calculation. For the dynamic ejecta in the NSBH cases (run
B1 - B3) and the winds we use the above described expansion model. For the NSBH cases we apply the parameters from the simulations of \cite{foucart14}. For
the entropy and electron fraction they provide results for individual trajectories that
are near $Y_e=0.06$ and $s= 2$ k$_{\rm B}$/nuc. We choose these values
in our models, but stress that the nucleosynthesis (and therefore the macronovae) in 
this regime is insensitive to the exact 
numbers\footnote{To illustrate this, have a look at Fig.~8 in \cite{korobkin12a}: 
{\em all} trajectories with $Y_e<0.15$ yield practically identical nucleosynthesis results.}.
The parameters for the dynamic ejecta are summarized in Table~\ref{tab:dynamic_ejecta}.
Since what we subsume under "winds" can have different physical origins, see the discussion
above, we vary the wind parameters in a wide range, see Table~\ref{tab:winds}.\\
Our baseline nucleosynthesis calculations are performed with a large nuclear reaction
network WinNet~\citep{winteler12,winteler12b} that is based on the BasNet
network~\citep{thielemann11}. It includes  5831
isotopes from nucleons up to $Z=111$ between the neutron drip line
and stability.  The reaction rates are from the compilation
of~\cite{rauscher00} for the finite range droplet model
(FRDM)~\citep{moeller95} and the weak interaction rates
(electron/positron captures and $\beta$-decays) are from ~\cite{fuller82} and \cite{langanke01}. In addition, the distribution of fission fragments from \cite{kodama75} is used as a default. Finally, we adopt neutron capture and neutron-induced fission rates of \cite{panov10} and $\beta$-delayed fission probabilities of \cite{panov05}.\\
The nucleosynthesis in the dynamic ejecta has been 
found to be extremely robust against variations of  
astrophysical parameters \citep{korobkin12a}, but the final abundance patterns 
are still sensitive to variations of the nuclear physics such as fission distributions,
beta-decay rates and nuclear mass formulae 
\citep{mendoza_temis15,eichler15,goriely15,barnes16a,wu16,mumpower16a}. 
For the dynamic ejecta
we  explore  the impact of the nuclear mass model on the nuclear heating
rate. Specifically, we compare the 31-parameter Duflo-Zuker (DZ31) mass formula 
\citep{duflo95} with the Finite Range Droplet Model (FRDM; \citealt{moeller95}).  The major 
effect in our context is a substantially enhanced trans-lead nuclei fraction for the 
DZ31-case, which leads to an increased nuclear heating rate via $\alpha$-decays 
\citep{barnes16a}. These comparisons are performed with 
the network described in detail in \cite{mendoza_temis15}.  
It includes 7360 nuclei between the proton and neutron 
drip-lines up to charge number $Z=110$.
For the nuclear masses we use experimental values from the Atomic Mass Evaluation \cite{wang12}  whenever available. Where available, we use experimental alpha, beta decay and spontaneous fission rates from the NUBASE \citep{audi12} and NuDat2 databases (http://www.nndc.bnl.gov/nudat2/) and theoretical predictions otherwise.
For the most relevant theoretical rates for the r-process, 
we employ the beta-decay rates of \cite{moeller03}, the neutron-capture and the reverse 
photo-dissociation rates of \cite{mendoza_temis15} for nuclei with $Z\leq 83$ for 
different mass models, 
the neutron-capture and neutron-induced fission rates of \cite{panov10} for $Z>83$
with FRDM mass model and Thomas-Fermi fission barriers,
and the beta-delayed and spontaneous fission rates of \cite{petermann12}. The fission fragment
distributions are taken from \cite{zinner07} calculated by the ABLA code, which
well reproduces available fission data and includes the possibility of neutron emission
before and after fission.

%----------------------------------------------------------------
\begin{figure*} %  figure placement: here, top, bottom, or page

\centerline{
  \includegraphics[width=8cm,angle=0]{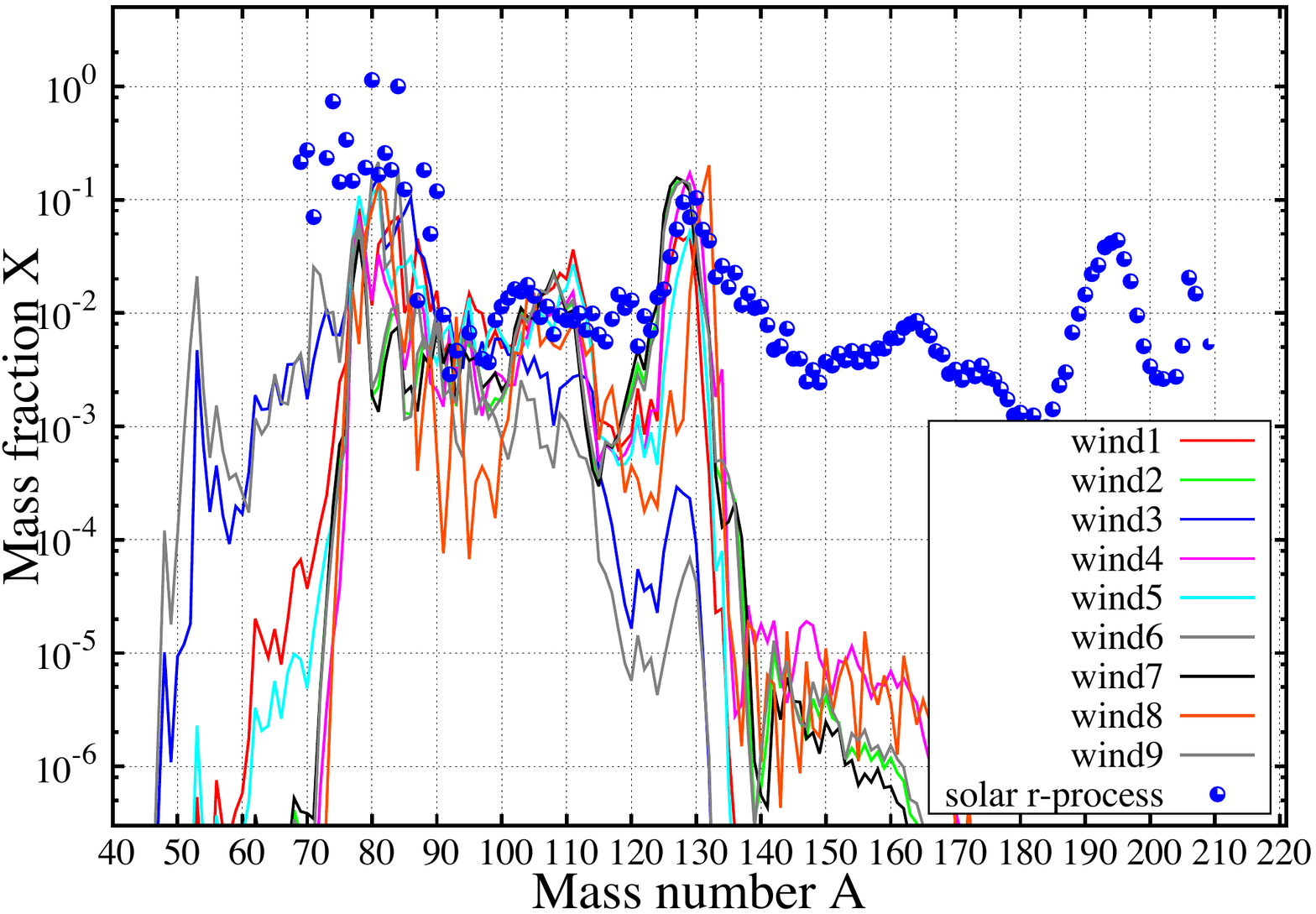} \hspace*{0.5cm}
  \includegraphics[width=8cm,angle=0]{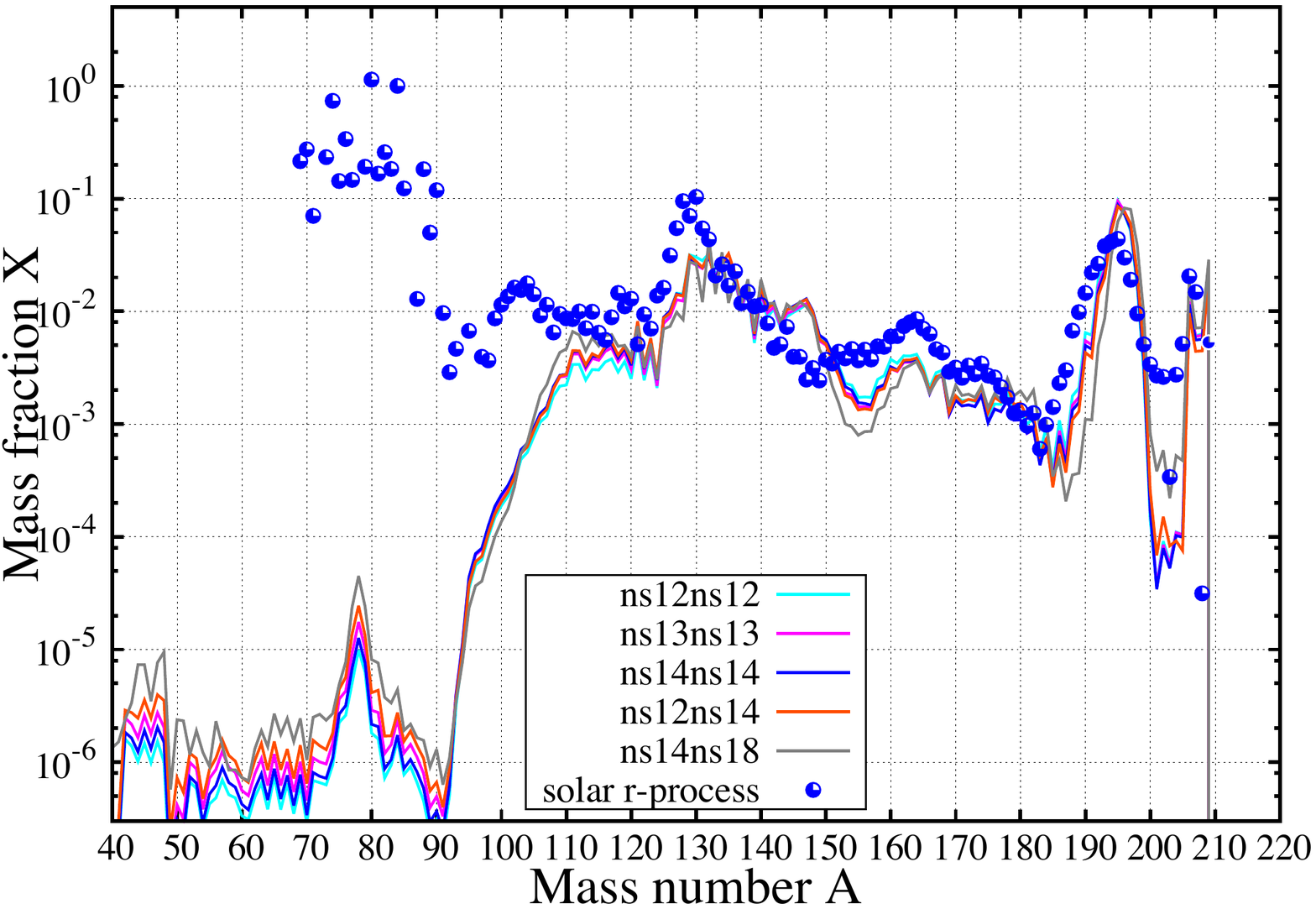}
  }
  \vspace*{-2.5cm}
  
   \caption{Resulting element distribution  for the  wind models (left) and dynamic ejecta (right). The blue
                 dots are the solar system r-process mass fractions.}
   \label{fig:all_nucleo}
\end{figure*}
%----------------------------------------------------------------

\subsection{Macronovae}
% basic idea and scaling laws
\subsubsection{Scaling relations}
The luminosity of a MN rises as more material becomes visible and once
all parts are transparent, one sees a decline in the luminosity dictated by 
the nuclear energy generation rate, $L\propto \dot{\epsilon}_{\rm nuc} m_{\rm ej}$. 
As a guidance for the following discussion, we provide the scaling laws that follow 
from simple arguments. After the neutron captures have ceased ($\sim 1$ s after 
merger) the heating rate can to a reasonable approximation be described by
a power law, $\dot{\epsilon}_{\rm nuc} \propto t^{-\alpha}$, with a power law index
$\alpha \approx 1.3$ \citep{metzger10b,roberts11,korobkin12a,hotokezaka16a}.
The peak emission is reached when expansion and diffusion times are comparable 
which yields \citep{grossman14a}
\be
t_{\rm peak} \approx 4.9 {\rm d} \left( \frac{\kappa}{10 \; {\rm cm^2/g}} 
\frac{m_{\rm ej}}{0.01 \; M_\odot} \frac{0.1 \; c}{v_{\rm ej, \infty}}\right)^{1/2}
\ee
for the peak time,
\be
L_{\rm peak}\approx 2.5 \times 10^{40} \frac{\rm erg}{\rm s} \left( \frac{v_{ ej, \infty}}{0.1 \; c}    
\frac{10 \; {\rm cm^2/g}}{\kappa}  \right)^{0.65} \left( \frac{m_{\rm ej}}{0.01 \; M_\odot}\right)^{0.35}
\label{eq:scaling_Lpeak_mass}
\ee
for the bolometric luminosity and
\be
T_{\rm eff} \approx 2200 \; {\rm K} \left(\frac{10 \; {\rm cm^2/g}}{\kappa}  \right)^{0.4125}  
\left(\frac{0.1 \; c}{v_{ ej, \infty}} \right)^{0.0875} \left(\frac{0.01 \; M_\odot}{m_{\rm ej}} \right)^{0.1615}
\label{eq:Teff}
\ee
for the effective temperature where  all exponents were evaluated with $\alpha=1.3$.

\subsubsection{Model 1}
\label{sec:MNmodel1}
As our macronova model 1 (hereafter 'MNmodel1') we employ a model 
similar to the one used in~\cite{grossman14a} with a spherically symmetric
homologously expanding radial density profile $\rho(v) \sim \rho_0
(1-v^2/v^2_{max})^3$. In this model, it is assumed that all macronova 
energy originates from the layer above the so-called diffusion surface,
defined as the surface for which the average diffusion time is equal to 
the time from the merger. Any photons traveling below this diffusion 
surface are considered trapped and therefore invisible. The total luminosity
is then taken to be the mass of this layer times the instantaneous 
radioactive nuclear heating rate; thermal emission from the layer 
is assumed to be lost to $PdV$ expansion work and therefore 
neglected. In this sense, our model provides a conservative lower 
bound for the macronova emission. The emitted spectrum has an 
effective temperature of the photosphere at optical depth 
$\tau_{\rm ph}=2/3$.\\ 
Practically, the models  inherit from the hydrodynamic calculation the
ejecta mass, the electron fraction and velocity and from the network 
calculation the instantaneous nuclear heating rate $\dot{\epsilon}_{\rm nuc}(t)$,
assuming that a fixed fraction ($f_{\rm tot}=0.5= $ const) thermalizes 
at all times while the rest is lost.
Note that in model 1 we always use  the FRDM mass model~\citep{moeller95}.\\
A major uncertainty in the prediction of macronova transients are the
opacities of the expanding r-process material, which unfortunately has the largest effect
on the spectral energy distribution (SED), see Eq.~(\ref{eq:Teff}). Based on atomic structure models,
\cite{kasen13a} argued that the opacity of expanding r-process material is dominated by
bound-bound-transitions from those ions that have the most complex valence electron 
structure. They found in particular that even small amounts lanthanides have a large 
impact on the opacities. For example, the neodymium opacities exceed those of iron
as long as their mass fraction $X_{\rm Nd}>10^{-4}$. From their studies based
on four species, they conclude that a gray opacity of $\approx 10$ cm$^2$/g
should be fairly effective for calculating bolometric light curves. Similar conclusions 
were reached in a study by \cite{tanaka13a}.
However, \cite{kasen13a} and, more recently,~\cite{fontes15}  point out that even this may be an underestimate, and that 
the true value could even be an order of magnitude larger.\\
We use  the sum of lanthanide and actinide  fraction, $X_{\rm lan}$ + $X_{\rm act}$, to decide which opacity value to use.
Whenever  it exceeds a limiting value of $10^{-3}$, we use as fiducial value for the opacity in the dynamic ejecta a value of $\kappa= 10$ cm$^2$/g, otherwise we use $\kappa= 1$ cm$^2$/g.  The dynamic ejecta show a very large lanthanide fraction of $X_{\rm lan}\approx 0.18$ while  this value varies  widely for the different wind cases, see column  seven in Table \ref{tab:winds}, but is in almost all cases below the threshold value.\\
Since the current knowledge is based on expensive atomic structure calculations of
so far only a few ions, the opacity value for strongly lanthanide-enriched material is likely subject to considerable uncertainties. We therefore
also explore  how  our brightest case, N3 with DZ31-mass model, would appear in the case  of a substantially larger opacity value
($\kappa= 100$ cm$^2$/g). 

\subsubsection{Model 2}
\label{sec:MNmodel2}
A major difference in macronova model 2 ('MNmodel2') is that  we use  time-dependent thermalization efficiencies. In addition, we can switch between
the FRDM and the DZ31 nuclear mass model. All MNmodel2 calculations are performed with the \cite{mendoza_temis15} network.
The effect of the DZ31 mass model is that a larger fraction of trans-lead nuclei are produced
and, as shown below, this has a substantial impact on the nuclear heating rate at times
around and after the macronova peak ($t > 1$ day). The thermalization efficiencies have
been explored in recent work \citep{hotokezaka16a,barnes16a}.  We apply here the thermalization
efficiencies based on simple analytical estimates from the latter work. The thermalization efficiency for photons is estimated as
\be
f_\gamma(t)= 1 - \exp\left(- \frac{1}{\eta_\gamma^2}\right),
\ee
while for the massive particles (electrons, $\alpha$-particles and fission products) it reads
\be
f_i(t)= \frac{\ln(1+2 \eta_i^2)}{2\eta_i^2}.
\ee
The quantity $\eta$ is the ratio of time after the merger and the thermalization time scale
of the considered particle, $\eta_k= t/\tau_k$. We use for the different thermalization time scales 
\bea
\tau_\gamma&=& 1.40 \;  m_5^{1/2} v_2^{-1} \; {\rm days}\label{eq:tau_gamma}\\
\tau_e&=& 7.40 \; m_5^{1/2} v_2^{-3/2} \left( \frac{0.5 \; \rm MeV}{E_e} \right)^{1/2}\; {\rm days}\\
\tau_\alpha&=& 7.74 \; m_5^{1/2} v_2^{-3/2} \left( \frac{6.0 \;  \rm MeV}{E_\alpha} \right)^{1/2}\; {\rm days}\\
\tau_{\rm fis}&=& 16.77 \; m_5^{1/2} v_2^{-3/2} \left( \frac{125.0 \; \rm MeV}{E_{\rm fis}} \right)^{1/2}\; {\rm days},\label{eq:tau_fis}
\eea
where $m_5\equiv m_{\rm ej}/(5 \times 10^{-3}$\msun) and $v_2\equiv v_{\rm ej}/(0.2 c)$. In the following, we 
evaluate the time scales $\tau_k$ at the characteristic scaling energies given above. The total thermalization
efficiency is then given by
\be
f_{\rm tot}(t)= \frac{\dot{\epsilon}_\beta(t) \left[\zeta_\gamma f_\gamma(t) + \zeta_e f_e(t) \right] + \dot{\epsilon}_\alpha(t) f_\alpha(t) 
                      + \dot{\epsilon}_{\rm fis}(t) f_{\rm fis}(t)}{\dot{\epsilon}_\beta(t) +  \dot{\epsilon}_\alpha(t) + \dot{\epsilon}_{\rm fis}(t)},
\ee
where we use $\zeta_\gamma= 0.45$ and $\zeta_e=0.2$ \citep{barnes16a}. The nuclear heating
rate that enters the macronova calculation is then 
\be
\dot{\epsilon}_{\rm heat}(t)= f_{\rm tot}  \left[\dot{\epsilon}_\beta(t) +  \dot{\epsilon}_\alpha(t) + \dot{\epsilon}_{\rm fis}(t)\right].
\label{eq:E_heat}
\ee

%----------------------------------------------------------------
\begin{figure*} %  figure placement: here, top, bottom, or page
 \centerline{
 \includegraphics[width=8cm,angle=0]{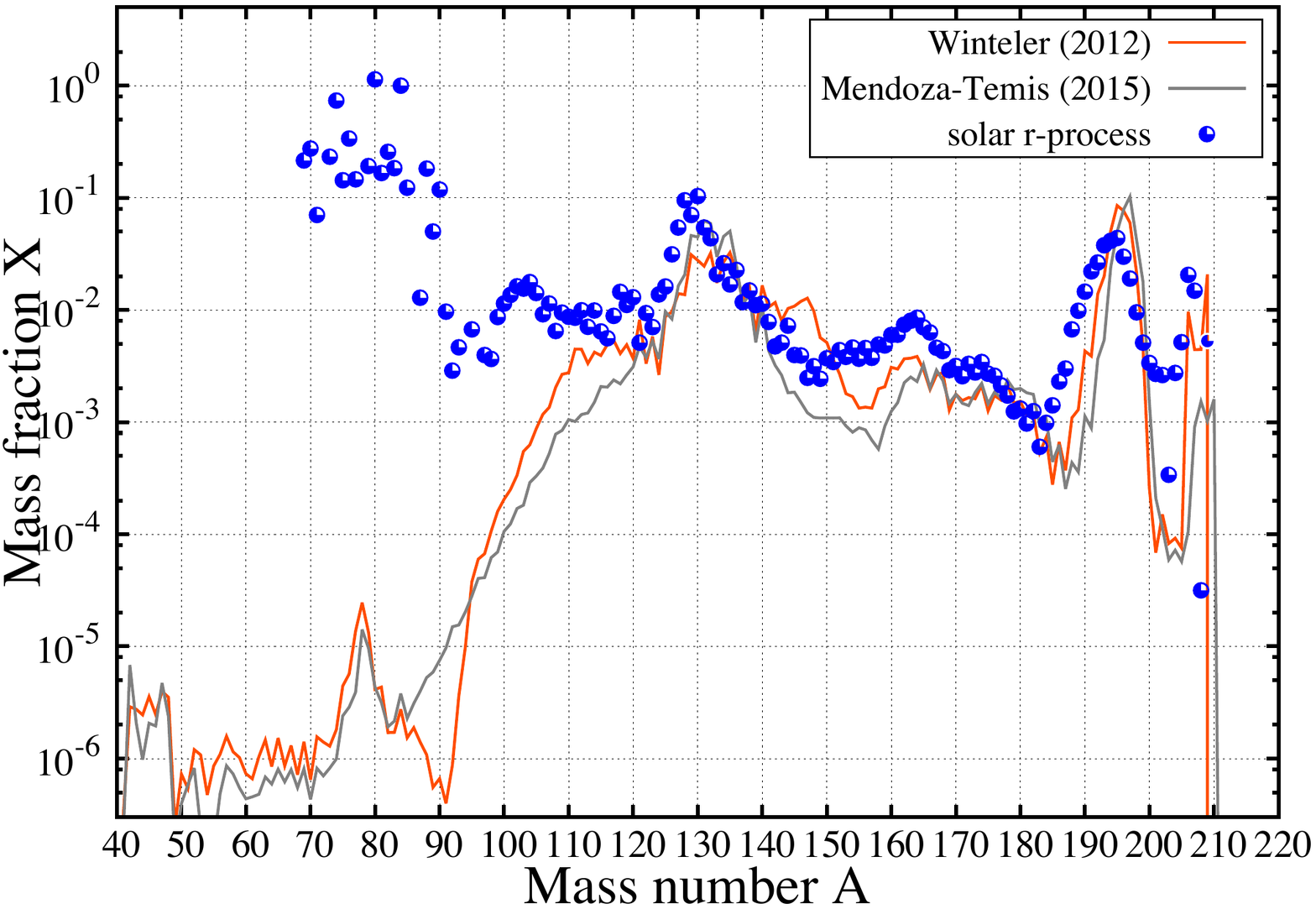}
 \includegraphics[width=8cm,angle=0]{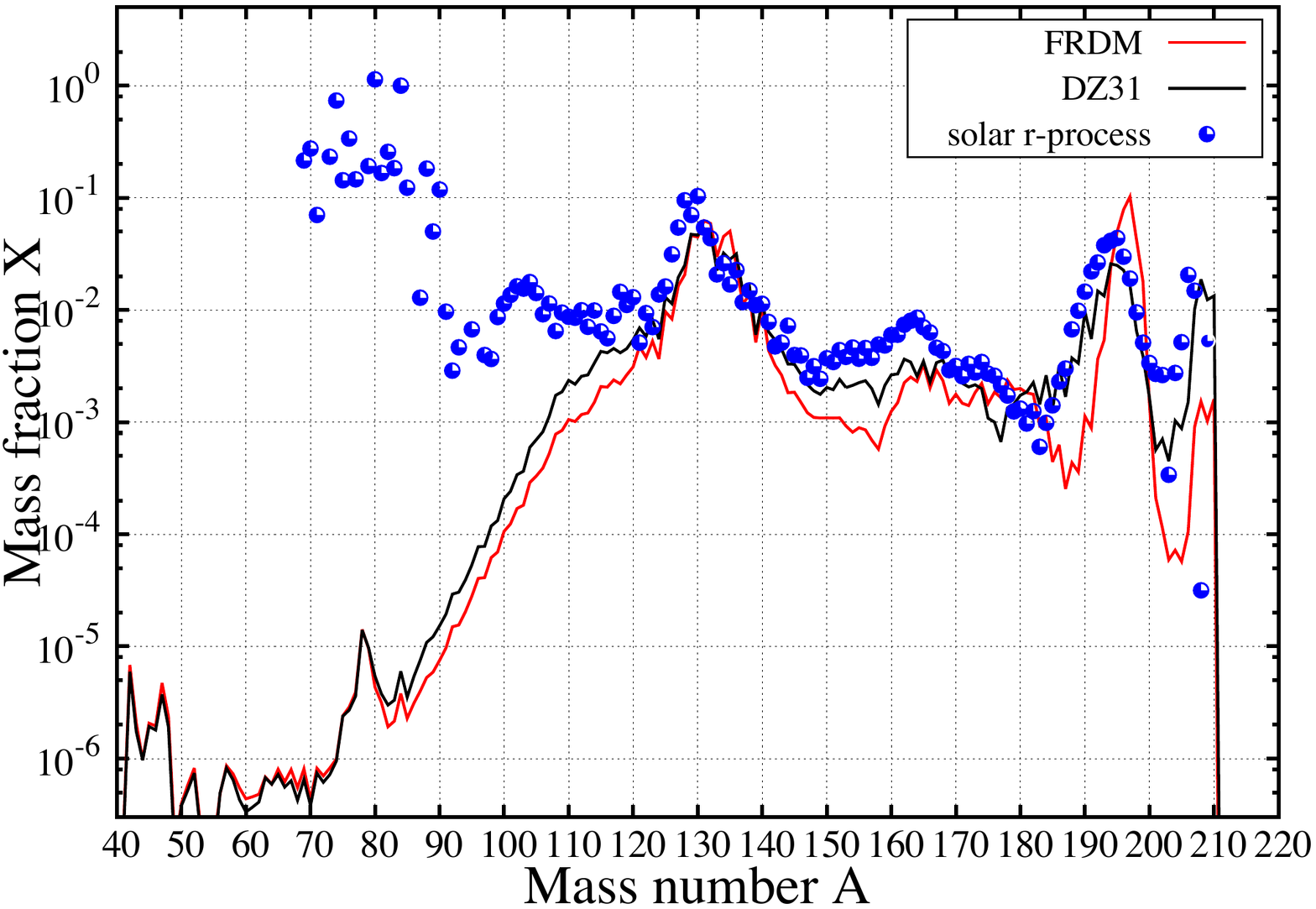}}
 \vspace*{-2.5cm}
   \caption{Left panel: comparison of the abundances for the run N4 (1.2 and 1.4 \msun; $t=$ 100 days) obtained 
   with the nuclear reaction networks of Winteler (2012) and of Mendoza-Temis et al. (2015), both using
   the FRDM nuclear mass model. Right panel: comparison of the results for the FRDM and DZ31 nuclear 
   mass model using the Mendoza-Temis et al. network. Note that the DZ31 mass formula is in much better agreement with the solar system abundances than FRDM, especially in the region around the "platinum peak" ($A\approx 195$).}
   \label{fig:network_comparison}
\end{figure*}
%----------------------------------------------------------------

\section{Results}
    
\subsection{Nucleosynthesis}
\begin{figure*} %  figure placement: here, top, bottom, or page
\centerline{
  \includegraphics[width=7cm,angle=-90]{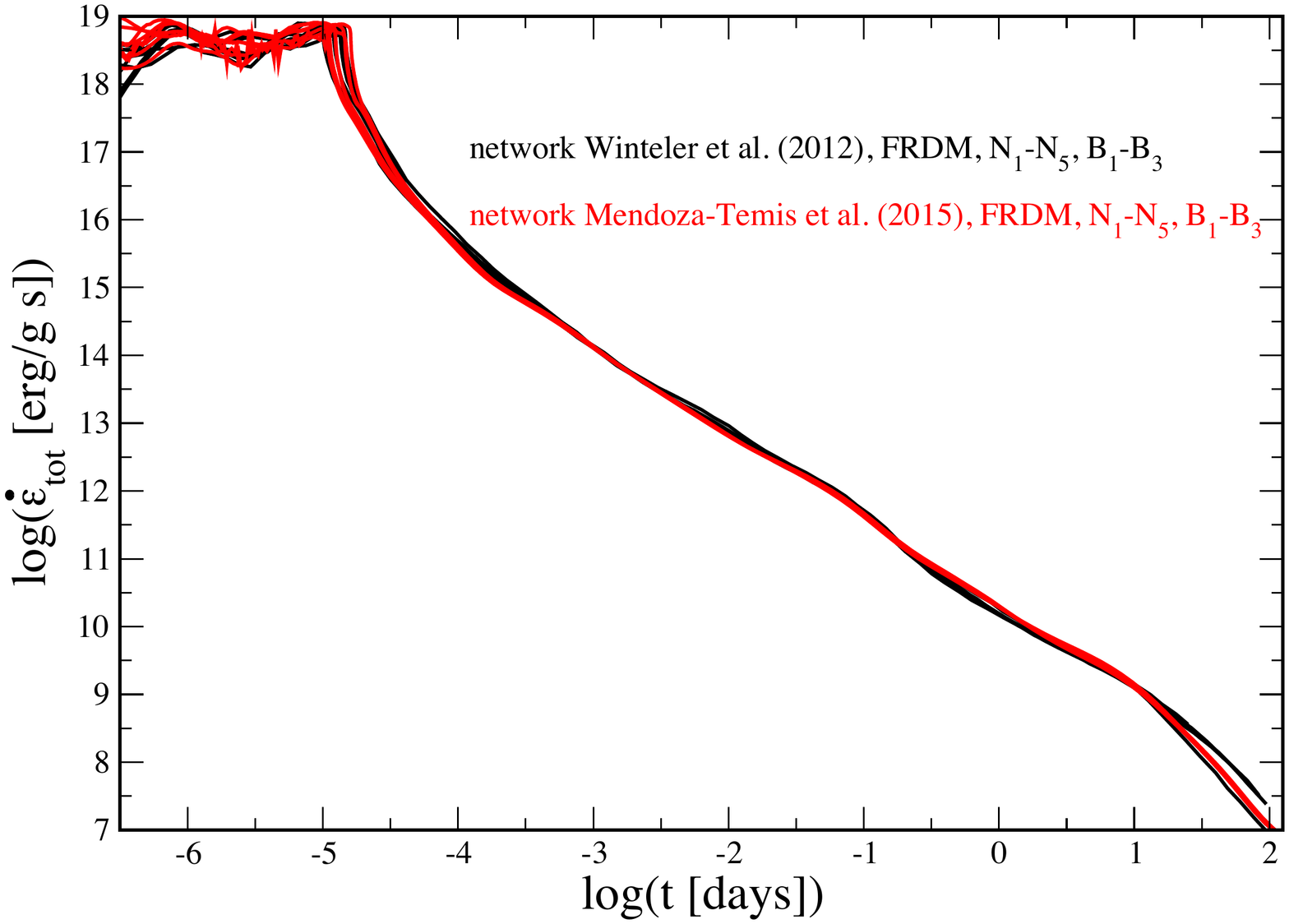}
  \hspace*{-0.8cm}
  \includegraphics[width=7cm,angle=-90]{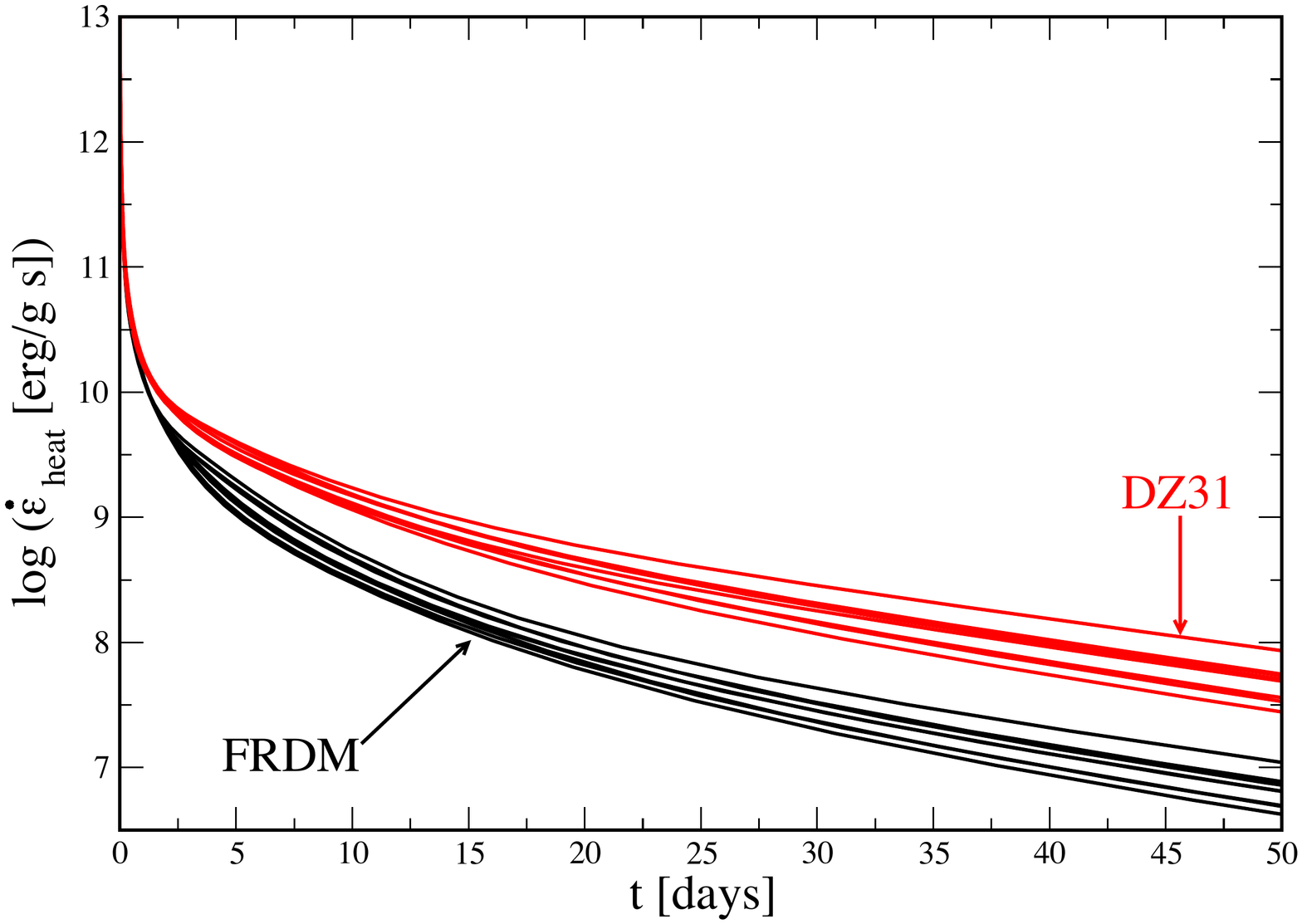} 
  }
   \caption{Left: Comparison of the total nuclear energy generation rates between the network of Winteler et al. (2012) and Mendoza-Temis et al. (2015), both using the FRDM mass model. The overall agreement between both networks is good over many orders of magnitude. Right: Net nuclear heating rates for the FRDM and DZ31 nuclear mass model (all runs
   N1-N5 and B1-B3; Mendoza-Temis et al. network). The DZ31 models yields consistently larger heating rates at late times ($t >  1$ day).}
   \label{fig:heating_rate}
\end{figure*}

% standard results
Unless mentioned otherwise, we refer to WinNet results with the FRDM mass model as a default.
From our simulations we find an electron 
distribution reaching up to 0.3, but with the 
majority of matter being  near 0.04. This results in a very  robust r-process pattern 
up to and beyond the third, "platinum" r-process peak near $A=195$, see right panel in 
Fig.~\ref{fig:all_nucleo}. For the neutrino-wind models the picture is different: here the resulting
abundance pattern is sensitive to the details and in particular to the electron fraction of a
thermodynamic trajectory. Consequently, the resulting pattern varies strongly between different
wind models. Due to their relative large electron fraction ($Y_e>0.25$), none of them reaches 
substantially beyond nucleon numbers of $A= 130$.\\
% network comparison
An exhaustive comparison of nuclear network results is beyond the intention of this paper.
Nevertheless, we perform a short comparison between the two networks \citep{winteler12,mendoza_temis15}
for one case (dynamic ejecta of run N4, NSNS binary with 1.2 and 1.4 \msun; $t=$ 100 days), simply in order to gauge by 
how much the nucleosynthesis results might be influenced by implementation details. In both cases we 
use the FRDM nuclear mass model. 
The final mass  fractions are shown in Fig.~\ref{fig:network_comparison}, left panel. The overall 
pattern agrees reasonably well, but there are noticeable differences in the regime from $A\approx 90$ to 170, 
due to the treatment of fission. We also briefly explore which impact  the nuclear mass model
has  on the final abundance pattern (run N4; Mendoza-Temis network; right panel of Fig.~\ref{fig:network_comparison}). A major difference
is the substantially higher mass fraction of trans-lead ($A>207$) nuclei when DZ31 is used. 
\cite{barnes16a} found that $\alpha-$decays of these nuclei have a substantial impact 
on the late-time lightcurve, see below. Overall, the DZ31 model shows a closer agreement 
with the solar r-process pattern,  especially around the third r-process peak ($A\approx 195$).\\
We also briefly compare the total nuclear energy generation rate resulting from both networks
(each time using FRDM, for all dynamic ejecta) in Fig.~\ref{fig:heating_rate}. There is good 
agreement between  both networks over many orders of magnitude, only at very late stages we
find for some trajectories  deviations of up to a factor of two. We leave more detailed comparisons 
of the effects of different input physics in the  networks to future work.
%----------------------------------------------------------------
%----------------------------------------------------------------

\subsection{Thermalization efficiency and nuclear heating rates}
\begin{figure*} %  figure placement: here, top, bottom, or page

\centerline{
  \includegraphics[width=7cm,angle=-90]{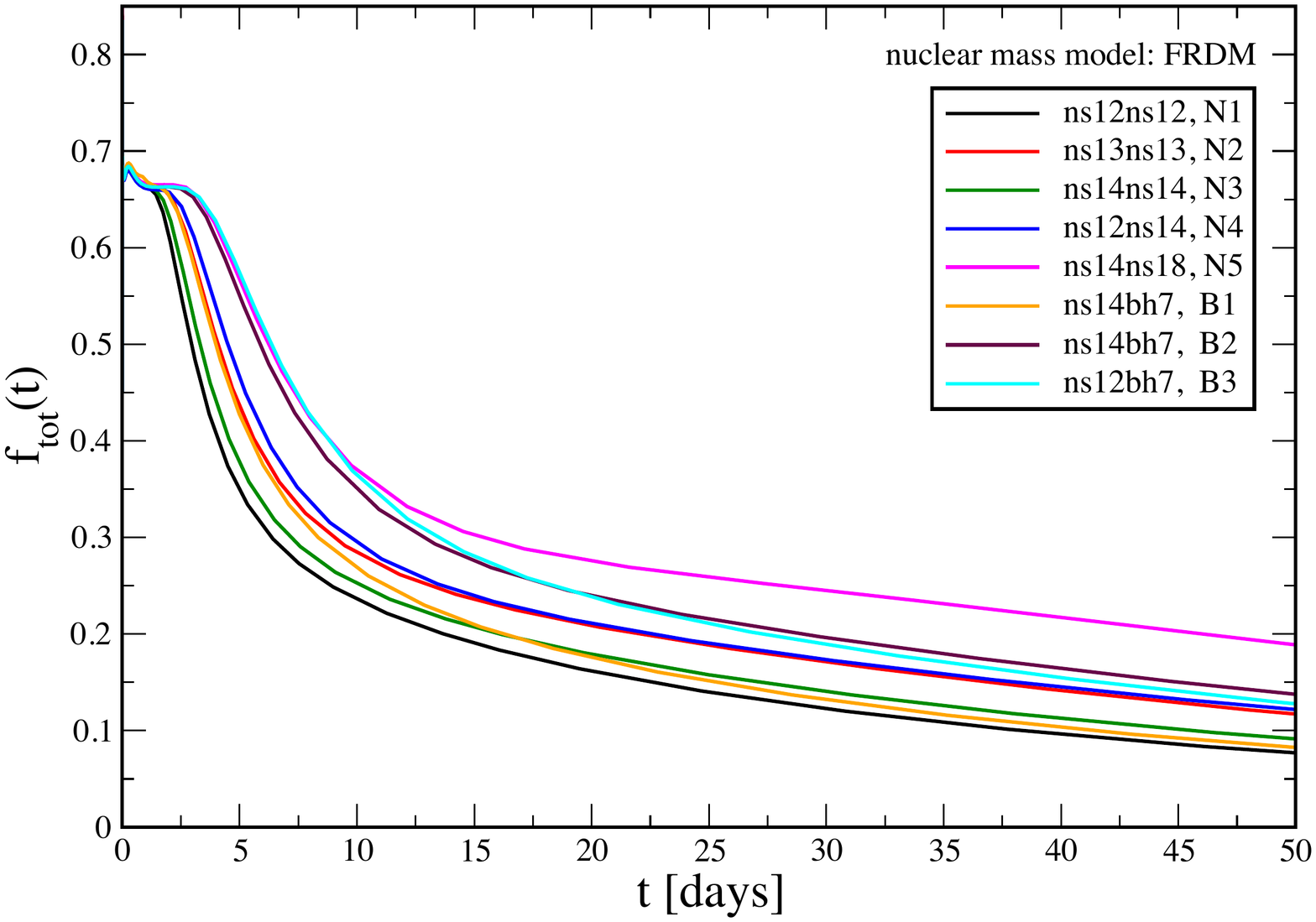} \hspace*{-1.5cm}
  \includegraphics[width=7cm,angle=-90]{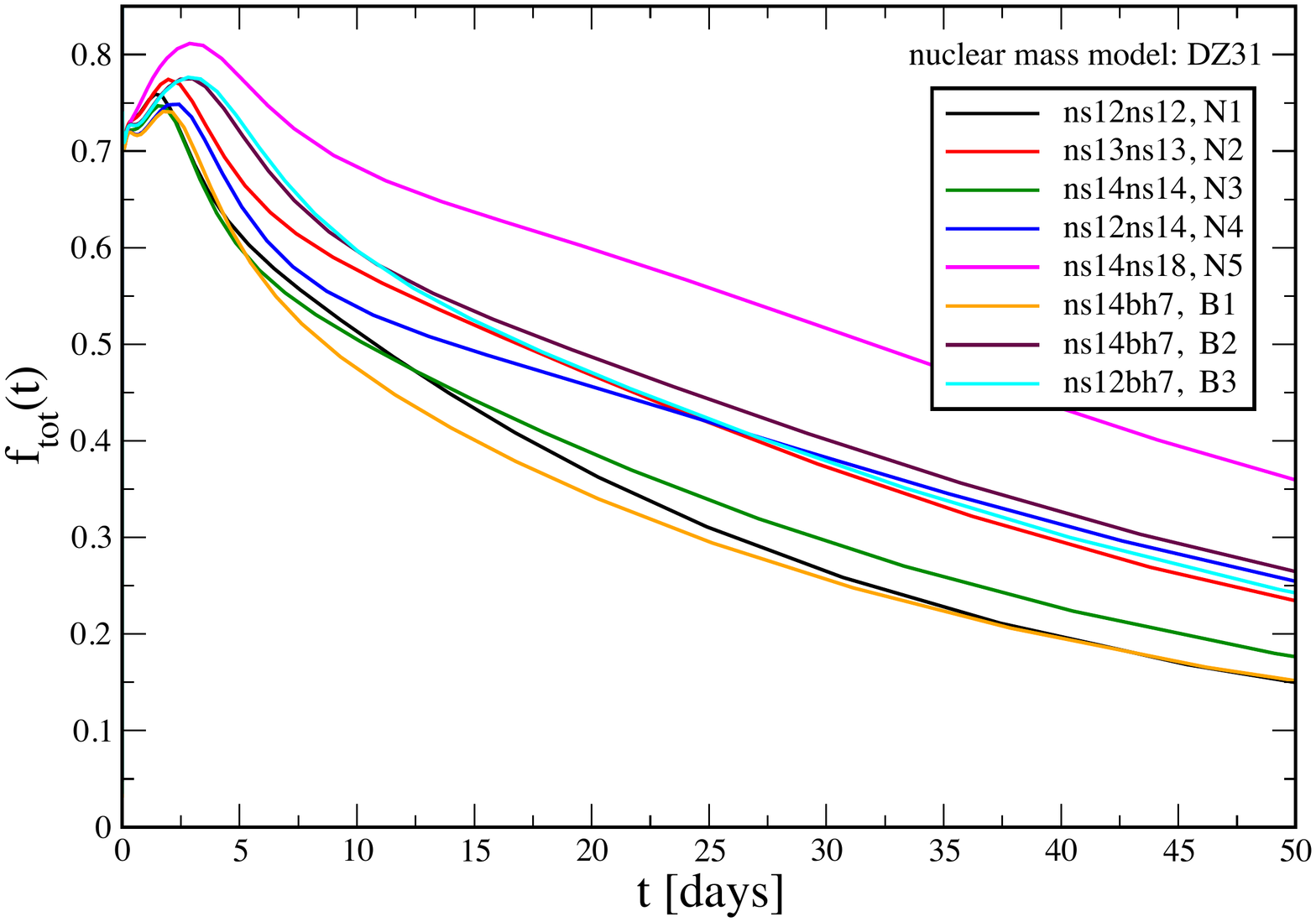}
  }
  \vspace*{0cm}
  
   \caption{Thermalization efficiencies for the different nuclear mass models.}
   \label{fig:thermalization}
\end{figure*}

Since the various reaction products thermalize on different time scales, see Eq.~(\ref{eq:tau_gamma})-(\ref{eq:tau_fis}),
the nuclear mass model has also an impact on the total thermalization efficiencies. Since $\alpha$-decays (due to translead
nuclei) are enhanced, we find noticeable differences in the overall efficiencies, see Fig.~\ref{fig:thermalization}. 
In all the cases $f_{\rm tot}$ remains approximately constant around $\sim 0.7$, but decreases substantially 
slower at late times (for $t > 3$ days) for the DZ31 mass 
model\footnote{Our results here differ somewhat from \cite{barnes16a}
in the sense that the ejecta are denser/slower. Therefore, the efficiencies can be 
higher in first few days and can actually increase if the $\alpha$-decays do so.}. 
%----------------------------------------------------------------
%----------------------------------------------------------------
Closely related,  for DZ31 the net heating rate, Eq.~(\ref{eq:E_heat}), is
significantly different,
see Fig.~\ref{fig:heating_rate}, right panel:  at late times ($t >  1$ day) 
it exceeds the FRDM-results by up to an order of magnitude. This has 
a serious impact on the resulting macronova lightcurves around the time 
of the peak emission.

\subsection{Optical and near-IR macronova lightcurves}
\begin{figure*} %  figure placement: here, top, bottom, or page
\centerline{
  \includegraphics[width=6.5cm,angle=0]{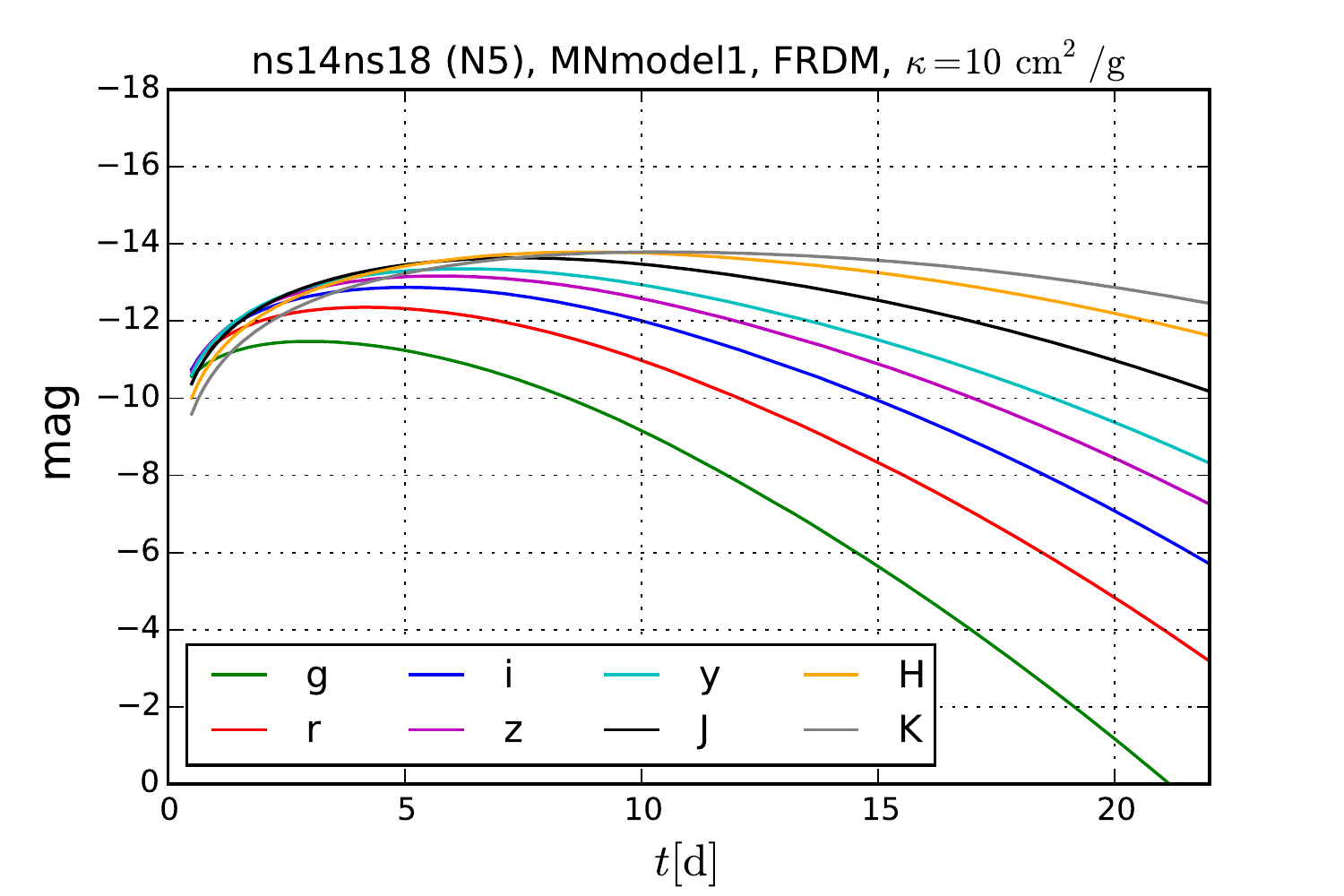}  \hspace*{-0.3cm}
  \includegraphics[width=6.5cm,angle=0]{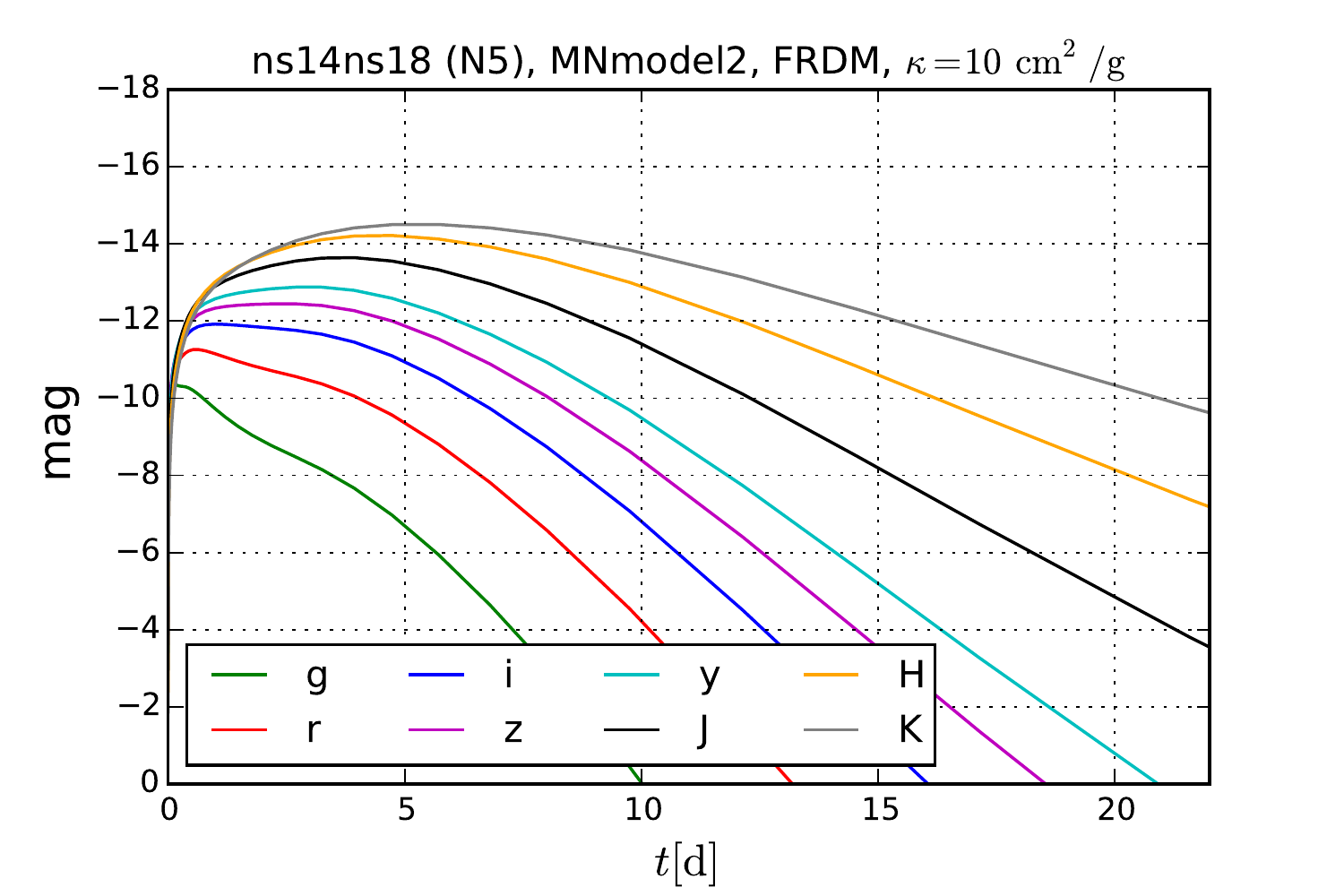}  \hspace*{-0.3cm}
  \includegraphics[width=6.5cm,angle=0]{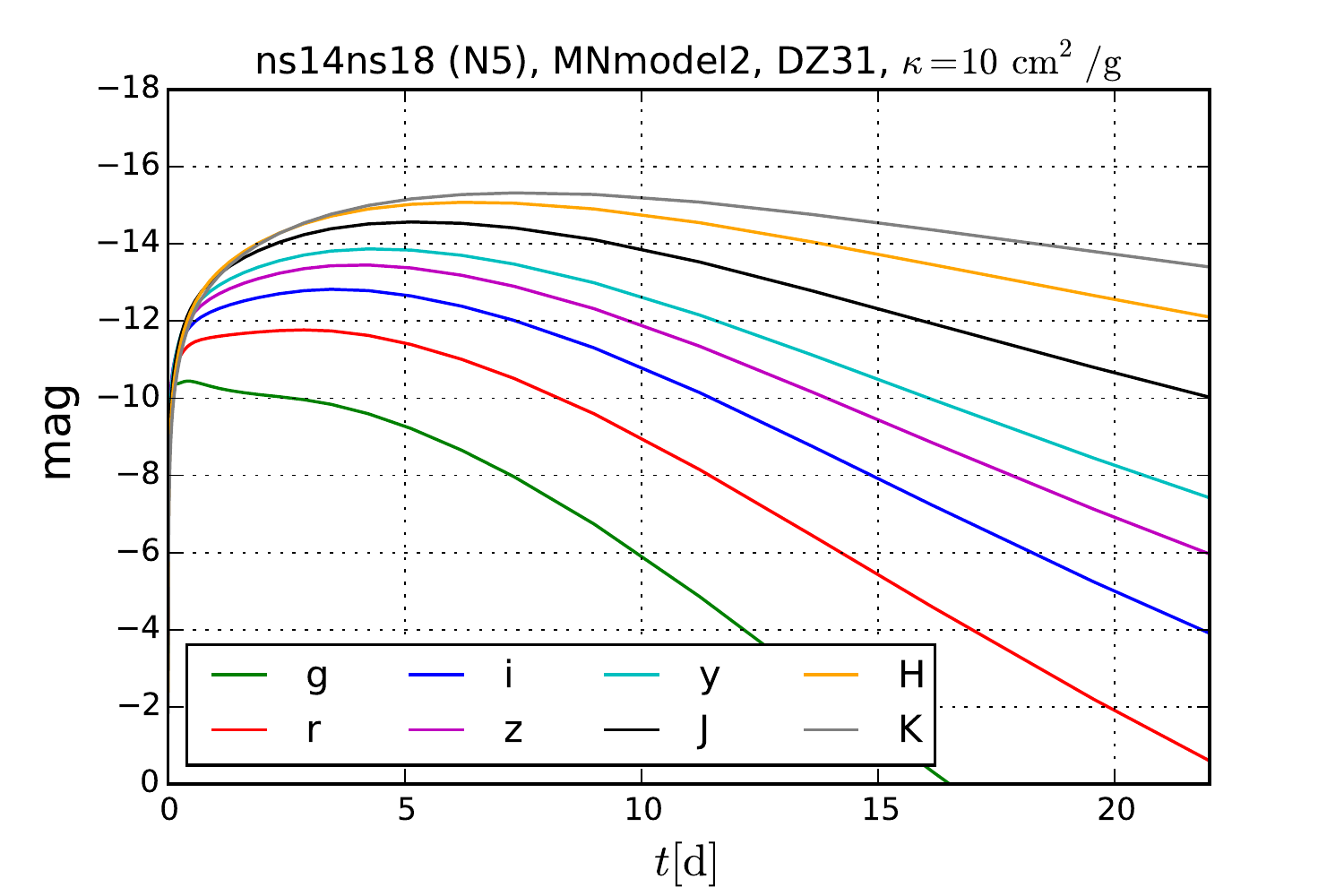}  
  } 
  
   \caption{Comparison of the light curves in different bands for the dynamic ejecta of the brightest
   NSNS binary system (simulation N5; 1.4 and 1.8 \msun). The left panel shows the results for
   macronova model 1 (Grossman et al. (2014), fixed thermalization efficiency $f_{\rm tot}=0.5$), the 
   other panels show the results for model 2 (time-dependent $f_{\rm tot}$) with the FRDM nuclear 
   mass model (middle) and the DZ31 nuclear mass formula (right). We could have used $m_{\rm ej,max}$ here to set an upper limit but the mass dependence in Eq.~(\ref{eq:scaling_Lpeak_mass}) only has a minor impact on the peak magnitude.  
   }
   \label{fig:nsns_model1_model2}
\end{figure*}

\begin{figure*} %  figure placement: here, top, bottom, or page
\centerline{
  \includegraphics[width=6.5cm,angle=0]{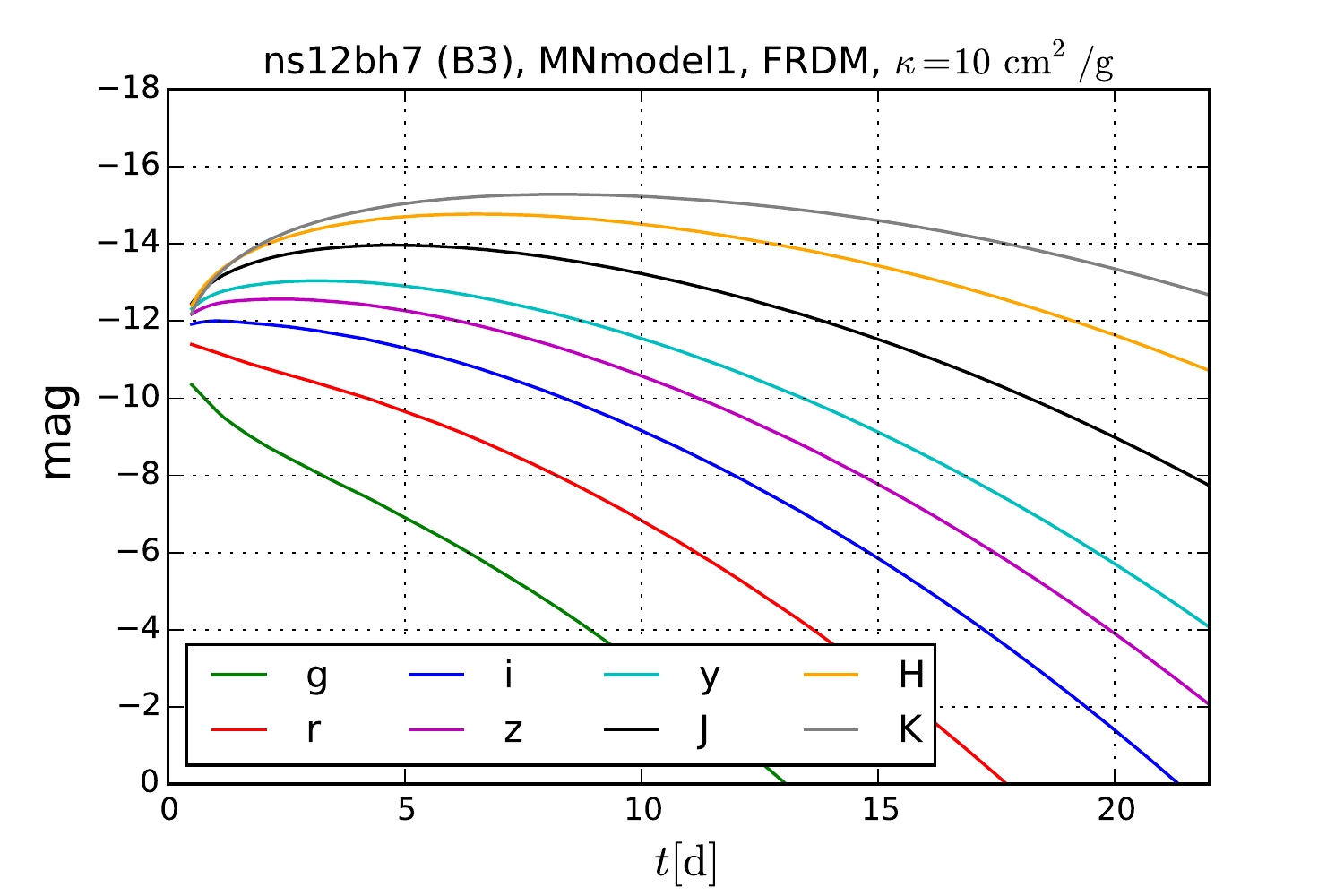}  \hspace*{-0.3cm}
  \includegraphics[width=6.5cm,angle=0]{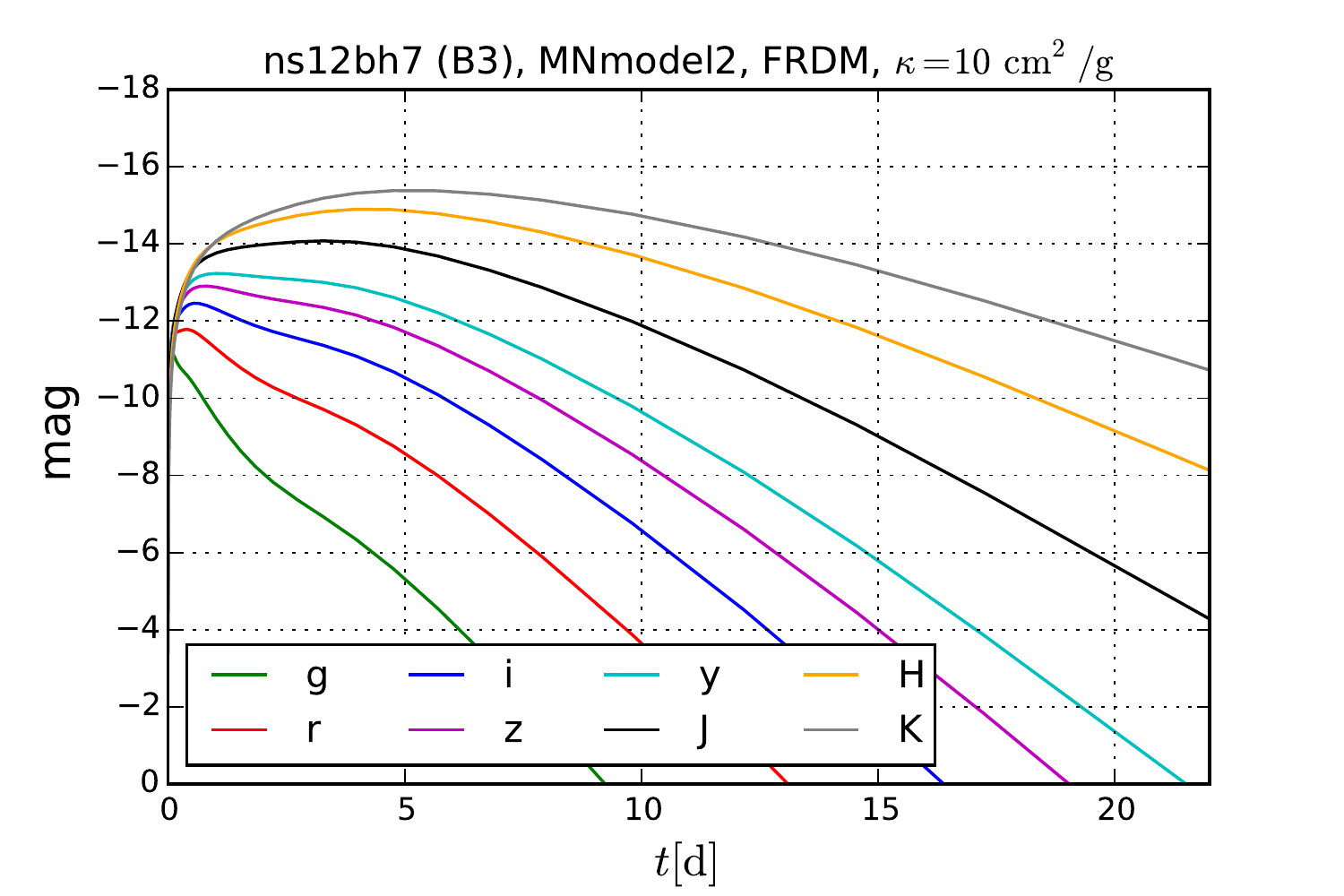}  \hspace*{-0.3cm}
  \includegraphics[width=6.5cm,angle=0]{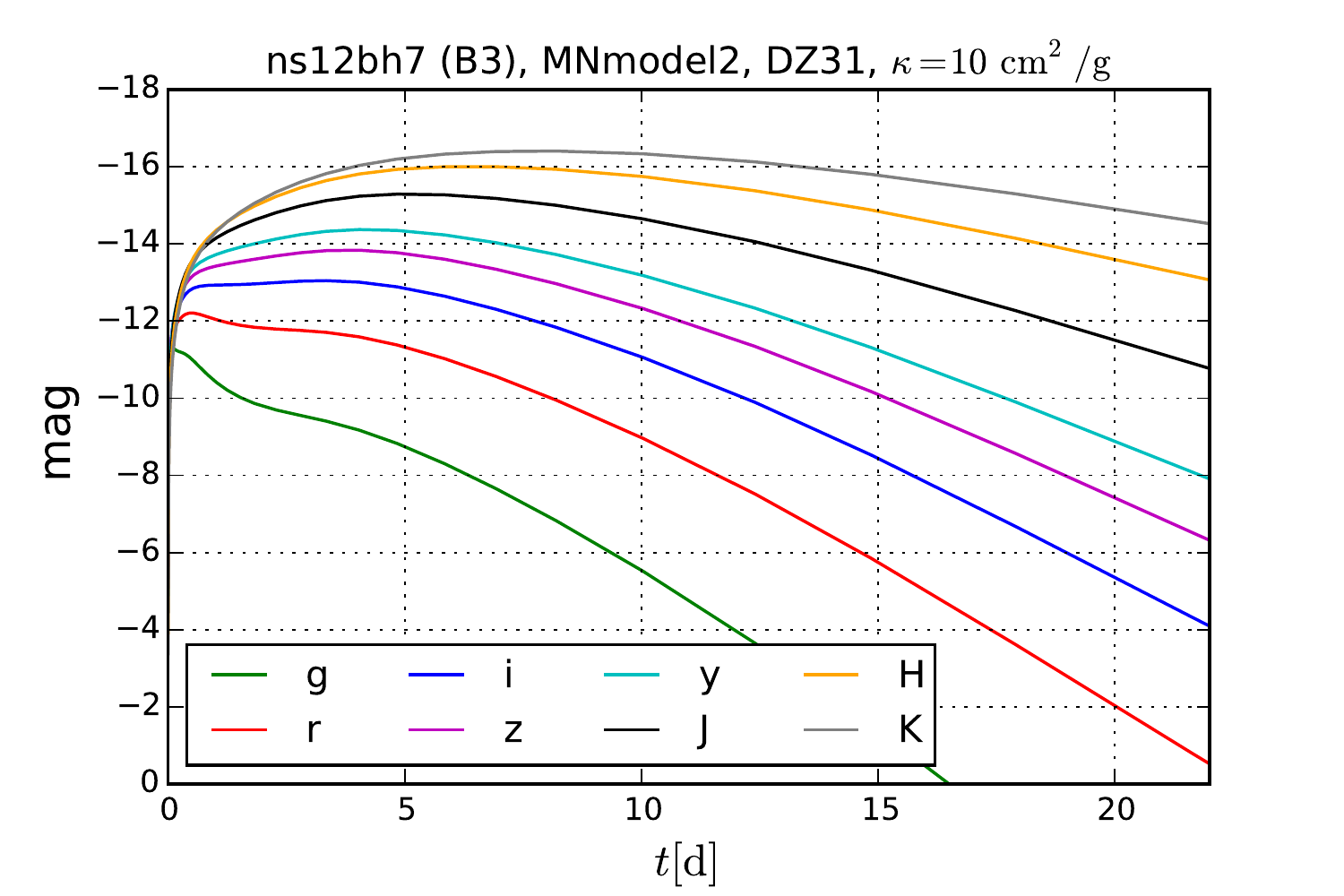}  
  }

   \caption{Comparison of the light curves in different bands for the dynamic ejecta of
   our brightest neutron star black hole system (simulation B3). The left panel shows the results for
   macronova model 1, the middle panel for model 2 with the FRDM nuclear mass model and
   the right panel shows the results of model 2 with the DZ31 nuclear mass formula.
   }
   \label{fig:nsbh_model1_model2}
\end{figure*}

One of the major objectives of our work is to explore the detectability of macronovae using current and future wide field-of-view optical and near-IR
facilities, either as a result of follow-up of LIGO triggers or through independent transient searches.
Figures \ref{fig:nsns_model1_model2} and  \ref{fig:nsbh_model1_model2} show the expected optical and near-IR lightcurves in absolute magnitudes for the dynamical ejecta of our brightest  NSNS (N5) and NSBH (B3) mergers, respectively.  Each time, we show the results from our macronova model 1 ('MNmodel1'; see Sec.~\ref{sec:MNmodel1}) and for model 2 ('MNmodel2'; see Sec.~\ref{sec:MNmodel2}), once for the FRDM and once for
the DZ31 nuclear mass model.
Throughout this work we use LSST $grizy$ filters and 2MASS $JHK$ and magnitudes in the AB-system. 

\begin{figure*} %  figure placement: here, top, bottom, or page    
 \centerline{
 \includegraphics[width=6.5cm,angle=0]{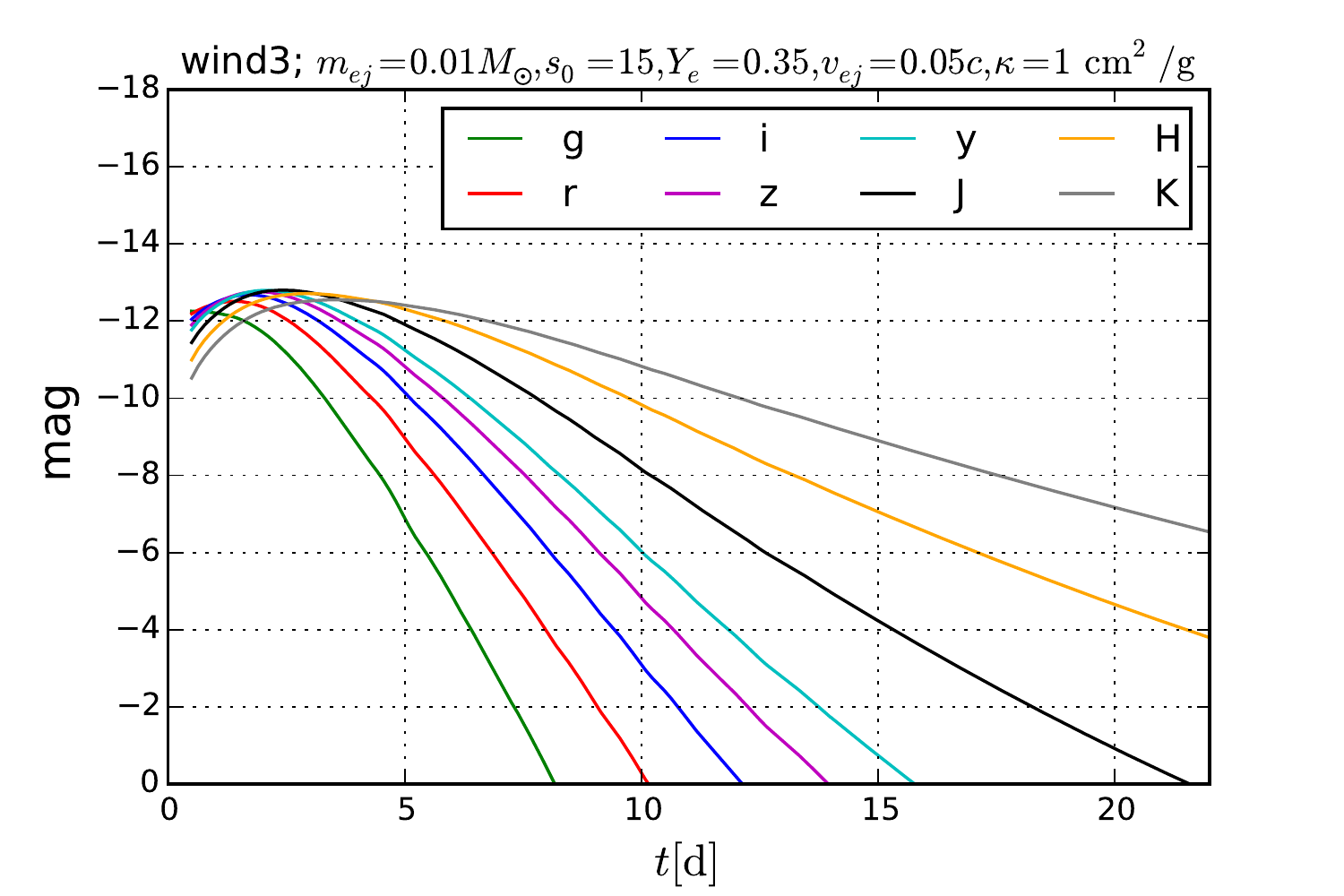} \hspace*{-0.3cm}
 \includegraphics[width=6.5cm,angle=0]{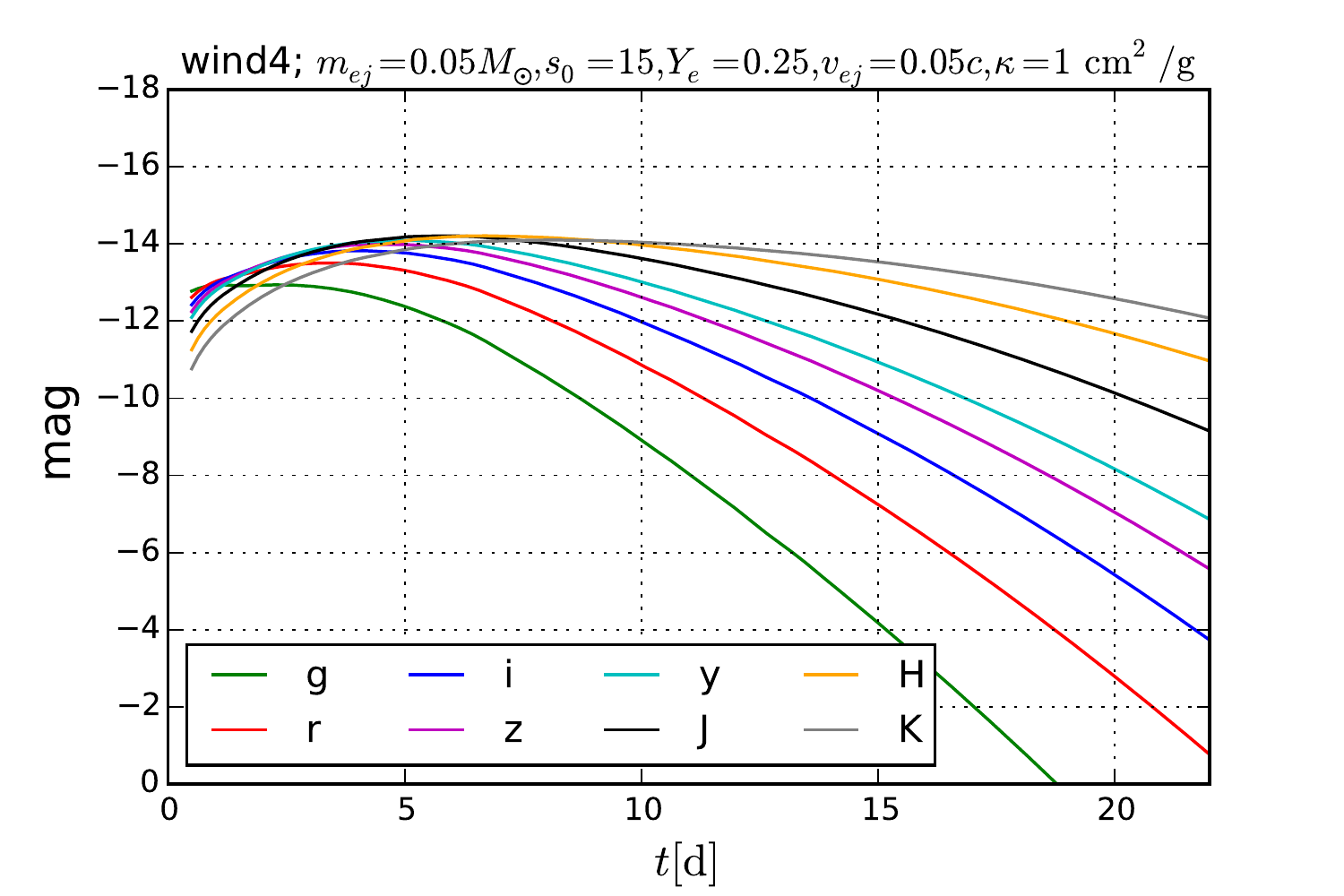} \hspace*{-0.3cm}
 \includegraphics[width=6.5cm,angle=0]{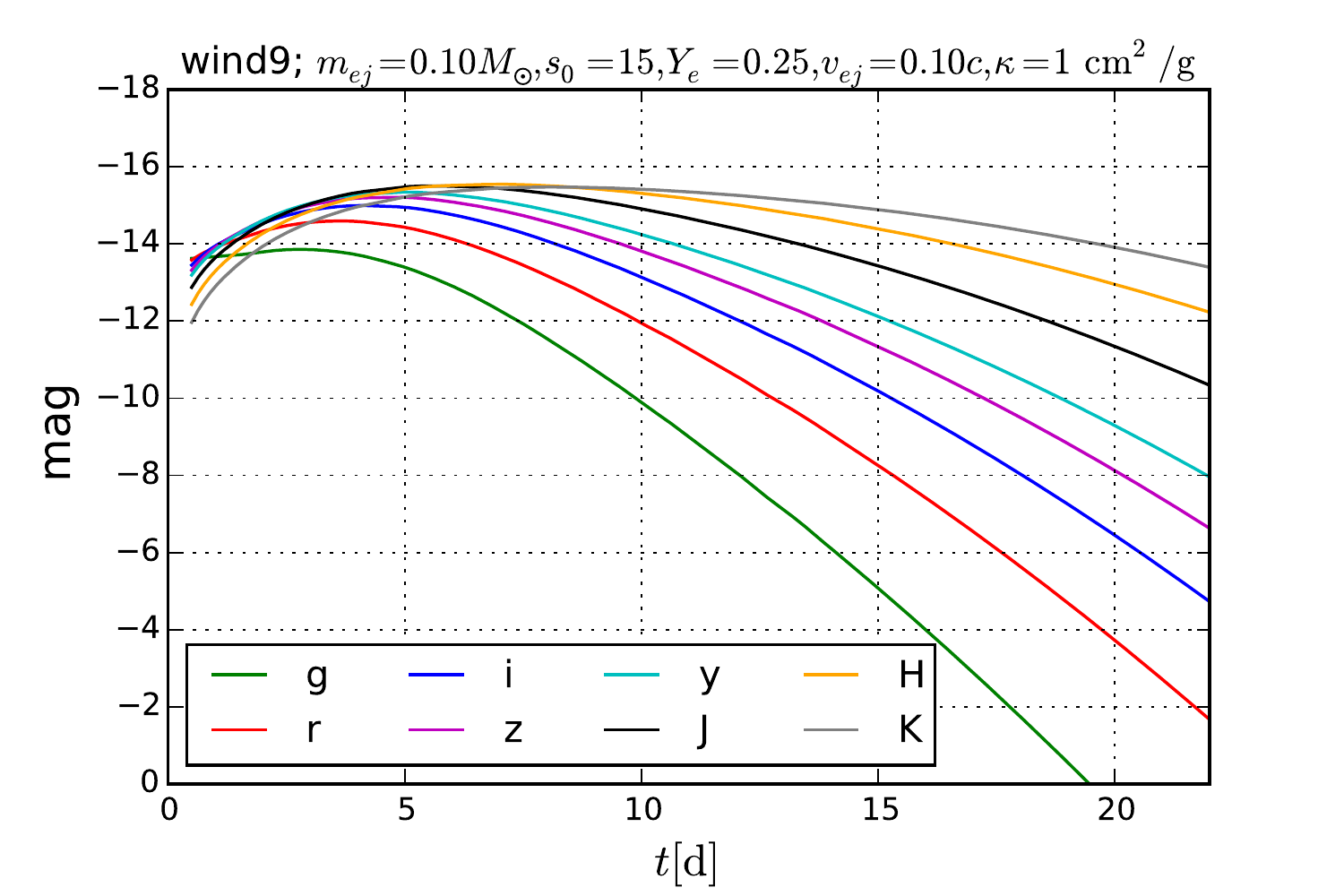}  }
  
   \caption{Lightcurves of selected wind models in different bands. The parameters of 'wind3' (left) are inspired by neutrino-driven winds in the aftermath of 
   a NSNS merger, those of 'wind4' (middle) by unbound torus material (stages IV, VI and IX in Fig.~\ref{fig:mass_loss}) and 'wind9' is similar to 'wind4', but with slightly more optimistic parameters.}
   \label{fig:MC_winds}
\end{figure*}

The dynamic ejecta in N5 has a mass ratio that deviates
substantially from  unity ($q=0.78$), but is consistent with the currently known mass ratios of
NSNS binaries. For example, the observed value of  J0453+1559 is $q=0.75$ \citep{martinez15} and the recently discovered  PSR J1913+1102 could have an even lower mass ratio
\citep{lazarus16}.
The left panel of Fig.~\ref{fig:nsns_model1_model2} shows the results for the macronova model 1, the middle panel shows MNmodel2 with the FRDM
mass formula and the right panel refers to MNmodel2 with DZ31. The general trends with a fainter and faster lightcurve in the bluer bands is apparent, while the near-infrared (NIR) 
lightcurves can stay bright for several weeks. We typically have $-11.5$
at peak in the $g$ band versus $-13.8$ in the $K$ band. The MNmodel2 results 
are about 0.7 magnitudes brighter at peak, but decay faster at later times. 
Both effects are
mainly due to the time variation of the thermalization efficiency, see Fig.~\ref{fig:thermalization}. As expected 
from the enhanced net heating rate at late times (see Fig.~\ref{fig:heating_rate}) the DZ31 mass model yields 
peak magnitudes that are another 0.8 magnitudes brighter than for the FRDM case. At the same time, both runs using MNmodel2 are significantly redder in the optical, being about one magnitude fainter at peak in the $g$ band. Additionally, their $g$-band lightcurves peak much earlier -- only half a day after the merger versus about three days for MNmodel1.\\
%----------------------------------------------------------------
%nsbh
Figure~\ref{fig:nsbh_model1_model2} shows the predictions for run B3 (1.2 \Msun NS and
a 7.0 \Msun BH with a dimensionless spin parameter $\chi=0.9$). The BH mass of 7 \Msun is close to the expected peak of the 
BH mass distribution \citep{oezel10}, but the spin is admittedly high. However,  if we are interested in NSBH systems that are able to launch
a short GRB, we need a large BH spin
($\chi \approx 0.9$ ) in the first place in order to form an accretion torus.
Otherwise the neutron star is essentially swallowed whole 
(see, for example, the discussion in Sec. 5.3 of \cite{rosswog15a} and references therein) and no GRB can be launched.  Moreover, measured BH spins in high-mass X-ray binary systems tend to have large spin values ($\chi >$ 0.85), and these systems are the likely progenitors of NSBH binaries
\citep{mcclintock14}.
\\
In the most favorable NSBH  case (MNmodel2, DZ31) a $K$-band peak magnitude 
brighter than $-16$ is reached.  The $g$ band reaches a similar brightness 
as for the NSNS case, but declines on a faster time scale for MNmodel1. Again, the objects are quite red, with the brightest magnitudes and more 
long lived light curves in the NIR.\\
%----------------------------------------------------------------
%--------------------------------------------------------------------------
Figure~\ref{fig:MC_winds} shows the lightcurves for selected wind models (MNmodel1). The
parameters of the first shown model (wind3) are guided by the simulations of the neutrino-driven
winds in the aftermath of a 1.4-1.4 \Msun merger by \cite{perego14b}, where the central remnant survives for at least
a few hundred milli-seconds before collapsing into a BH. With only very few available wind 
simulations, we consider this for now as representative for neutrino-driven winds from a NSNS 
merger. We consider the parameters of the wind model shown in the second panel 
(wind4)
as representative
for unbound accretion disk material 
\citep{fernandez13b,just15,wu16}. The last shown wind model (wind9) is a rather extreme case where 0.1 \Msun
is ejected at 0.1c, while still having a low opacity due to the large electron fraction ($Y_e= 0.25$; the lanthanide fraction in this case is $\sim 3.5 \times 10^{-5}$, see column six in Tab.~\ref{tab:winds}).\\
For an efficient comparison, we have summarised the  basic properties of the lightcurves for all models considered in Tab.~\ref{tab:lc_peak_width}. In addition to the peak magnitudes in LSST $r$ and 2MASS $J$ band, we calculated the time for which the lightcurve is within 1~magnitude of peak in those filters. All the 
calculated light curves  available at:  \texttt{http://snova.fysik.su.se/transient-rates/}.

\begin{table*}
 \centering
 \begin{minipage}{140mm}
\begin{tabular}{@{}lrrrrc@{}}
\multicolumn{5}{c}{\bf Lightcurve parameters}\\
\hline
     run          & $r_{\rm max}$ & $\Delta{t_{1{\rm mag}}} \; (r)$ & $J_{\rm max}$ & $\Delta{t_{1{\rm mag}}} \; (J)$ & comment\\
\hline
\multicolumn{6}{c}{\bf MNmodel1, FRDM}\\
\hline
ns12ns12 (N1) & -11.24 &  3.59 & -12.75 &  6.34 &\\
ns13ns13 (N2) & -11.31 &  4.64 & -12.91 &  7.79 \\
ns14ns14 (N3) & -11.21 &  3.94 & -12.72 &  6.80 \\
ns12ns14 (N4) & -11.48 &  5.49 & -13.05 &  8.78 \\
ns14ns18 (N5) & -12.36 &  8.13 & -13.64 & 11.59 \\
ns14bh7 (B1)  & -11.21 &  2.46 & -13.51 &  7.49 &\\
ns14bh7 (B2)  & -11.17 &  3.28 & -13.62 &  9.26 &\\
ns12bh7 (B3)  & -11.40 &  2.65 & -13.97 & 10.08 &\\
wind1         & -13.09 &  2.96 & -13.24 &  5.60 &\\
wind2         & -13.11 &  4.14 & -13.52 &  7.79 &\\
wind3         & -12.51 &  2.57 & -12.80 &  4.47 &\\
wind4         & -13.51 &  6.66 & -14.21 &  9.82 &\\
wind5         & -13.30 &  4.75 & -13.73 &  8.40 &\\
wind6         & -13.50 &  4.96 & -13.99 &  6.82 &\\
wind7         & -14.37 &  5.45 & -15.17 &  8.46 &\\
wind8         & -13.60 & 11.02 & -13.46 &  9.68 &\\
wind9         & -14.60 &  6.70 & -15.50 &  9.62 &\\
wind10        & -13.44 &  2.89 & -14.07 &  6.33 &\\
wind11        & -11.76 &  2.30 & -14.09 &  5.43 & $\kappa=10 \; {\rm cm^2/g}$\\
wind12        & -12.29 &  0.93 & -14.26 &  3.63 & $\kappa=10 \; {\rm cm^2/g}$\\
wind13        & -12.41 &  8.22 & -12.41 & 11.02 &\\
wind14        & -13.33 &  5.27 & -13.82 &  9.91 &\\
wind15        & -12.40 &  8.56 & -12.44 & 10.93 &\\
wind16        & -13.36 &  5.77 & -13.91 & 11.97 &\\
wind17        & -14.63 &  7.90 & -15.67 & 10.95 &\\
wind18        & -11.34 &  5.51 & -11.34 &  7.69 &\\
wind19        & -14.18 &  6.88 & -13.91 & 12.30 &\\
wind20        & -15.08 &  4.56 & -16.33 &  7.20 &\\
wind21        & -15.24 &  4.85 & -16.23 &  7.77 &\\
\hline
\multicolumn{6}{c}{\bf MNmodel2, FRDM}\\
\hline
ns12ns12 (N1) & -11.28 &  1.72 & -12.97 &  3.71 &\\
ns12ns14 (N4) & -11.31 &  2.18 & -13.15 &  5.35 &\\
ns13ns13 (N2) & -11.24 &  2.07 & -13.06 &  4.87 &\\
ns14ns14 (N3) & -11.23 &  1.87 & -12.93 &  4.07 &\\
ns14ns18 (N5) & -11.27 &  3.39 & -13.65 &  6.85 &\\
ns14bh7 (B1)  & -11.69 &  1.55 & -13.69 &  5.08 &\\
ns14bh7 (B2)  & -11.61 &  1.77 & -13.79 &  6.67 &\\
ns12bh7 (B3)  & -11.78 &  1.42 & -14.08 &  7.01 &\\
\hline
\multicolumn{6}{c}{\bf MNmodel2, DZ31}\\
\hline
ns12ns12 (N1) & -11.65 &  3.30 & -13.71 &  6.03 &\\
ns12ns14 (N4) & -11.61 &  4.48 & -13.95 &  7.49 &\\
ns13ns13 (N2) & -11.67 &  4.59 & -13.99 &  7.31 &\\
ns14ns14 (N3) & -11.59 &  3.42 & -13.64 &  6.22 &\\
ns14ns18 (N5) & -11.77 &  6.63 & -14.57 &  9.60 &\\
ns14bh7 (B1)  & -12.06 &  3.41 & -14.53 &  7.66 &\\
ns14bh7 (B2)  & -12.04 &  5.60 & -14.92 &  9.64 &\\
ns12bh7 (B3)  & -12.21 &  5.24 & -15.29 & 10.30 &\\
ns12bh7 (B3)  &  -8.32 &  0.22 & -10.80 &  7.76 & $\kappa=100 \; {\rm cm^2/g}$\\
\end{tabular}
\caption{Lightcurve parameters in LSST $r$ and 2MASS $J$ bands. $r_{\rm max}$ and $J_{\rm max}$ are the absolute AB-magnitude at maximum, $\Delta{t_{1{\rm mag}}}$ is the time (in days), during which the lightcurve is within 1~magnitude of peak brightness.
The dynamic ejecta for the NSNS and NSBH models have $\kappa=10~{\rm cm^2/g}$
whereas the wind models have  $\kappa=1~{\rm cm^2/g}$
unless their lanthanide fraction should exceed a threshold value of $X_{\rm lan}> 10^{-3}$. These few cases are indicated in the table. We also explore in one case (last line) the hypothetical possibility that the opacities of lanthanide-rich material should be substantially larger ($=100$ cm$^2$/g).
}
\label{tab:lc_peak_width}
\end{minipage}
\end{table*}

%----------------------------------------------------------------

%------------------------wind lightcurves----------------------------- 

 \subsection{Detection feasibility}
Next, we estimate the expected number of observable macronovae by integrating the
expected rate of neutron star mergers (NSMs) $R_{\textrm{NSM}}$ over the comoving volume in which the resulting macronova is observable. 
The expected number  of macronovae
observable then becomes:
\begin{equation}
  \label{eq:mn-number}
  n_{\mathrm{MN}}=\int\limits_{z<z_\mathrm{max}}
  R_{\mathrm{NSM}}\left(1+z\right)^{-1}\textrm{d}V_{C},
\end{equation}
where $\textrm{d}V_{C}$ is the comoving volume element and $z_\mathrm{max}$ is the maximum redshift at which the macronova is brighter than $m_{\textrm{lim}}$.
For $R_{\textrm{NSM}}$ we  adopted
an "informed best guess"
value of $300~\textrm{yr}^{-1}
\textrm{Gpc}^{-3}$, see Fig.~\ref{fig:rate_constraints}. 
To study the detectability, the lightcurves in observer-frame 
$grizJHK$
were calculated from the time-dependent SEDs of each run
 using the Python package \texttt{sncosmo} \citep{Barbary:11938} to account for spectral redshift. 
Calculations of cosmological distances and volumes were performed using the
Python package \texttt{astropy} \citep{2013A&A...558A..33A}. 
We assumed a cosmology with H$_0=67~\rm{km~s^{-1}~Mpc^{-1}}$ and
$\Omega_{m}$ = 0.307 \citep{ade15}.

%-------------------------------------------------------------------------- 

\begin{figure*} %  figure placement: here, top, bottom, or page    
\center{
 \includegraphics[width=8.8cm,angle=0]{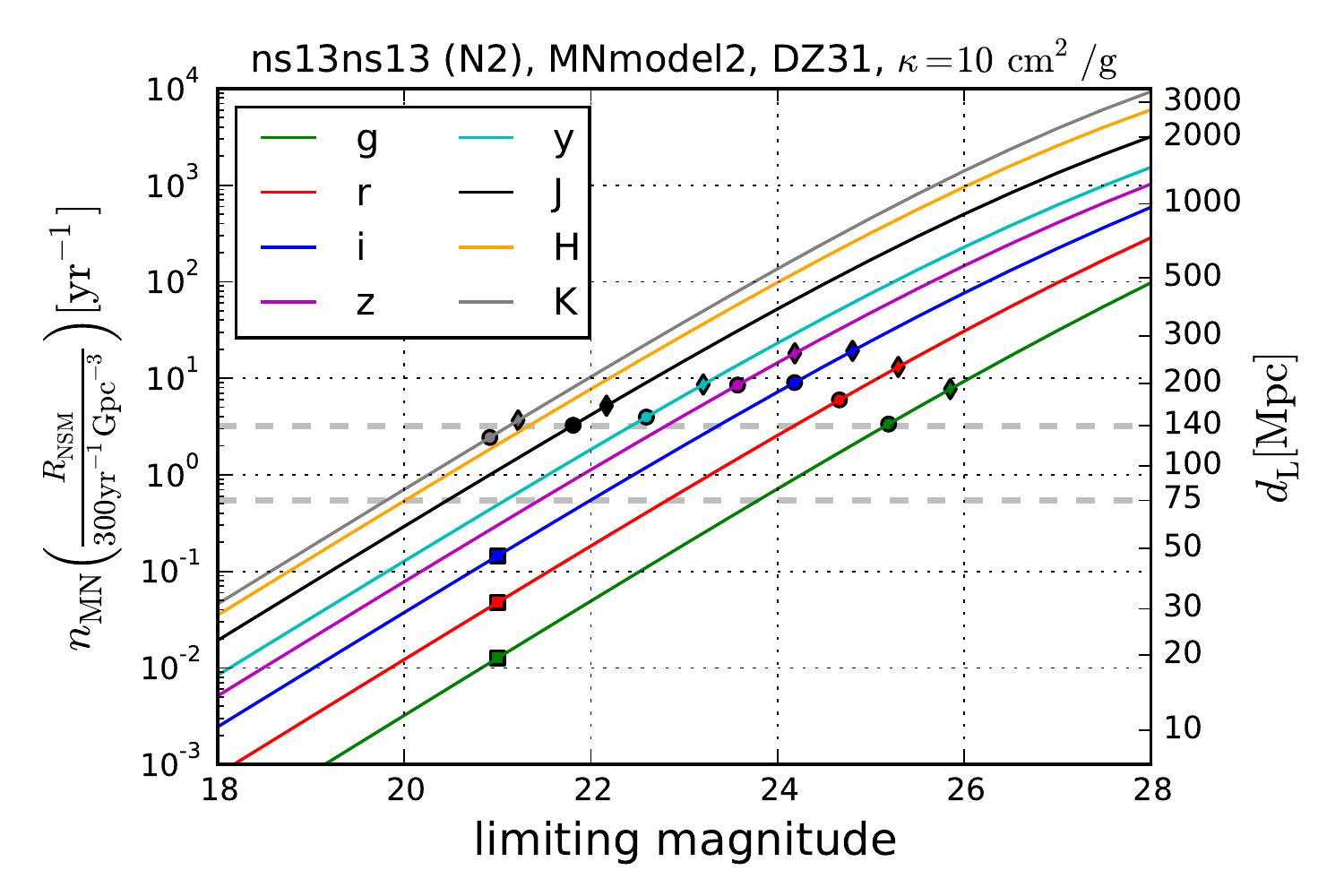} %\hspace*{-0.3cm}
 \includegraphics[width=8.8cm,angle=0]{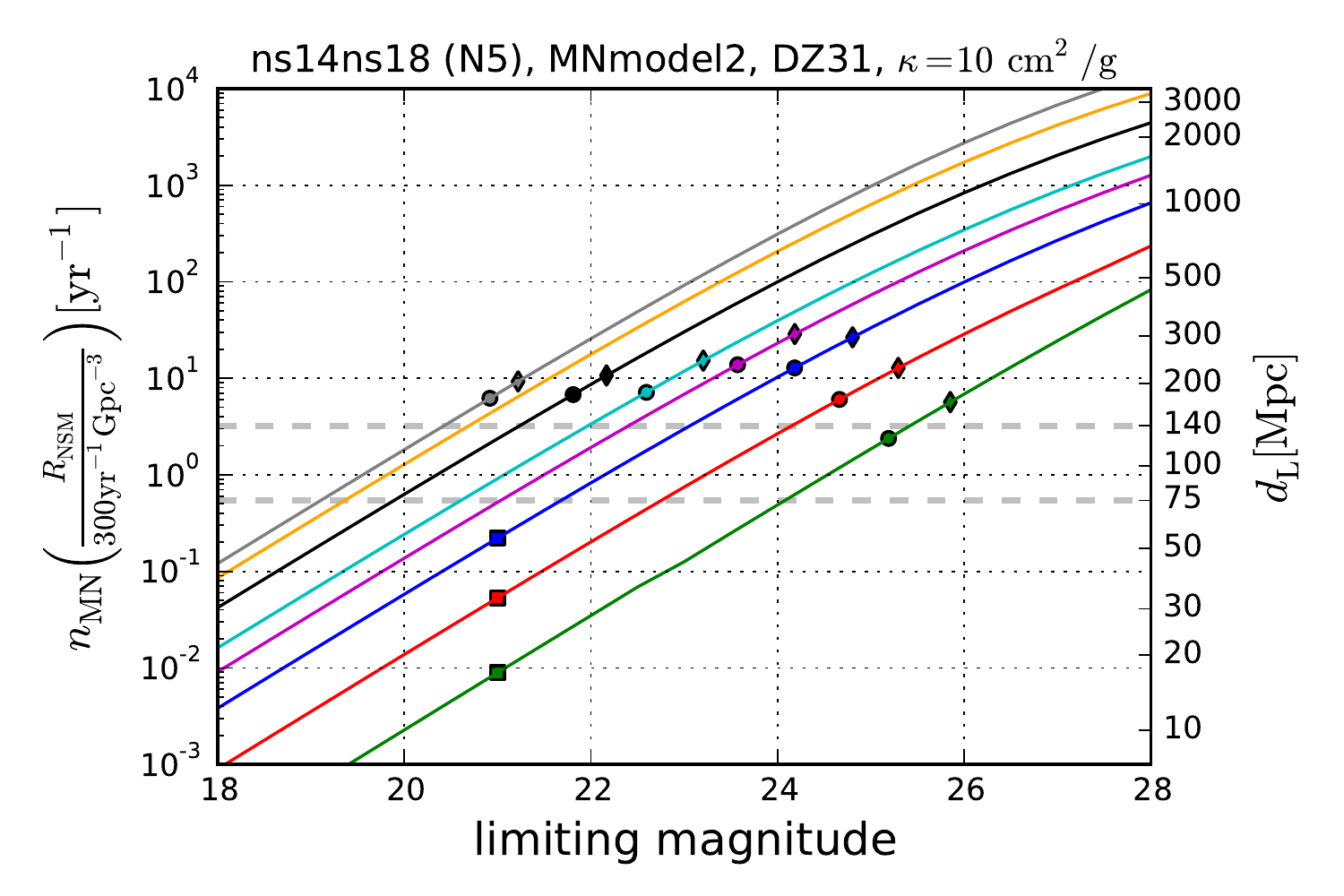} }
  
    %\vspace*{-5.cm}
      
   \center{
 \includegraphics[width=8.8cm,angle=0]{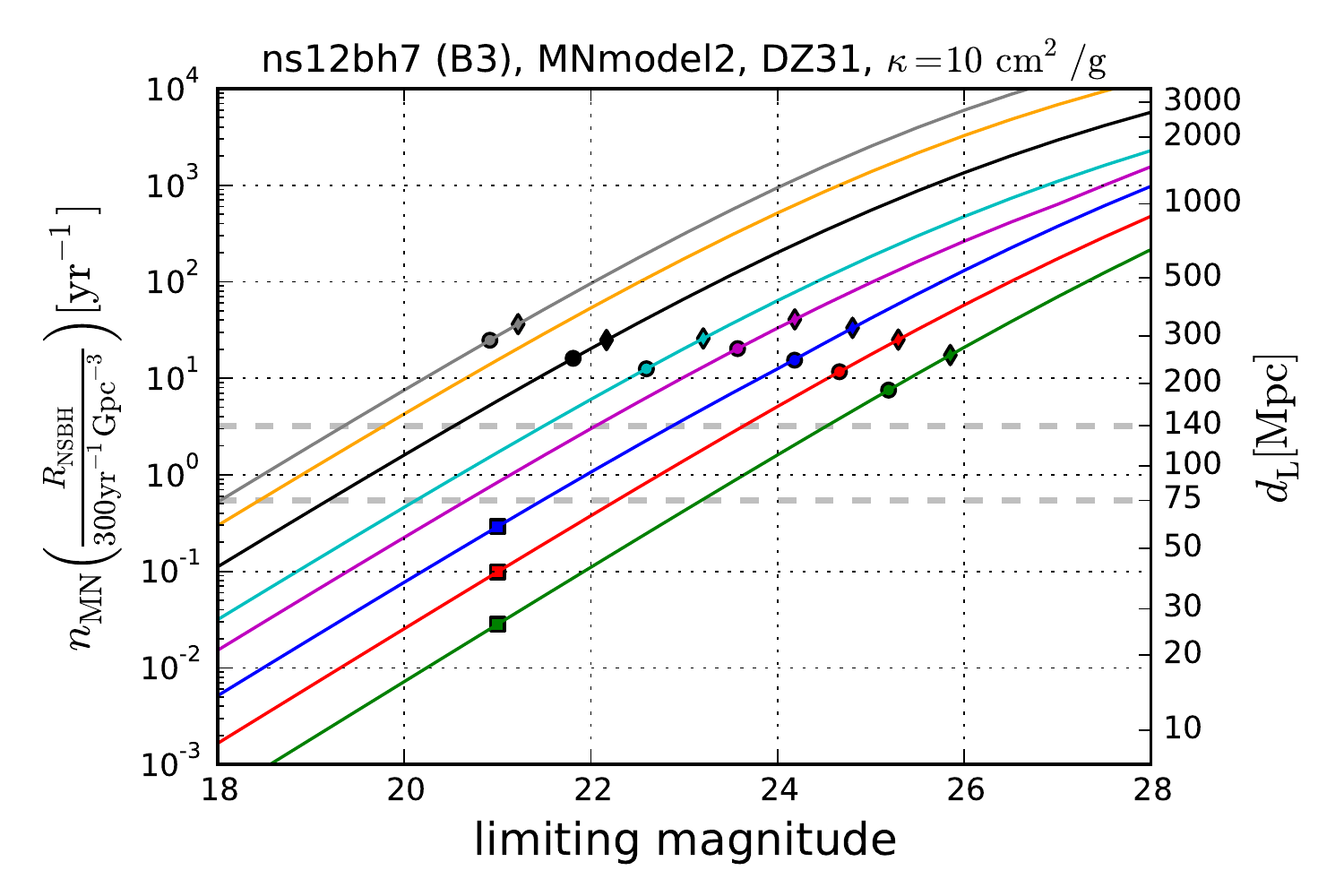} %\hspace*{-0.3cm}
 \includegraphics[width=8.8cm,angle=0]{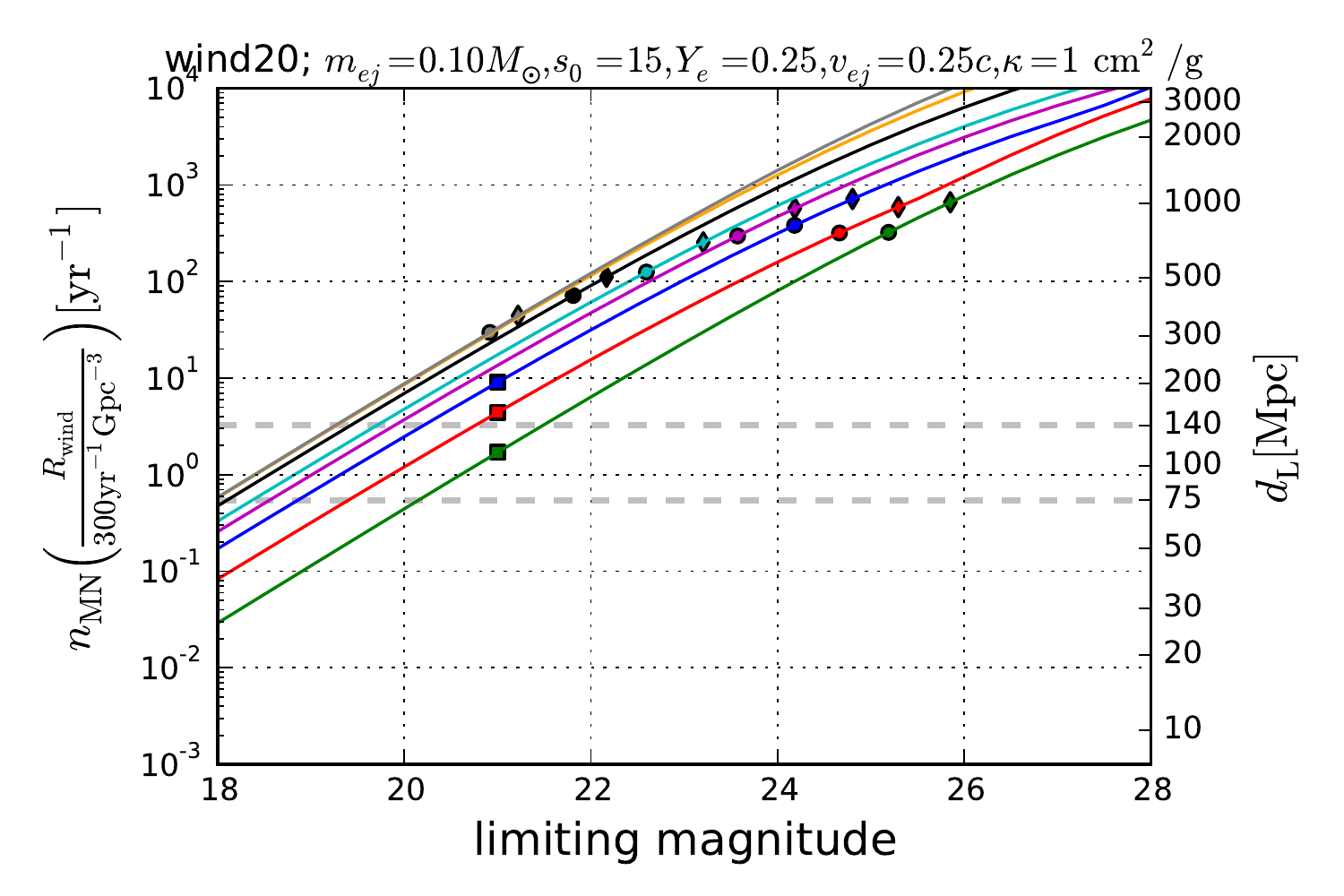}  }   
  
   \caption{Expected number of MNe for selected bright models for each of the three categories (for the dynamic ejecta we use the brighter results coming from the DZ31 mass formula). The markers show the expected depths for a 60-second (circle) and 180-second (diamond) exposures with VISTA ($J$\&$K$ band) or LSST ($grizy$). Square markers show the expected numbers for a depth of 21~mag in $gri$ as expected for ZTF.
  On the right-hand side the y-axis shows the luminosity distance up to which the NS merger rate was integrated to obtain $n_\mathrm{MN}$. The gray dashed lines show the LIGO range for discoveries of gravitational wave signals from NS-NS mergers (75~Mpc) and NS-BH mergers (140~Mpc). Note that  the results are scaled
  to the  "best guess" for the NSNS merger rate of
  300 yr$^{-1}$ Gpc$^{-3}$.
  }
   \label{fig:rate}
\end{figure*}

\begin{figure}  
\center{
 \includegraphics[width=8.8cm,angle=0]{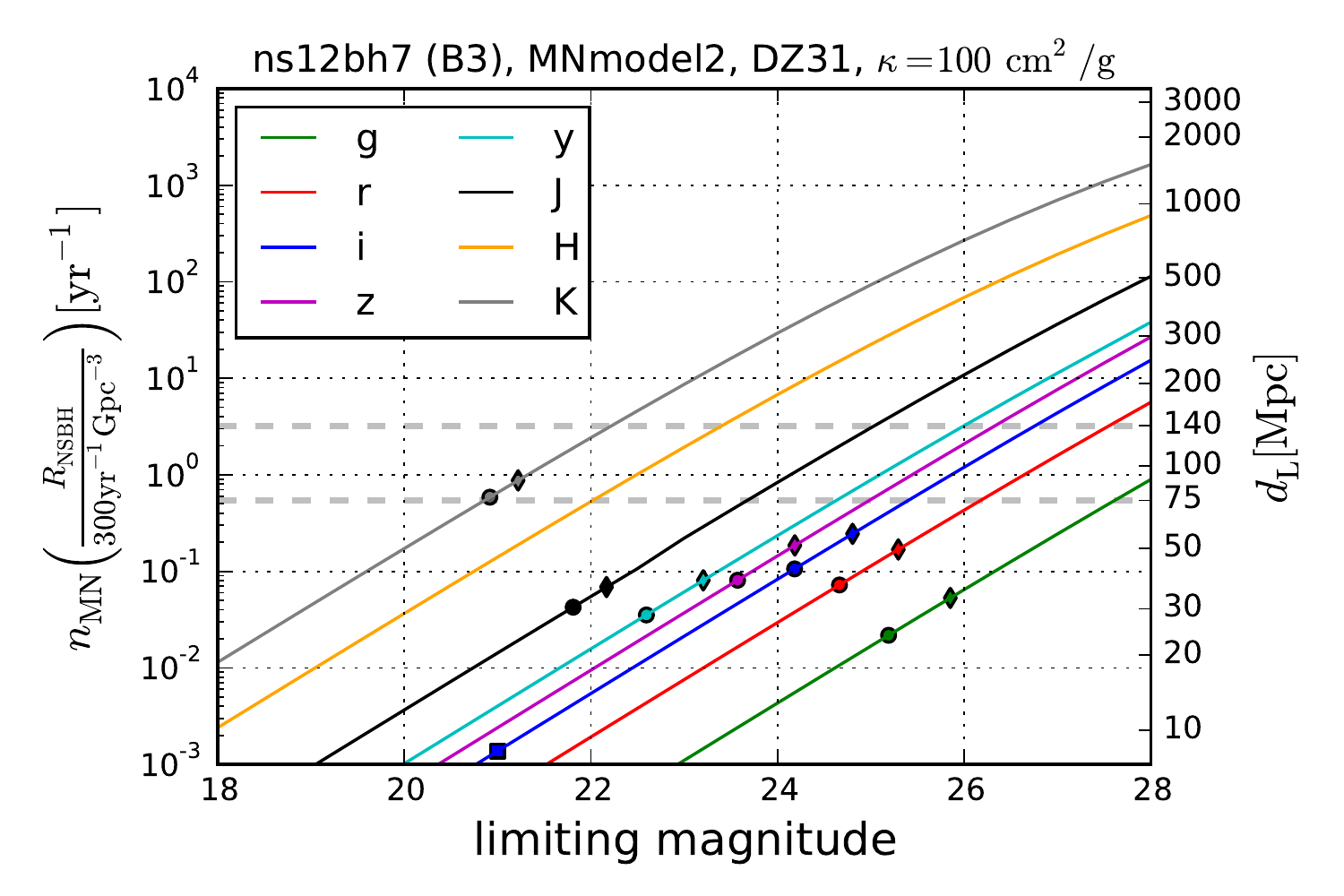}
 }
   \caption{Expected number of MNe for the brightest NS-BH merger model (B3) and the hypothetical case that the effective opacity should be as  large as $\kappa=100 \; {\rm cm^2/g}$ (using the results from the DZ31 mass formula). The markers show the expected depths as in Fig. \ref{fig:rate}. }
   \label{fig:rate-b3-kappa100}
\end{figure}

The four panels in Figure~\ref{fig:rate} show representative model predictions  for the expected number of MNe per year as a function of limiting AB magnitude over the whole sky, normalised to a nominal volumetric rate $R_{NSM} = 300$ yr$^{-1}$Gpc$^{-3}$ (left
axis) and the corresponding reach in distance (axis on the right).  Dashed lines at 75 Mpc and 140 Mpc show the quoted sensitivity limit for GW detections for 
LIGO run \citep{martinov16,abbott_LRR}
for NSNS and
NSBH respectively. 
We also indicate the limiting magnitudes of 60 and 180-second exposures for VISTA and LSST in $J$ and $K$~band and $grizy$, respectively. When determining the limiting magnitudes, we used the ESO Infrared Exposure Time Calculator\footnote{\texttt{https://www.eso.org/observing/etc/}} 
for VISTA  and a Python exposure time calculator for 
LSST\footnote{\texttt{https://github.com/lsst-sims/exposure-time-calc}},
assuming a target signal-to-noise ratio of 5. We also indicate the expected limiting magnitudes for the Zwicky Transient Facility (ZTF) $gri \sim$ 21. 

Here we have concentrated on a selection of the available models (N2, N5, B3 with DZ31 and wind20). Given the assumed rate of NSNS mergers, we can deduce the number of observable events for any given survey. We have done the calculations for these two given existing/upcoming facilities, but the community can use this for any of their favorite surveys. These calculations are also provided  under   \texttt{http://snova.fysik.su.se/transient-rates/}.

Figure~\ref{fig:rate} also allows to read off the relative efficiency 
to detect NSNS mergers in the different passbands. The lightcurves 
show that more flux is expected for longer wavelengths, indicating that 
near-IR surveys would be particularly suitable. 
However, given the difficulty for ground based instruments to observe at these wavelengths, it is reassuring that also optical telescopes have a good
sensitivity for macronova detections. The figure shows that in this particular case, one minute exposures in the optical bands ($i$) can be as efficient in detecting the merger. 

Fig.~\ref{fig:rate} is based on the best currently available opacity information.
The opacities for dynamic ejecta, however, have just begun to be explored and
are therefore not very accurately known. We therefore explore also the hypothetical
case that the dynamic ejecta opacities should be as large as 100 cm$^2$/g, see
Fig.~\ref{fig:rate-b3-kappa100}. If such extreme opacities should be realized in nature, this would substantially deteriorate the detection prospect for the
dynamic ejecta.

The above feasibility study shows that there are indeed prospects for finding optical and NIR emission from macronovae with large telescopes. More than a handful of events would be expected each year, distributed over the entire sky. 
In fact, if a survey such as LSST would allow high enough cadence (about once per night) in multiple filters, one could even discover macronovae without a trigger.
Such un-triggered searches have already started, but are hampered by both 
difficulties in subtracting faint transients on potentially bright host 
galaxies, and by the background of supernovae \citep{doctor16}.
The most interesting prospects for finding them clearly comes in
connection with triggers from either GW signals or high-energy
emission (Gamma-ray bursts).

\subsection{Follow-up of LIGO triggers}

The optical follow-up of gravitational wave signals is currently a large
effort within the transient community \citep{abbott16c}.
We can now relate our model predictions with ongoing
surveys such as iPTF \citep{kasliwal16} 
and Pan-STARRS \citep{smartt16b}. 
These are typically conducted in the $r$ band, reaching limiting 
magnitudes of $\sim20-21$. This will be typical also for the upcoming very wide area 
search with ZTF.
For the most recent GW signals, at 
a distance of 400 Mpc, the detectability is illustrated 
by \citet[][their Fig. 9]{smartt16a}.

To such large distances, none of our models for dynamical ejecta becomes bright
enough to be detectable.
Even the most optimistic wind models are below the detection threshold of \cite{smartt16a} at these distances.
However, note that the current LIGO configuration can detect
NSNS mergers only out to 75 Mpc \citep{martinov16}. 
For such a distance, the wind model would be easily detected by the search such as that by \cite{smartt16a}, 
and also the more optimistic dynamical ejecta models (e.g., N5 MNmodel2 DZ31) 
could be detected by ZTF \citep{kasliwal16} if nearby enough.
With the rate of such events assumed above, we would expect 
about
one such event per year, and not all may be observable by optical telescopes. 

The larger telescopes studied here (LSST/VISTA) will more easily detect any such counterpart to a GW trigger, and as the sensitivity of LIGO for NSNS mergers increases from 75 to 200 Mpc, we will go from one event every second year to about 10 events per year, and these events would indeed be within range for these telescopes.

In this comparison we also note that the timescales for the
brighter models is several days, and not the very steep declines
envisioned by some groups 
\citep[see for example Fig. 3 in][and our discussion in the introduction]{kasliwal16}. 
This is of great importance in determining the observing strategy, and our
models suggest that a nightly cadence would be suitable, with several nights needed to establish variability. On the other hand, pursuing the search for much
more than 2 weeks, as in  \cite{smartt16a}, 
seems not to be warranted by these models.

We thus conclude that forthcoming surveys with the 8-meter LSST  ($\sim$10 sq.~deg field of view) and existing 4-meter VISTA ($\sim$2 sq.~deg) will be able to search for GW-EM counterparts for any expected trigger from LIGO involving at least one neutron star with relatively short exposures per field. With such surveys, search areas of hundreds of sq. deg may be searched in a single night. 

The  feasibility of detection of coincident events is therefore promising. For more nearby mergers within $\sim$30 Mpc, for
most of the models considered also surveys like ZTF would have an excellent chance of capturing the lighcturves of macronovae in multiple filters. 

  \subsection{GRB triggers for the macronova candidate}

Another way to get a trigger for a macronova event is via high-energy emission.
The best case for a macronova so far comes from \citep{tanvir13a}. The detection of the short  GRB 130603B is currently the best macronova candidate, the GRB had a fast decaying optical afterglow and was well placed within the host galaxy \citep{ugarte13} at redshift $z=0.356$. The Hubble Space Telescope (HST) imaging of \citep{tanvir13a} detected an $H$-band candidate at $25.73\pm 0.20$ (F160W) at 7 rest-frame days (9.5 in observer frame) past GRB. Simultaneous optical (F606W) observations did not detect the source down to 28.25 (95\%).

Comparison to all models shows which ones are bright enough to explain this event as a macronova. In Fig.~\ref{fig:wind_param_dm}, upper panel,
we display the magnitude in the mass-velocity plane (see bottom panel same figure) for MNmodel1 and FRDM (top) and MNmodel2 and DZ31 (bottom). The colors of the symbols indicate the magnitude difference with respect to the Tanvir et al. detection and thus show that several models are within a factor of 3 of the required luminosity (orange and red symbols). Therefore, if we interpret this event as a MN, we 
note that the opacities can not be very different from those used here.  
Even the otherwise most promising models (e.g. model B3) would not be able to account
for these observations if the effective opacities should be as large as
$\kappa=100~\textrm{cm}^2/\textrm{g}$).

\begin{figure*}     
 \center{
 \includegraphics[width=5.5cm,angle=0]{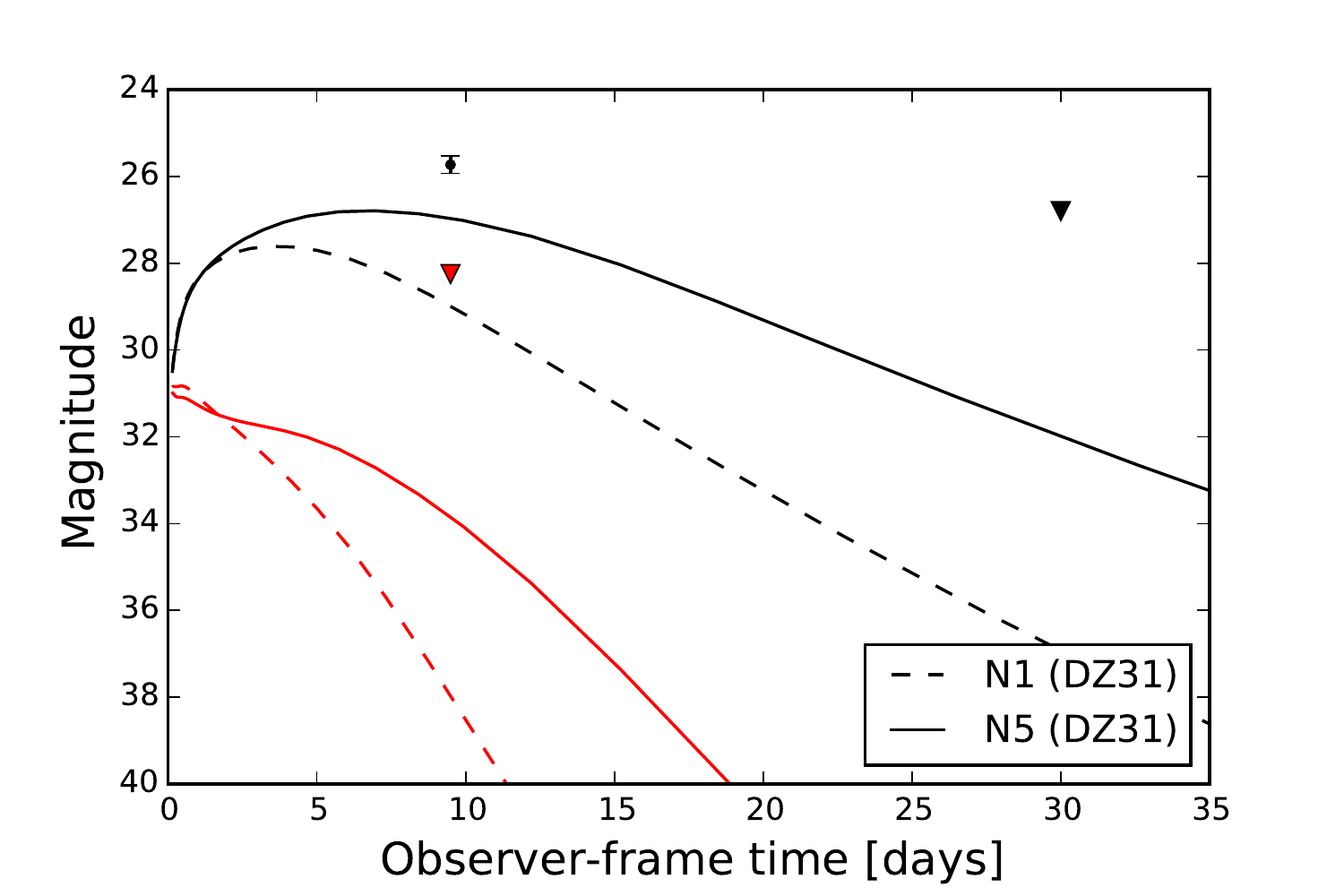} \hspace*{-0.3cm}
 \includegraphics[width=5.5cm,angle=0]{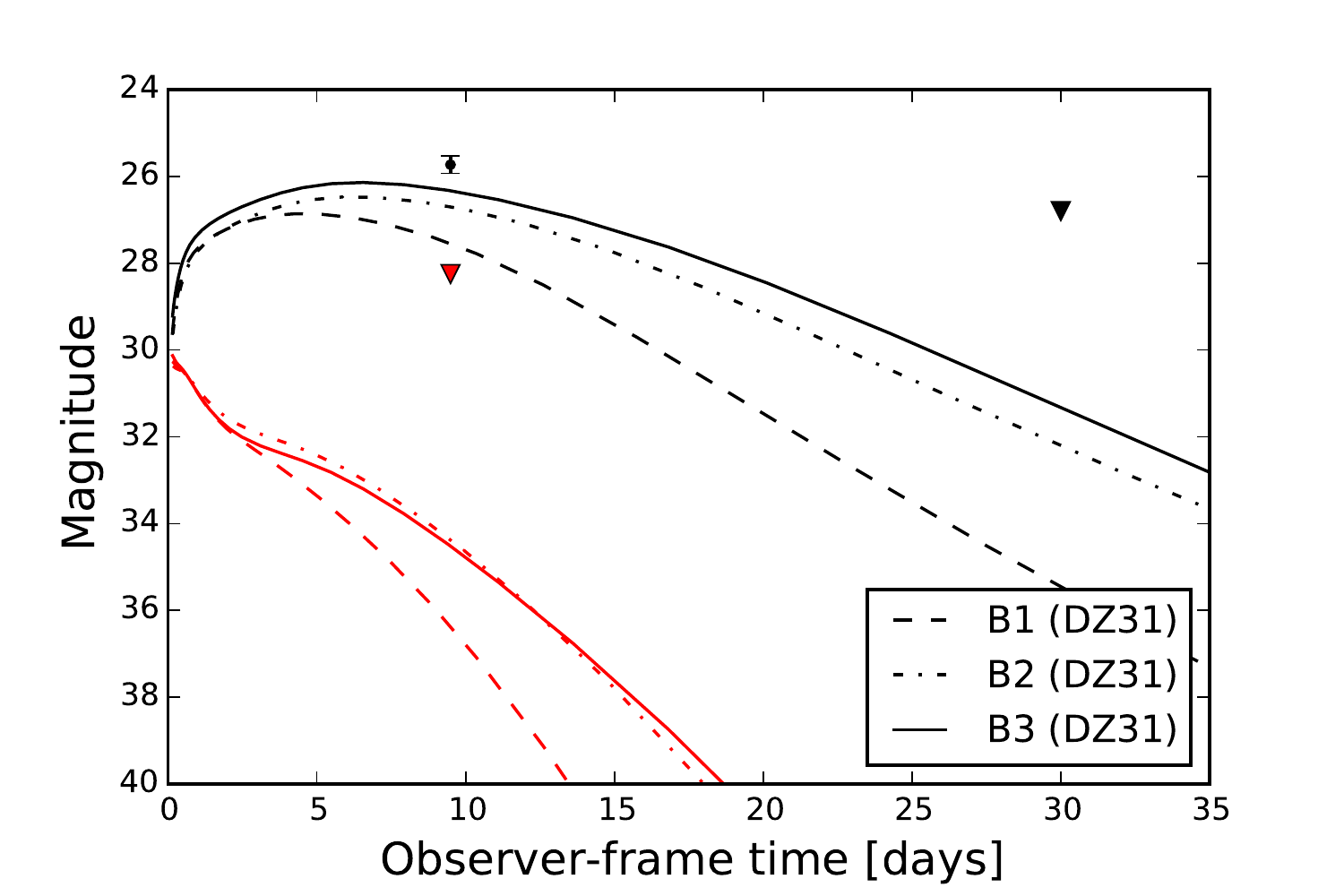} \hspace*{-0.3cm}
 \includegraphics[width=5.5cm,angle=0]{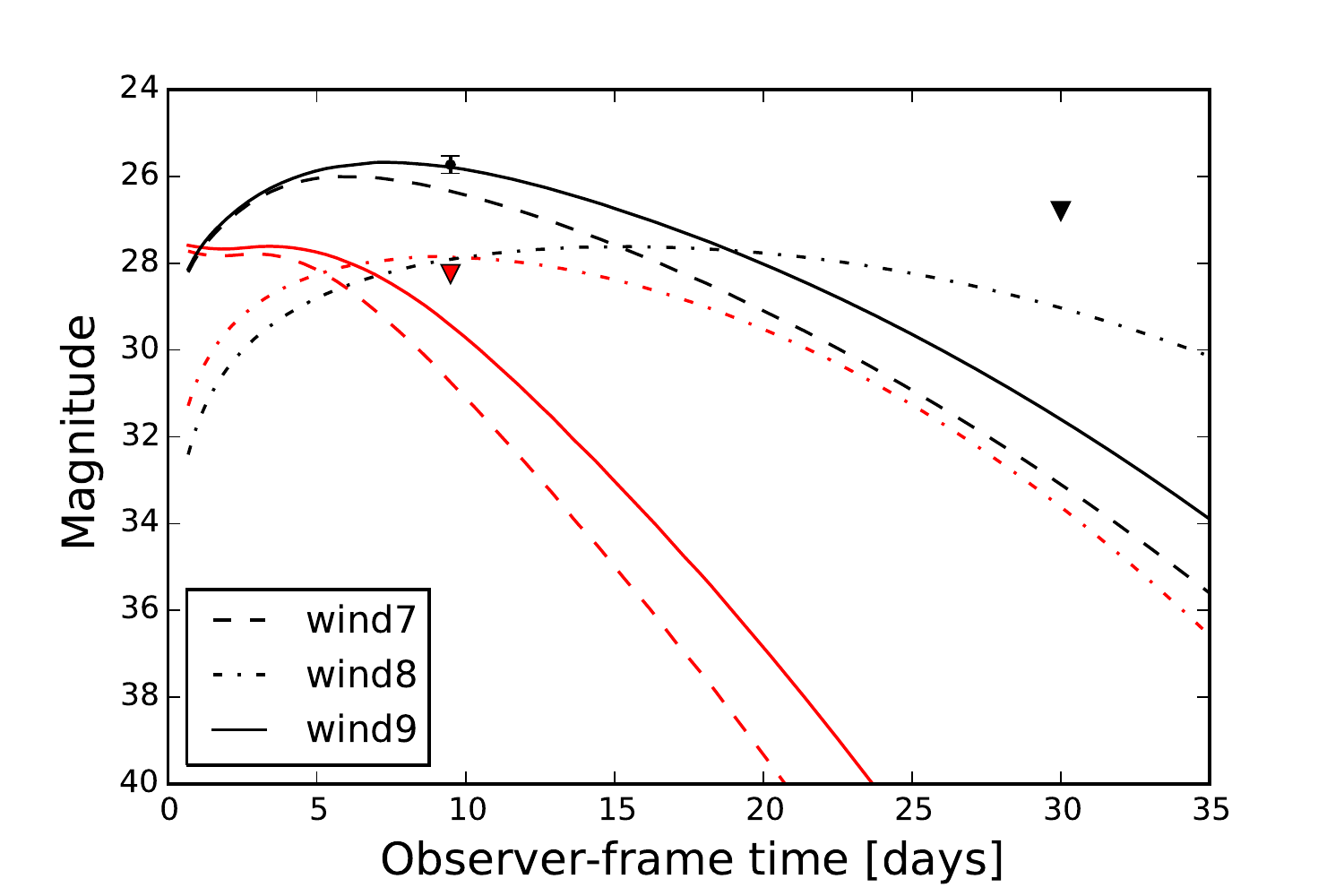}  }
  
   \caption{Lightcurves for selected models transformed to the redshift of GRB 130603B ($z=0.356$) in HST-band F160W (black) and F606W (red). The down-pointing triangles correspond to 95\% upper limits. 
   }
   \label{fig:lcs-hst}
\end{figure*}

\begin{figure} 
 \hspace*{-0.5cm}\includegraphics[width=9cm,angle=0]{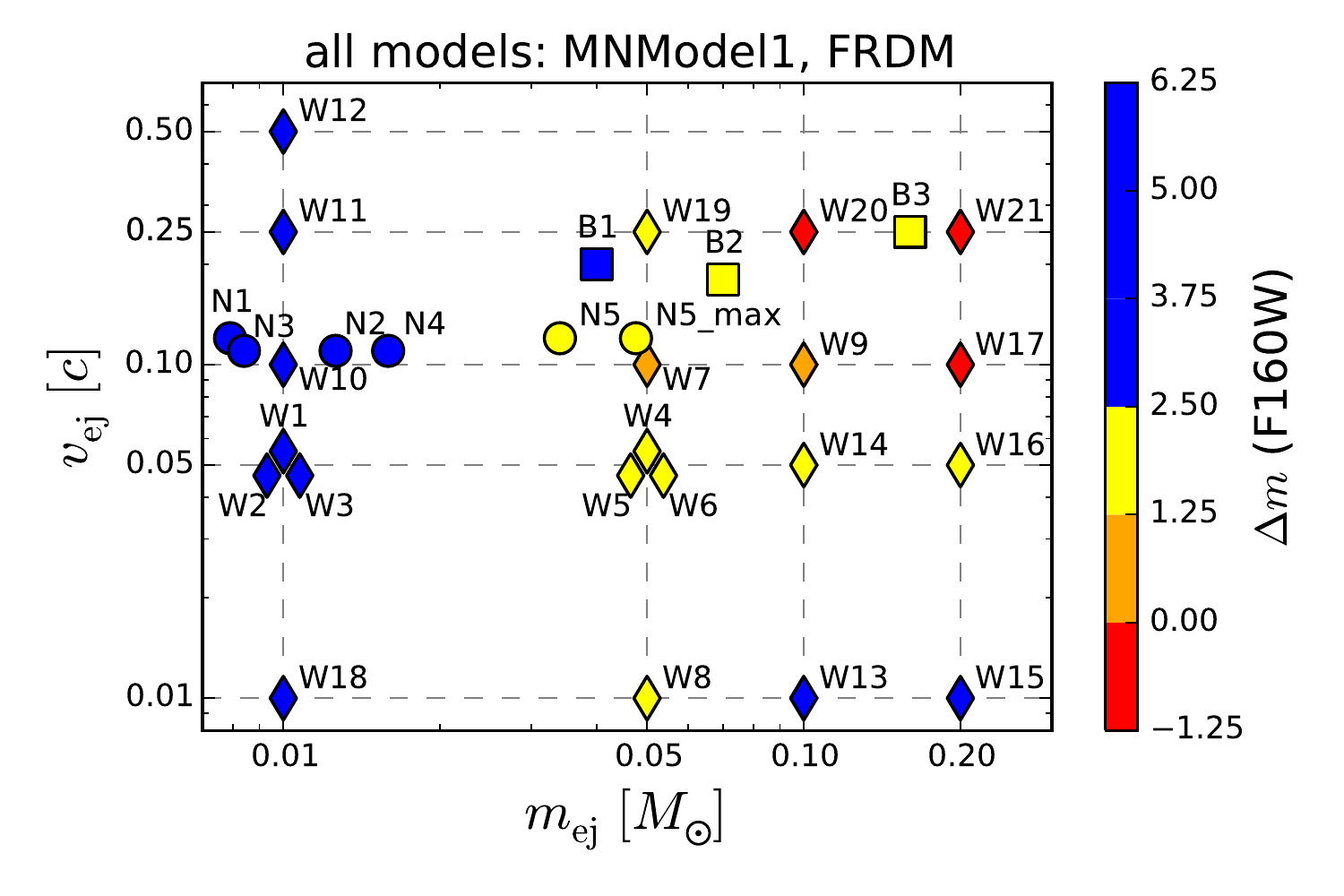}
 \hspace*{-0.5cm}\includegraphics[width=9cm,angle=0]{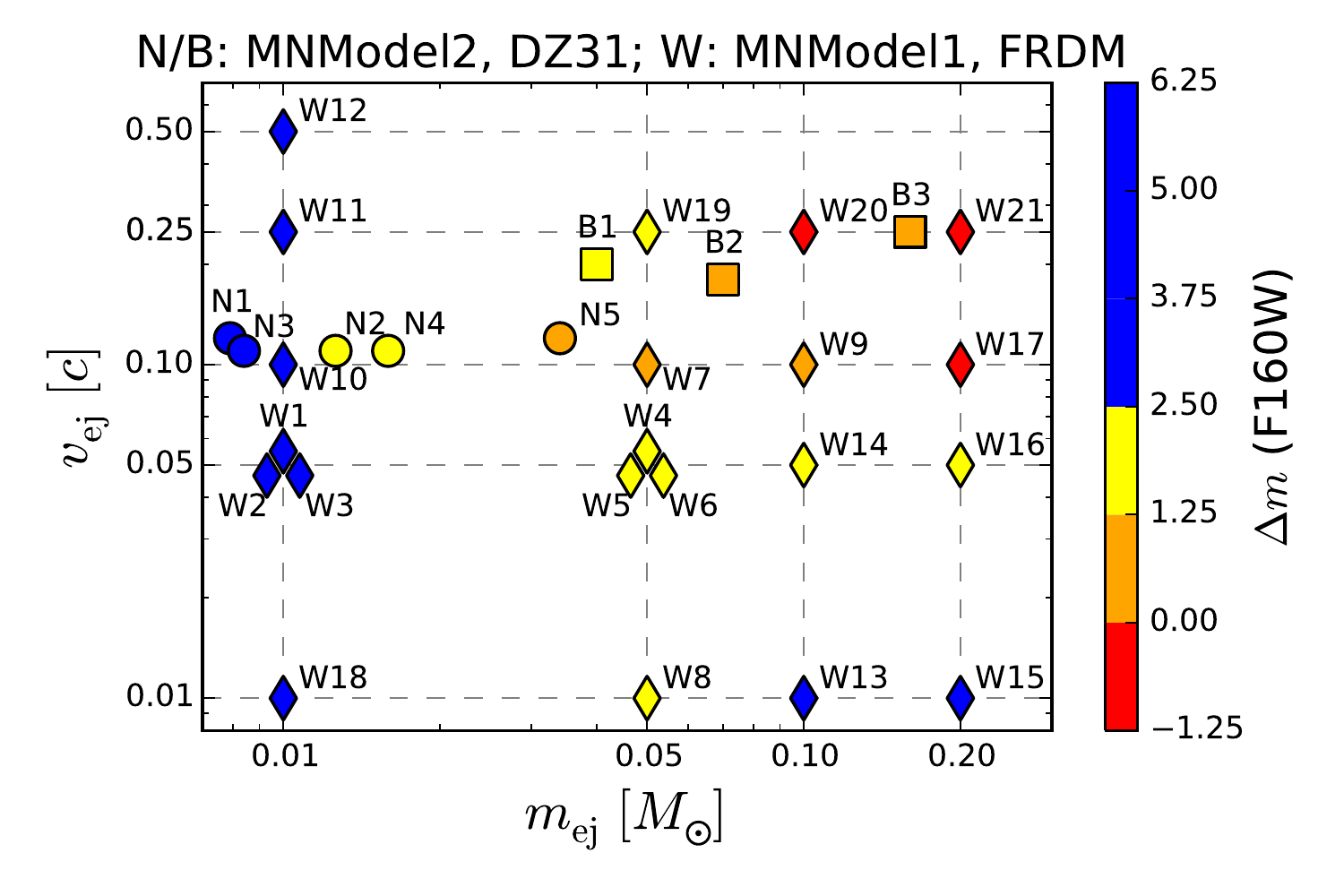}
 \caption{Comparison of our model predictions (same 
 parameter space in $m_{\rm ej}-v_{\rm ej}$ as 
 Fig.~\ref{fig:wind_param_space}) with  the H-band (F160W) observation in the aftermath of  GRB 130603B \citep{tanvir13a}. The wind models are shown as diamonds, NS-NS mergers as circles and NS-BH mergers as squares. Note that W1--3 and W4--6 correspond to models with the same mass and velocity, but with different electron fractions, see table~\ref{tab:winds}.  
 All wind models are calculated with MNmodel1 and FRDM,  the dynamic
 ejecta results for NSNS mergers (labeled "N*") and NSBH mergers (labeled "B*")  are  once calculated with  MNmodel1 and FRDM (upper panel) and
 once with  MNmodel2 and  DZ31 (lower panel).
}
\label{fig:wind_param_dm}
\end{figure}

The wind models provide sufficiently bright lightcurves for the macronovae. Most notably wind model 9 matches the $H$-band detection perfectly while model 7 is only $\sim~0.5$~mag too faint, which is not significant given the model uncertainties, see the right panel of Fig.~\ref{fig:lcs-hst}. 
The parameters of these models are close to what is expected for unbound accretion torus material (see stage IV, VI and IX in Fig.\ref{fig:mass_loss}).
The $R$-band upper limit, on the other hand, constrains the wind used for model 8, for which 27.81 is predicted. Perhaps even more interesting is that the brighter dynamical ejecta models also provide light curves that match this event, see left and middle panels of  Fig.~\ref{fig:lcs-hst}. 
%%%%%%%%%%%%%%%%%%%%%%%%%%%%%%%%%%%%%%%%%

\section{Summary and discussion}
%
% What have we done?
%
The major aim of this study was to explore the detectability 
of radioactively powered electromagnetic transients in the 
aftermath of compact binary mergers.
We have performed simulations of neutron star mergers to extract
hydrodynamic trajectories of the ejected material. These have
been complemented by trajectories representing neutron star black hole
mergers and various forms of wind outflows that have been treated in a
parametrized form. What we denote collectively as  'winds' can have
different physical origins such as neutrino-driven outflows or the
late-time release of accretion torus material.\\
Along these trajectories nuclear reaction network calculations were 
performed to extract the radioactive heating rates which, in turn,
serve as input for the subsequent macronova modeling. 
We have compared two different reaction networks
\citep{winteler12,mendoza_temis15} to ensure the robustness 
of our results.  Since the  r-process path for very 
neutron-rich ejecta meanders  through a region  of the nuclear chart
where no experimental information is available such calculations
are based  on theoretical nuclear models. We have therefore 
explored the difference between two frequently used nuclear mass
models, the Finite Range Droplet Model (FRDM) \citep{moeller95} 
and the 31-parameter model (DZ31) of \cite{duflo95}.
For the aspects explored in this manuscript the main differences 
for the observability come from different predictions for the amount 
of $\alpha$-decaying trans-lead nuclei.
We have further compared the results of two macronova models. 
The first one \citep{grossman14a} uses the FRDM mass model and 
assumes a fixed thermalization efficiency of 50\%. In
our second, and more sophisticated model, we use time-dependent 
thermalization efficiencies based on the recent work of 
\cite{barnes16a} and we allow to switch
between the FRDM and the DZ31 mass model.
We have calculated the resulting macronova lightcurves in 
different bands for eight representative dynamic ejecta cases, 
see Tab.~\ref{tab:dynamic_ejecta}, and for 21 wind models, see Tab.~\ref{tab:winds}.\\
%
% What are the results?
%
We find that the nuclear mass model has a decisive impact 
on the results. In terms of nucleosynthesis, the DZ31 model 
reproduces the solar-system r-process abundances around the 
platinum peak (A=195) with substantially higher accuracy
than the FRDM model.  It is worth emphasizing, however,
that also other  nuclear properties  such as decay half-lives
and fission yields have substantial uncertainties. With DZ31
one finds substantially larger amounts of
trans-lead material and its $\alpha$-decay 
yields nuclear heating rates that are, at the times relevant 
for the macronova emission, about an order of magnitude 
larger than in the FRDM case, see Fig.~\ref{fig:heating_rate}, 
right panel.
In addition, the thermalization efficiencies in the DZ31 
cases are increased by factors of a few. As a result, the 
DZ31 macronova light curves are substantially brighter
than those obtained with FRDM.\\
We find that the more promising dynamic ejecta models from 
NSNS mergers reach K-band peak magnitudes in excess of $-15$ 
(model N5 with a mass ratio close to the observed NSNS binary 
J0453+1559; see right panel of Fig.~\ref{fig:nsns_model1_model2}), 
while the brightest NSBH dynamic ejecta model 
reaches values beyond $-16$. The wind models with parameters
inspired by neutrino-driven winds from an NSNS-merger peak 
below $-13$, while models mimicking unbound torus matter
\citep{just15,wu16} reach peak values of $-14$ and 
-- with only slightly increased mass and velocity parameters --
they can reach close to $-16$ (Fig.~\ref{fig:MC_winds}).\\
Since more flux is expected at longer wavelengths, near-IR 
surveys are particularly suitable for macronova detections. 
We note, however, that 
exposures in optical bands 
(i) may be as efficient in detecting such mergers. Since the 
transients are generally faint and other sources evolving on 
these time scales are abundant, the best detection chances 
come from events that are triggered by either GW-signals or 
GRB-emission. 
For our adopted "best guess merger rate" of 300 Gpc$^{-3}$ 
yr$^{-1}$ larger telescopes such as LSST or VISTA should 
detect of order 10 events per year. For mergers within 30 Mpc
ZTF would have an excellent chance of capturing macronovae 
in multiple filters.\\
We have also explored which of our models could be plausible 
explanations for the best-to-date macronova candidate, the transient
observed in the aftermath of GRB 130603B \citep{tanvir13a,berger13b}. 
Adopting the DZ31 models as our standard, several of the explored 
cases get close to this observation. Both dynamic ejecta
models (the non-equal mass NSNS merger model N5 and the NSBH 
models B2 and B3) and several of our wind models (wind7, wind9, 
wind17, wind20 and wind21) produce transients with similar properties.
We note, however, that none of our dynamic ejecta models would be able 
to account for this observation, for the hypothetical case that the 
effective opacities would be substantially larger than our adopted 
value of 10 cm$^2$/g \citep{kasen13a,barnes13a,tanaka13a}.\\
% Where are the caveats?
The results presented here are based on our currently best 
macronova models. These involve a chain of different 
calculations where each has its own challenges. For example, 
the calculations involve gravity, hydrodynamics, neutrino
and nuclear reactions, thermalization efficiencies and an 
approximate treatment of radiation. 
For several of these ingredients there  are substantial
uncertainties, for example for the nuclear equation 
of the state, the nuclear physics near the neutron-dripline 
and the involved matter opacities. Moreover, we note that our 
current models are not using proper radiative transfer 
calculations. While the discussed results represent the current 
status of our models, they certainly can and will be improved 
in future.\\

\noindent Lightcurves,  expected rates and redshift distributions are available
under:  {\bf http://snova.fysik.su.se/transient-rates/}.

\section*{Acknowledgements}
It is a great pleasure to acknowledge stimulating discussions
with many collaborators, in particular Almudena Arcones, Jennifer 
Barnes, Brian Metzger, Albino
Perego and Friedrich-Karl Thielemann.
This work has been supported by the Swedish Research 
Council (VR) under grant 621-2012-4870. It has further
been supported by  the CompStar network, COST Action MP1304, 
and  by the  Knut and Alice Wallenberg Foundation. 
MRW acknowledges support from the Villum Foundation 
(Project No. 13164) and the Danish National Research 
Foundation (DNRF91) and GMP was partly supported by the
Deutsche Forschungsgemeinschaft through contract SFB 1245 and 
the BMBF-Verbundforschungsprojekt 05P15RDFN1.
The simulations for this 
paper were performed on the facilities of the North-German 
Supercomputing Alliance (HLRN).

\bibliographystyle{mn2e}
\bibliography{astro_SKR}

\hyphenation{Post-Script Sprin-ger}
\begin{thebibliography}{}

\bibitem[\protect\citeauthoryear{{Abadie}, {Abbott}, {Abbott}, {Abernathy},
  {Accadia}, {Acernese}, {Adams}, {Adhikari}, {Ajith}, {Allen} \& et
  al.}{{Abadie} et~al.}{2010}]{abadie10}
{Abadie} J.,  {Abbott} B.~P.,  {Abbott} R.,  {Abernathy} M.,  {Accadia} T.,
  {Acernese} F.,  {Adams} C.,  {Adhikari} R.,  {Ajith} P.,  {Allen} B.,    et
  al. 2010, Classical and Quantum Gravity, 27, 173001

\bibitem[\protect\citeauthoryear{{Abbott}, {Abbott}, {Abbott}, {Abernathy},
  {Acernese}, {Ackley}, {Adams}, {Adams}, {Addesso}, {Adhikari} \& et
  al.}{{Abbott} et~al.}{2016c}]{abbott16c}
{Abbott} B.~P.,  {Abbott} R.,  {Abbott} T.~D.,  {Abernathy} M.~R.,  {Acernese}
  F.,  {Ackley} K.,  {Adams} C.,  {Adams} T.,  {Addesso} P.,  {Adhikari} R.~X.,
     et al. 2016c, \apjl, 826, L13

\bibitem[\protect\citeauthoryear{{Abbott}, {Abbott}, {Abbott}, {Abernathy},
  {Acernese}, {Ackley}, {Adams}, {Adams}, {Addesso}, {Adhikari} \& et
  al.}{{Abbott} et~al.}{2016a}]{abbott16a}
{Abbott} B.~P.,  {Abbott} R.,  {Abbott} T.~D.,  {Abernathy} M.~R.,  {Acernese}
  F.,  {Ackley} K.,  {Adams} C.,  {Adams} T.,  {Addesso} P.,  {Adhikari} R.~X.,
     et al. 2016a, Physical Review Letters, 116, 061102

\bibitem[\protect\citeauthoryear{{Abbott}, {Abbott}, {Abbott}, {Abernathy},
  {Acernese}, {Ackley}, {Adams}, {Adams}, {Addesso}, {Adhikari} \& et
  al.}{{Abbott} et~al.}{2016b}]{abbott16h}
{Abbott} B.~P.,  {Abbott} R.,  {Abbott} T.~D.,  {Abernathy} M.~R.,  {Acernese}
  F.,  {Ackley} K.,  {Adams} C.,  {Adams} T.,  {Addesso} P.,  {Adhikari} R.~X.,
     et al. 2016b, arXiv:1607.07456

\bibitem[\protect\citeauthoryear{Abbott, Collaboration \& Collaboration}{Abbott
  et~al.}{2016}]{abbott_LRR}
Abbott B.~P.,  Collaboration L.~S.,    Collaboration V.,  2016, Living Reviews
  in Relativity, 19

\bibitem[\protect\citeauthoryear{{Antoniadis}, {Freire}, {Wex}, {Tauris},
  {Verbiest} \& {Whelan}}{{Antoniadis} et~al.}{2013}]{antoniadis13}
{Antoniadis} J.,  {Freire} P.~C.~C.,  {Wex} N.,  {Tauris} T.~M.,  {Verbiest}
  J.~P.~W.,    {Whelan} D.~G.,  2013, Science, 340, 448

\bibitem[\protect\citeauthoryear{{Arnould}, {Goriely} \& {Takahashi}}{{Arnould}
  et~al.}{2007}]{arnould07}
{Arnould} M.,  {Goriely} S.,    {Takahashi} K.,  2007, Phys. Reports, 450, 97

\bibitem[\protect\citeauthoryear{{Astropy Collaboration}, {Robitaille},
  {Tollerud}, {Greenfield}, {Droettboom}, {Bray}, {Aldcroft}, {Davis},
  {Ginsburg}, {Price-Whelan}, {Kerzendorf}, {Conley}, {Crighton} \& {et
  al.}}{{Astropy Collaboration} et~al.}{2013}]{2013A&A...558A..33A}
{Astropy Collaboration} {Robitaille} T.~P.,  {Tollerud} E.~J.,  {Greenfield}
  P.,  {Droettboom} M.,  {Bray} E.,  {Aldcroft} T.,  {Davis} M.,  {Ginsburg}
  A.,  {Price-Whelan} A.~M.,  {Kerzendorf} W.~E.,  {Conley} A.,  {Crighton} N.,
     {et al.} 2013, \aap, 558, A33

\bibitem[\protect\citeauthoryear{Audi, Kondev, Wang, Pfeiffer, Sun, Blachot \&
  MacCormick}{Audi et~al.}{2012}]{audi12}
Audi G.,  Kondev F.,  Wang M.,  Pfeiffer B.,  Sun X.,  Blachot J.,
  MacCormick M.,  2012, Chinese Physics C, 36, 1157

\bibitem[\protect\citeauthoryear{Barbary}{Barbary}{2014}]{Barbary:11938}
Barbary K.,  2014

\bibitem[\protect\citeauthoryear{{Barnes} \& {Kasen}}{{Barnes} \&
  {Kasen}}{2013}]{barnes13a}
{Barnes} J.,  {Kasen} D.,  2013, ApJ, 775, 18

\bibitem[\protect\citeauthoryear{{Barnes}, {Kasen}, {Wu} \&
  {Martinez-Pinedo}}{{Barnes} et~al.}{2016}]{barnes16a}
{Barnes} J.,  {Kasen} D.,  {Wu} M.-R.,    {Martinez-Pinedo} G.,  2016, ApJ,
  829, 110

\bibitem[\protect\citeauthoryear{{Baumgarte}, {Shapiro} \&
  {Shibata}}{{Baumgarte} et~al.}{2000}]{baumgarte00}
{Baumgarte} T.~W.,  {Shapiro} S.~L.,    {Shibata} M.,  2000, ApJL, 528, L29

\bibitem[\protect\citeauthoryear{{Bauswein}, {Ardevol Pulpillo}, {Janka} \&
  {Goriely}}{{Bauswein} et~al.}{2014}]{bauswein14b}
{Bauswein} A.,  {Ardevol Pulpillo} R.,  {Janka} H.-T.,    {Goriely} S.,  2014,
  ApJL, 795, L9

\bibitem[\protect\citeauthoryear{{Bauswein}, {Baumgarte} \& {Janka}}{{Bauswein}
  et~al.}{2013}]{bauswein13b}
{Bauswein} A.,  {Baumgarte} T.~W.,    {Janka} H.-T.,  2013, Physical Review
  Letters, 111, 131101

\bibitem[\protect\citeauthoryear{{Bauswein}, {Goriely} \& {Janka}}{{Bauswein}
  et~al.}{2013}]{Bauswein13a}
{Bauswein} A.,  {Goriely} S.,    {Janka} H.-T.,  2013, ApJ, 773, 78

\bibitem[\protect\citeauthoryear{{Belczynski}, {Repetto}, {Holz},
  {O'Shaughnessy}, {Bulik}, {Berti}, {Fryer} \& {Dominik}}{{Belczynski}
  et~al.}{2016}]{belczynski16a}
{Belczynski} K.,  {Repetto} S.,  {Holz} D.~E.,  {O'Shaughnessy} R.,  {Bulik}
  T.,  {Berti} E.,  {Fryer} C.,    {Dominik} M.,  2016, ApJ, 819, 108

\bibitem[\protect\citeauthoryear{{Beloborodov}}{{Beloborodov}}{2008}]{beloborodov08}
{Beloborodov} A.~M.,  2008, in {M.~Axelsson} ed., American Institute of Physics
  Conference Series Vol.~1054 of American Institute of Physics Conference
  Series, {Hyper-accreting black holes}.
pp 51--70

\bibitem[\protect\citeauthoryear{{Berger}, {Fong} \& {Chornock}}{{Berger}
  et~al.}{2013}]{berger13b}
{Berger} E.,  {Fong} W.,    {Chornock} R.,  2013, ApJL, 774, L23

\bibitem[\protect\citeauthoryear{{Bildsten} \& {Cutler}}{{Bildsten} \&
  {Cutler}}{1992}]{bildsten92}
{Bildsten} L.,  {Cutler} C.,  1992, ApJ, 400, 175

\bibitem[\protect\citeauthoryear{{Chruslinska}, {Belczynski}, {Bulik} \&
  {Gladysz}}{{Chruslinska} et~al.}{2016}]{chruslinska16}
{Chruslinska} M.,  {Belczynski} K.,  {Bulik} T.,    {Gladysz} W.,  2016, ArXiv
  e-prints

\bibitem[\protect\citeauthoryear{{Ciolfi} \& {Siegel}}{{Ciolfi} \&
  {Siegel}}{2015}]{ciolfi15}
{Ciolfi} R.,  {Siegel} D.~M.,  2015, ApJL, 798, L36

\bibitem[\protect\citeauthoryear{{de Ugarte Postigo}, {Th{\"o}ne}, {Rowlinson},
  {Garcia-Benito}, {Levan}, {Gorosabel}, {Goldoni}, {Schulze}, {Zafar},
  {Wiersema}, {Sanchez-Ramirez}, {Melandri} \& {D'Avanzo}}{{de Ugarte Postigo}
  et~al.}{2014}]{ugarte13}
{de Ugarte Postigo} A.,  {Th{\"o}ne} C.~C.,  {Rowlinson} A.,  {Garcia-Benito}
  R.,  {Levan} A.~J.,  {Gorosabel} J.,  {Goldoni} P.,  {Schulze} S.,  {Zafar}
  T.,  {Wiersema} K.,  {Sanchez-Ramirez} R.,  {Melandri} A.,    {D'Avanzo}
  2014, \aap, 563, A62

\bibitem[\protect\citeauthoryear{{Deaton}, {Duez}, {Foucart}, {O'Connor},
  {Ott}, {Kidder}, {Muhlberger}, {Scheel} \& {Szilagyi}}{{Deaton}
  et~al.}{2013}]{deaton13a}
{Deaton} M.~B.,  {Duez} M.~D.,  {Foucart} F.,  {O'Connor} E.,  {Ott} C.~D.,
  {Kidder} L.~E.,  {Muhlberger} C.~D.,  {Scheel} M.~A.,    {Szilagyi} B.,
  2013, ApJ, 776, 47

\bibitem[\protect\citeauthoryear{{Demorest}, {Pennucci}, {Ransom}, {Roberts} \&
  {Hessels}}{{Demorest} et~al.}{2010}]{demorest10}
{Demorest} P.~B.,  {Pennucci} T.,  {Ransom} S.~M.,  {Roberts} M.~S.~E.,
  {Hessels} J.~W.~T.,  2010, Nature, 467, 1081

\bibitem[\protect\citeauthoryear{{Dessart}, {Ott}, {Burrows}, {Rosswog} \&
  {Livne}}{{Dessart} et~al.}{2009}]{dessart09}
{Dessart} L.,  {Ott} C.~D.,  {Burrows} A.,  {Rosswog} S.,    {Livne} E.,  2009,
  ApJ, 690, 1681

\bibitem[\protect\citeauthoryear{{Doctor}, {Kessler}, {Chen}, {Farr}, {Finley},
  {Foley}, {Goldstein} \& {Wester}}{{Doctor} et~al.}{2016}]{doctor16}
{Doctor} Z.,  {Kessler} R.,  {Chen} H.~Y.,  {Farr} B.,  {Finley} D.~A.,
  {Foley} R.~J.,  {Goldstein} D.~A.,    {Wester} W.,  2016, ArXiv e-prints

\bibitem[\protect\citeauthoryear{{Duez}, {Foucart}, {Kidder}, {Ott} \&
  {Teukolsky}}{{Duez} et~al.}{2010}]{duez10b}
{Duez} M.~D.,  {Foucart} F.,  {Kidder} L.~E.,  {Ott} C.~D.,    {Teukolsky}
  S.~A.,  2010, Classical and Quantum Gravity, 27, 114106

\bibitem[\protect\citeauthoryear{{Duflo} \& {Zuker}}{{Duflo} \&
  {Zuker}}{1995}]{duflo95}
{Duflo} J.,  {Zuker} A.~P.,  1995, Phys. Rev. C, 52, R23

\bibitem[\protect\citeauthoryear{Eichler, Livio, Piran \& Schramm}{Eichler
  et~al.}{1989}]{eichler89}
Eichler D.,  Livio M.,  Piran T.,    Schramm D.~N.,  1989, Nature, 340, 126

\bibitem[\protect\citeauthoryear{{Eichler}, {Arcones}, {Kelic}, {Korobkin},
  {Langanke}, {Marketin}, {Martinez-Pinedo}, {Panov}, {Rauscher}, {Rosswog},
  {Winteler}, {Zinner} \& {Thielemann}}{{Eichler} et~al.}{2015}]{eichler15}
{Eichler} M.,  {Arcones} A.,  {Kelic} A.,  {Korobkin} O.,  {Langanke} K.,
  {Marketin} T.,  {Martinez-Pinedo} G.,  {Panov} I.,  {Rauscher} T.,  {Rosswog}
  S.,  {Winteler} C.,  {Zinner} N.~T.,    {Thielemann} F.-K.,  2015, ApJ, 808,
  30

\bibitem[\protect\citeauthoryear{{Fernandez} \& {Metzger}}{{Fernandez} \&
  {Metzger}}{2013a}]{fernandez13b}
{Fernandez} R.,  {Metzger} B.~D.,  2013a, MNRAS, 435, 502

\bibitem[\protect\citeauthoryear{{Fernandez} \& {Metzger}}{{Fernandez} \&
  {Metzger}}{2013b}]{fernandez13a}
{Fernandez} R.,  {Metzger} B.~D.,  2013b, ApJ, 763, 108

\bibitem[\protect\citeauthoryear{{Fernandez} \& {Metzger}}{{Fernandez} \&
  {Metzger}}{2015}]{fernandez16}
{Fernandez} R.,  {Metzger} B.~D.,  2015, ArXiv e-prints

\bibitem[\protect\citeauthoryear{{Fernandez}, {Quataert}, {Schwab}, {Kasen} \&
  {Rosswog}}{{Fernandez} et~al.}{2015}]{fernandez15}
{Fernandez} R.,  {Quataert} E.,  {Schwab} J.,  {Kasen} D.,    {Rosswog} S.,
  2015, MNRAS, 449, 390

\bibitem[\protect\citeauthoryear{{Fontes}, {Fryer}, {Hungerford}, {Hakel},
  {Colgan}, {Kilcrease} \& {Sherrill}}{{Fontes} et~al.}{2015}]{fontes15}
{Fontes} C.~J.,  {Fryer} C.~L.,  {Hungerford} A.~L.,  {Hakel} P.,  {Colgan} J.,
   {Kilcrease} D.~P.,    {Sherrill} M.~E.,  2015, High Energy Density Physics,
  16, 53

\bibitem[\protect\citeauthoryear{{Foucart}}{{Foucart}}{2012}]{foucart12}
{Foucart} F.,  2012, Phys. Rev. D, 86, 124007

\bibitem[\protect\citeauthoryear{{Foucart}, {Deaton}, {Duez}, {Kidder},
  {MacDonald}, {Ott}, {Pfeiffer}, {Scheel}, {Szilagyi} \&
  {Teukolsky}}{{Foucart} et~al.}{2013}]{foucart13}
{Foucart} F.,  {Deaton} M.~B.,  {Duez} M.~D.,  {Kidder} L.~E.,  {MacDonald} I.,
   {Ott} C.~D.,  {Pfeiffer} H.~P.,  {Scheel} M.~A.,  {Szilagyi} B.,
  {Teukolsky} S.~A.,  2013, Phys. Rev. D, 87, 084006

\bibitem[\protect\citeauthoryear{{Foucart}, {Deaton}, {Duez}, {O'Connor},
  {Ott}, {Haas}, {Kidder}, {Pfeiffer}, {Scheel} \& {Szilagyi}}{{Foucart}
  et~al.}{2014}]{foucart14}
{Foucart} F.,  {Deaton} M.~B.,  {Duez} M.~D.,  {O'Connor} E.,  {Ott} C.~D.,
  {Haas} R.,  {Kidder} L.~E.,  {Pfeiffer} H.~P.,  {Scheel} M.~A.,    {Szilagyi}
  B.,  2014, Phys. Rev. D, 90, 024026

\bibitem[\protect\citeauthoryear{Freiburghaus, Rosswog \&
  Thielemann}{Freiburghaus et~al.}{1999}]{freiburghaus99b}
Freiburghaus C.,  Rosswog S.,    Thielemann F.-K.,  1999, ApJ, 525, L121

\bibitem[\protect\citeauthoryear{{Fuller}, {Fowler} \& {Newman}}{{Fuller}
  et~al.}{1982}]{fuller82}
{Fuller} G.~M.,  {Fowler} W.~A.,    {Newman} M.~J.,  1982, ApJS, 48, 279

\bibitem[\protect\citeauthoryear{{Giacomazzo}, {Perna}, {Rezzolla}, {Troja} \&
  {Lazzati}}{{Giacomazzo} et~al.}{2013}]{giacomazzo13}
{Giacomazzo} B.,  {Perna} R.,  {Rezzolla} L.,  {Troja} E.,    {Lazzati} D.,
  2013, ApJL, 762, L18

\bibitem[\protect\citeauthoryear{{Gondek-Rosinska}, {Kowalska}, {Villain},
  {Ansorg} \& {Kucaba}}{{Gondek-Rosinska} et~al.}{2016}]{gondek16}
{Gondek-Rosinska} D.,  {Kowalska} I.,  {Villain} L.,  {Ansorg} M.,    {Kucaba}
  M.,  2016, ArXiv e-prints

\bibitem[\protect\citeauthoryear{{Goriely}}{{Goriely}}{2015}]{goriely15}
{Goriely} S.,  2015, European Physical Journal A, 51, 22

\bibitem[\protect\citeauthoryear{{Grossman}, {Korobkin}, {Rosswog} \&
  {Piran}}{{Grossman} et~al.}{2014}]{grossman14a}
{Grossman} D.,  {Korobkin} O.,  {Rosswog} S.,    {Piran} T.,  2014, MNRAS, 439,
  757

\bibitem[\protect\citeauthoryear{{Hotokezaka}, {Kiuchi}, {Kyutoku},
  {Muranushi}, {Sekiguchi}, {Shibata} \& {Taniguchi}}{{Hotokezaka}
  et~al.}{2013}]{hotokezaka13b}
{Hotokezaka} K.,  {Kiuchi} K.,  {Kyutoku} K.,  {Muranushi} T.,  {Sekiguchi}
  Y.-i.,  {Shibata} M.,    {Taniguchi} K.,  2013, Phys. Rev. D, 88, 044026

\bibitem[\protect\citeauthoryear{{Hotokezaka}, {Kiuchi}, {Kyutoku}, {Okawa},
  {Sekiguchi}, {Shibata} \& {Taniguchi}}{{Hotokezaka}
  et~al.}{2013}]{hotokezaka13a}
{Hotokezaka} K.,  {Kiuchi} K.,  {Kyutoku} K.,  {Okawa} H.,  {Sekiguchi} Y.-i.,
  {Shibata} M.,    {Taniguchi} K.,  2013, Phys. Rev. D, 87, 024001

\bibitem[\protect\citeauthoryear{{Hotokezaka}, {Kyutoku}, {Okawa}, {Shibata} \&
  {Kiuchi}}{{Hotokezaka} et~al.}{2011}]{hotokezaka11}
{Hotokezaka} K.,  {Kyutoku} K.,  {Okawa} H.,  {Shibata} M.,    {Kiuchi} K.,
  2011, Phys. Rev. D, 83, 124008

\bibitem[\protect\citeauthoryear{{Hotokezaka}, {Wanajo}, {Tanaka}, {Bamba},
  {Terada} \& {Piran}}{{Hotokezaka} et~al.}{2016}]{hotokezaka16a}
{Hotokezaka} K.,  {Wanajo} S.,  {Tanaka} M.,  {Bamba} A.,  {Terada} Y.,
  {Piran} T.,  2016, MNRAS, 459, 35

\bibitem[\protect\citeauthoryear{{Jin}, {Li}, {Cano}, {Covino}, {Fan} \&
  {Wei}}{{Jin} et~al.}{2015}]{jin15}
{Jin} Z.-P.,  {Li} X.,  {Cano} Z.,  {Covino} S.,  {Fan} Y.-Z.,    {Wei} D.-M.,
  2015, ApJL, 811, L22

\bibitem[\protect\citeauthoryear{{Just}, {Bauswein}, {Pulpillo}, {Goriely} \&
  {Janka}}{{Just} et~al.}{2015}]{just15}
{Just} O.,  {Bauswein} A.,  {Pulpillo} R.~A.,  {Goriely} S.,    {Janka} H.-T.,
  2015, MNRAS, 448, 541

\bibitem[\protect\citeauthoryear{{Kaplan}, {Ott}, {O'Connor}, {Kiuchi},
  {Roberts} \& {Duez}}{{Kaplan} et~al.}{2014}]{kaplan14}
{Kaplan} J.~D.,  {Ott} C.~D.,  {O'Connor} E.~P.,  {Kiuchi} K.,  {Roberts} L.,
   {Duez} M.,  2014, ApJ, 790, 19

\bibitem[\protect\citeauthoryear{{Kasen}, {Badnell} \& {Barnes}}{{Kasen}
  et~al.}{2013}]{kasen13a}
{Kasen} D.,  {Badnell} N.~R.,    {Barnes} J.,  2013, ApJ, 774, 25

\bibitem[\protect\citeauthoryear{{Kasliwal}, {Cenko}, {Singer}, {Corsi}, {Cao},
  {Barlow}, {Bhalerao}, {Bellm}, {Cook}, {Duggan}, {Ferretti}, {Frail}
  et~al.,}{{Kasliwal} et~al.}{2016}]{kasliwal16}
{Kasliwal} M.~M.,  {Cenko} S.~B.,  {Singer} L.~P.,  {Corsi} A.,  {Cao} Y.,
  {Barlow} T.,  {Bhalerao} V.,  {Bellm} E.,  {Cook} D.,  {Duggan} G.~E.,
  {Ferretti} R.,  {Frail} D.~A.,    et~al., 2016, ApJL, 824, L24

\bibitem[\protect\citeauthoryear{{Kastaun} \& {Galeazzi}}{{Kastaun} \&
  {Galeazzi}}{2015}]{kastaun15}
{Kastaun} W.,  {Galeazzi} F.,  2015, Phys. Rev. D, 91, 064027

\bibitem[\protect\citeauthoryear{{Kochanek}}{{Kochanek}}{1992}]{kochanek92}
{Kochanek} C.~S.,  1992, ApJ, 398, 234

\bibitem[\protect\citeauthoryear{{Kodama} \& {Takahashi}}{{Kodama} \&
  {Takahashi}}{1975}]{kodama75}
{Kodama} T.,  {Takahashi} K.,  1975, Nuclear Physics A, 239, 489

\bibitem[\protect\citeauthoryear{{Korobkin}, {Rosswog}, {Arcones} \&
  {Winteler}}{{Korobkin} et~al.}{2012}]{korobkin12a}
{Korobkin} O.,  {Rosswog} S.,  {Arcones} A.,    {Winteler} C.,  2012, MNRAS,
  426, 1940

\bibitem[\protect\citeauthoryear{{Kulkarni}}{{Kulkarni}}{2005}]{kulkarni05}
{Kulkarni} S.~R.,  2005, ArXiv Astrophysics e-prints

\bibitem[\protect\citeauthoryear{{Kyutoku}, {Ioka} \& {Shibata}}{{Kyutoku}
  et~al.}{2013}]{kyutoku13}
{Kyutoku} K.,  {Ioka} K.,    {Shibata} M.,  2013, \prd, 88, 041503

\bibitem[\protect\citeauthoryear{{Langanke} \& {Martinez-Pinedo}}{{Langanke} \&
  {Martinez-Pinedo}}{2001}]{langanke01}
{Langanke} K.,  {Martinez-Pinedo} G.,  2001, Atomic Data and Nuclear Data
  Tables, 79, 1

\bibitem[\protect\citeauthoryear{Lattimer \& Schramm}{Lattimer \&
  Schramm}{1974}]{lattimer74}
Lattimer J.~M.,  Schramm D.~N.,  1974, ApJ, (Letters), 192, L145

\bibitem[\protect\citeauthoryear{{Lazarus}, {Freire}, {Allen}, {Aulbert} \&
  {Zhu}}{{Lazarus} et~al.}{2016}]{lazarus16}
{Lazarus} P.,  {Freire} P.~C.~C.,  {Allen} B.,  {Aulbert} C.,    {Zhu} W.~W.,
  2016, ApJ, 831, 150

\bibitem[\protect\citeauthoryear{{Lehner}, {Liebling}, {Palenzuela},
  {Caballero}, {O'Connor}, {Anderson} \& {Neilsen}}{{Lehner}
  et~al.}{2016}]{lehner16a}
{Lehner} L.,  {Liebling} S.~L.,  {Palenzuela} C.,  {Caballero} O.~L.,
  {O'Connor} E.,  {Anderson} M.,    {Neilsen} D.,  2016, Classical and Quantum
  Gravity, 33, 184002

\bibitem[\protect\citeauthoryear{{Li} \& {Paczy{\'n}ski}}{{Li} \&
  {Paczy{\'n}ski}}{1998}]{li98}
{Li} L.-X.,  {Paczy{\'n}ski} B.,  1998, ApJL, 507, L59

\bibitem[\protect\citeauthoryear{{Martin}, {Perego}, {Arcones}, {Thielemann},
  {Korobkin} \& {Rosswog}}{{Martin} et~al.}{2015}]{martin15}
{Martin} D.,  {Perego} A.,  {Arcones} A.,  {Thielemann} F.-K.,  {Korobkin} O.,
    {Rosswog} S.,  2015, ApJ, 813, 2

\bibitem[\protect\citeauthoryear{{Martinez}, {Stovall}, {Freire}, {Deneva},
  {Jenet}, {McLaughlin}, {Bagchi}, {Bates} \& {Ridolfi}}{{Martinez}
  et~al.}{2015}]{martinez15}
{Martinez} J.~G.,  {Stovall} K.,  {Freire} P.~C.~C.,  {Deneva} J.~S.,  {Jenet}
  F.~A.,  {McLaughlin} M.~A.,  {Bagchi} M.,  {Bates} S.~D.,    {Ridolfi} A.,
  2015, ApJ, 812, 143

\bibitem[\protect\citeauthoryear{{Martynov}, {Hall}, {Abbott}, {Abbott},
  {Abbott}, {Adams}, {Adhikari}, {Anderson}, {Anderson}, {Arai} \& et
  al.}{{Martynov} et~al.}{2016}]{martinov16}
{Martynov} D.~V.,  {Hall} E.~D.,  {Abbott} B.~P.,  {Abbott} R.,  {Abbott}
  T.~D.,  {Adams} C.,  {Adhikari} R.~X.,  {Anderson} R.~A.,  {Anderson} S.~B.,
  {Arai} K.,    et al. 2016, Phys. Rev. D, 93, 112004

\bibitem[\protect\citeauthoryear{{McClintock}, {Narayan} \&
  {Steiner}}{{McClintock} et~al.}{2014}]{mcclintock14}
{McClintock} J.~E.,  {Narayan} R.,    {Steiner} J.~F.,  2014, Space Science
  Review, 183, 295

\bibitem[\protect\citeauthoryear{{McMillan}}{{McMillan}}{2011}]{mcmillan11a}
{McMillan} P.~J.,  2011, MNRAS, 414, 2446

\bibitem[\protect\citeauthoryear{{Mendoza-Temis}, {Martinez-Pinedo},
  {Langanke}, {Bauswein} \& {Janka}}{{Mendoza-Temis}
  et~al.}{2015}]{mendoza_temis15}
{Mendoza-Temis} J.,  {Martinez-Pinedo} G.,  {Langanke} K.,  {Bauswein} A.,
  {Janka} H.-T.,  2015, Phys. Rev. C, 92, 055805

\bibitem[\protect\citeauthoryear{{Metzger}}{{Metzger}}{2016}]{metzger16a}
{Metzger} B.~D.,  2016, ArXiv e-prints

\bibitem[\protect\citeauthoryear{{Metzger}, {Bauswein}, {Goriely} \&
  {Kasen}}{{Metzger} et~al.}{2015}]{metzger15a}
{Metzger} B.~D.,  {Bauswein} A.,  {Goriely} S.,    {Kasen} D.,  2015, MNRAS,
  446, 1115

\bibitem[\protect\citeauthoryear{{Metzger} \& {Berger}}{{Metzger} \&
  {Berger}}{2012}]{metzger12a}
{Metzger} B.~D.,  {Berger} E.,  2012, ApJ, 746, 48

\bibitem[\protect\citeauthoryear{{Metzger}, {Martinez-Pinedo}, {Darbha},
  {Quataert}, {Arcones}, {Kasen}, {Thomas}, {Nugent}, {Panov} \&
  {Zinner}}{{Metzger} et~al.}{2010}]{metzger10b}
{Metzger} B.~D.,  {Martinez-Pinedo} G.,  {Darbha} S.,  {Quataert} E.,
  {Arcones} A.,  {Kasen} D.,  {Thomas} R.,  {Nugent} P.,  {Panov} I.~V.,
  {Zinner} N.~T.,  2010, MNRAS, 406, 2650

\bibitem[\protect\citeauthoryear{{Metzger}, {Piro} \& {Quataert}}{{Metzger}
  et~al.}{2008}]{metzger08a}
{Metzger} B.~D.,  {Piro} A.~L.,    {Quataert} E.,  2008, MNRAS, 390, 781

\bibitem[\protect\citeauthoryear{M\"oller, Nix, Myers \& Swiatecki}{M\"oller
  et~al.}{1995}]{moeller95}
M\"oller P.,  Nix J.~R.,  Myers W.~D.,    Swiatecki W.~J.,  1995, At. Data
  Nucl. Data Tables, 59, 185

\bibitem[\protect\citeauthoryear{Moller, Pfeiffer \& Kratz}{Moller
  et~al.}{2003}]{moeller03}
Moller P.,  Pfeiffer B.,    Kratz K.-L.,  2003, Phys. Rev., C67, 055802

\bibitem[\protect\citeauthoryear{{Monaghan}}{{Monaghan}}{2005}]{monaghan05}
{Monaghan} J.~J.,  2005, Reports on Progress in Physics, 68, 1703

\bibitem[\protect\citeauthoryear{{Mumpower}, {Surman}, {McLaughlin} \&
  {Aprahamian}}{{Mumpower} et~al.}{2016}]{mumpower16a}
{Mumpower} M.~R.,  {Surman} R.,  {McLaughlin} G.~C.,    {Aprahamian} A.,  2016,
  Progress in Particle and Nuclear Physics, 86, 86

\bibitem[\protect\citeauthoryear{{Murguia-Berthier}, {Montes}, {Ramirez-Ruiz},
  {De Colle} \& {Lee}}{{Murguia-Berthier} et~al.}{2014}]{murguia14}
{Murguia-Berthier} A.,  {Montes} G.,  {Ramirez-Ruiz} E.,  {De Colle} F.,
  {Lee} W.~H.,  2014, ApJL, 788, L8

\bibitem[\protect\citeauthoryear{{Murguia-Berthier}, {Ramirez-Ruiz}, {Montes},
  {De Colle}, {Rezzolla}, {Rosswog}, {Takami}, {Perego} \&
  {Lee}}{{Murguia-Berthier} et~al.}{2016}]{murguia16}
{Murguia-Berthier} A.,  {Ramirez-Ruiz} E.,  {Montes} G.,  {De Colle} F.,
  {Rezzolla} L.,  {Rosswog} S.,  {Takami} K.,  {Perego} A.,    {Lee} W.~H.,
  2016, ArXiv e-prints

\bibitem[\protect\citeauthoryear{{Oechslin}, {Janka} \& {Marek}}{{Oechslin}
  et~al.}{2007}]{oechslin07a}
{Oechslin} R.,  {Janka} H.,    {Marek} A.,  2007, A \& A, 467, 395

\bibitem[\protect\citeauthoryear{{{\"O}zel}, {Psaltis}, {Narayan} \&
  {McClintock}}{{{\"O}zel} et~al.}{2010}]{oezel10}
{{\"O}zel} F.,  {Psaltis} D.,  {Narayan} R.,    {McClintock} J.~E.,  2010, ApJ,
  725, 1918

\bibitem[\protect\citeauthoryear{{Panov}, {Kolbe}, {Pfeiffer}, {Rauscher},
  {Kratz} \& {Thielemann}}{{Panov} et~al.}{2005}]{panov05}
{Panov} I.~V.,  {Kolbe} E.,  {Pfeiffer} B.,  {Rauscher} T.,  {Kratz} K.-L.,
  {Thielemann} F.-K.,  2005, Nuclear Physics A, 747, 633

\bibitem[\protect\citeauthoryear{{Panov}, {Korneev}, {Rauscher},
  {Martinez-Pinedo}, {Keli{\'c}-Heil}, {Zinner} \& {Thielemann}}{{Panov}
  et~al.}{2010}]{panov10}
{Panov} I.~V.,  {Korneev} I.~Y.,  {Rauscher} T.,  {Martinez-Pinedo} G.,
  {Keli{\'c}-Heil} A.,  {Zinner} N.~T.,    {Thielemann} F.-K.,  2010, A \& A,
  513, A61

\bibitem[\protect\citeauthoryear{{Perego}, {Rosswog}, {Cabez{\'o}n},
  {Korobkin}, {K{\"a}ppeli}, {Arcones} \& {Liebend{\"o}rfer}}{{Perego}
  et~al.}{2014}]{perego14b}
{Perego} A.,  {Rosswog} S.,  {Cabez{\'o}n} R.~M.,  {Korobkin} O.,
  {K{\"a}ppeli} R.,  {Arcones} A.,    {Liebend{\"o}rfer} M.,  2014, MNRAS, 443,
  3134

\bibitem[\protect\citeauthoryear{Petermann, Langanke, Martinez-Pinedo, Panov,
  Reinhard \& Thielemann}{Petermann et~al.}{2012}]{petermann12}
Petermann I.,  Langanke K.,  Martinez-Pinedo G.,  Panov I.~V.,  Reinhard P.~G.,
     Thielemann F.~K.,  2012, Eur. Phys. J., A48, 122

\bibitem[\protect\citeauthoryear{{Petrillo}, {Dietz} \& {Cavaglia}}{{Petrillo}
  et~al.}{2013}]{petrillo13}
{Petrillo} C.,  {Dietz} A.,    {Cavaglia} M.,  2013, ApJ, 767, 140

\bibitem[\protect\citeauthoryear{{Planck Collaboration}, {Ade}, {Aghanim},
  {Arnaud}, {Ashdown}, {Aumont}, {Baccigalupi}, {Banday}, {Barreiro},
  {Bartlett} \& et al.}{{Planck Collaboration} et~al.}{2016}]{ade15}
{Planck Collaboration} {Ade} P.~A.~R.,  {Aghanim} N.,  {Arnaud} M.,  {Ashdown}
  M.,  {Aumont} J.,  {Baccigalupi} C.,  {Banday} A.~J.,  {Barreiro} R.~B.,
  {Bartlett} J.~G.,    et al. 2016, \aap, 594, A13

\bibitem[\protect\citeauthoryear{{Price}}{{Price}}{2012}]{price12a}
{Price} D.~J.,  2012, Journal of Computational Physics, 231, 759

\bibitem[\protect\citeauthoryear{{Radice}, {Galeazzi}, {Lippuner}, {Roberts},
  {Ott} \& {Rezzolla}}{{Radice} et~al.}{2016}]{radice16a}
{Radice} D.,  {Galeazzi} F.,  {Lippuner} J.,  {Roberts} L.~F.,  {Ott} C.~D.,
  {Rezzolla} L.,  2016, \mnras, 460, 3255

\bibitem[\protect\citeauthoryear{{Rauscher} \& {Thielemann}}{{Rauscher} \&
  {Thielemann}}{2000}]{rauscher00}
{Rauscher} T.,  {Thielemann} F.-K.,  2000, Atomic Data and Nuclear Data Tables,
  75, 1

\bibitem[\protect\citeauthoryear{{Roberts}, {Kasen}, {Lee} \&
  {Ramirez-Ruiz}}{{Roberts} et~al.}{2011}]{roberts11}
{Roberts} L.~F.,  {Kasen} D.,  {Lee} W.~H.,    {Ramirez-Ruiz} E.,  2011, ApJL,
  736, L21

\bibitem[\protect\citeauthoryear{{Rosswog}}{{Rosswog}}{2005}]{rosswog05a}
{Rosswog} S.,  2005, ApJ, 634, 1202

\bibitem[\protect\citeauthoryear{Rosswog}{Rosswog}{2009}]{rosswog09b}
Rosswog S.,  2009, New Astronomy Reviews, 53, 78

\bibitem[\protect\citeauthoryear{{Rosswog}}{{Rosswog}}{2013}]{rosswog13b}
{Rosswog} S.,  2013, Royal Society of London Philosophical Transactions Series
  A, 371, 20272

\bibitem[\protect\citeauthoryear{{Rosswog}}{{Rosswog}}{2015a}]{rosswog15b}
{Rosswog} S.,  2015a, MNRAS, 448, 3628

\bibitem[\protect\citeauthoryear{{Rosswog}}{{Rosswog}}{2015b}]{rosswog15c}
{Rosswog} S.,  2015b, Living Reviews of Computational Astrophysics (2015), 1

\bibitem[\protect\citeauthoryear{{Rosswog}}{{Rosswog}}{2015c}]{rosswog15a}
{Rosswog} S.,  2015c, International Journal of Modern Physics D, 24, 1530012

\bibitem[\protect\citeauthoryear{{Rosswog} \& {Liebend{\"o}rfer}}{{Rosswog} \&
  {Liebend{\"o}rfer}}{2003}]{rosswog03a}
{Rosswog} S.,  {Liebend{\"o}rfer} M.,  2003, MNRAS, 342, 673

\bibitem[\protect\citeauthoryear{Rosswog, Liebend\"orfer, Thielemann, Davies,
  Benz \& Piran}{Rosswog et~al.}{1999}]{rosswog99}
Rosswog S.,  Liebend\"orfer M.,  Thielemann F.-K.,  Davies M.,  Benz W.,
  Piran T.,  1999, A \&\ A, 341, 499

\bibitem[\protect\citeauthoryear{{Rosswog}, {Piran} \& {Nakar}}{{Rosswog}
  et~al.}{2013}]{rosswog13a}
{Rosswog} S.,  {Piran} T.,    {Nakar} E.,  2013, MNRAS, 430, 2585

\bibitem[\protect\citeauthoryear{{Ruffert} \& {Janka}}{{Ruffert} \&
  {Janka}}{2001}]{ruffert01}
{Ruffert} M.,  {Janka} H.-T.,  2001, A\&A, 380, 544

\bibitem[\protect\citeauthoryear{{S}chaback \& {W}endland}{{S}chaback \&
  {W}endland}{2006}]{schaback06}
{S}chaback R.,  {W}endland H.,  2006, {A}cta {N}umer., 15, 543

\bibitem[\protect\citeauthoryear{Shen, Toki, Oyamatsu \& Sumiyoshi}{Shen
  et~al.}{1998}]{shen98a}
Shen H.,  Toki H.,  Oyamatsu K.,    Sumiyoshi K.,  1998, Nuclear Physics, A
  637, 435

\bibitem[\protect\citeauthoryear{{Shibata} \& {Taniguchi}}{{Shibata} \&
  {Taniguchi}}{2006}]{shibata06c}
{Shibata} M.,  {Taniguchi} K.,  2006, Phys. Rev. D, 73, 064027

\bibitem[\protect\citeauthoryear{{Siegel}, {Ciolfi} \& {Rezzolla}}{{Siegel}
  et~al.}{2014}]{siegel14a}
{Siegel} D.~M.,  {Ciolfi} R.,    {Rezzolla} L.,  2014, ApJL, 785, L6

\bibitem[\protect\citeauthoryear{{Smartt}, {Chambers}, {Smith}, {Huber},
  {Young}, {Cappellaro}, {Wright}, {Coughlin}, {Schultz}, {Denneau},
  {Flewelling}, {Heinze} et~al.,}{{Smartt} et~al.}{2016}]{smartt16a}
{Smartt} S.~J.,  {Chambers} K.~C.,  {Smith} K.~W.,  {Huber} M.~E.,  {Young}
  D.~R.,  {Cappellaro} E.,  {Wright} D.~E.,  {Coughlin} M.,  {Schultz}
  A.~S.~B.,  {Denneau} L.,  {Flewelling} H.,  {Heinze} A.,    et~al., 2016,
  MNRAS, 462, 4094

\bibitem[\protect\citeauthoryear{{Smartt}, {Chambers}, {Smith}, {Huber},
  {Young}, {Cappellaro}, {Wright} et~al.,}{{Smartt} et~al.}{2016}]{smartt16b}
{Smartt} S.~J.,  {Chambers} K.~C.,  {Smith} K.~W.,  {Huber} M.~E.,  {Young}
  D.~R.,  {Cappellaro} E.,  {Wright} D.~E.,    et~al., 2016, MNRAS, 462, 4094

\bibitem[\protect\citeauthoryear{{Springel}}{{Springel}}{2010}]{springel10a}
{Springel} V.,  2010, ARAA, 48, 391

\bibitem[\protect\citeauthoryear{{Takami}, {Rezzolla} \& {Baiotti}}{{Takami}
  et~al.}{2014}]{takami14}
{Takami} K.,  {Rezzolla} L.,    {Baiotti} L.,  2014, ArXiv e-prints

\bibitem[\protect\citeauthoryear{{Tanaka} \& {Hotokezaka}}{{Tanaka} \&
  {Hotokezaka}}{2013}]{tanaka13a}
{Tanaka} M.,  {Hotokezaka} K.,  2013, ApJ, 775, 113

\bibitem[\protect\citeauthoryear{{Tanvir}, {Levan}, {Fruchter}, {Hjorth},
  {Hounsell}, {Wiersema} \& {Tunnicliffe}}{{Tanvir} et~al.}{2013}]{tanvir13a}
{Tanvir} N.~R.,  {Levan} A.~J.,  {Fruchter} A.~S.,  {Hjorth} J.,  {Hounsell}
  R.~A.,  {Wiersema} K.,    {Tunnicliffe} R.~L.,  2013, Nature, 500, 547

\bibitem[\protect\citeauthoryear{{Thielemann}, {Arcones}, {K{\"a}ppeli},
  {Liebend{\"o}rfer}, {Rauscher}, {Winteler}, {Fr{\"o}hlich}, {Dillmann},
  {Fischer}, {Martinez-Pinedo}, {Langanke}, {Farouqi}, {Kratz}, {Panov} \&
  {Korneev}}{{Thielemann} et~al.}{2011}]{thielemann11}
{Thielemann} F.-K.,  {Arcones} A.,  {K{\"a}ppeli} R.,  {Liebend{\"o}rfer} M.,
  {Rauscher} T.,  {Winteler} C.,  {Fr{\"o}hlich} C.,  {Dillmann} I.,  {Fischer}
  T.,  {Martinez-Pinedo} G.,  {Langanke} K.,  {Farouqi} K.,  {Kratz} K.-L.,
  {Panov} I.,    {Korneev} I.~K.,  2011, Progress in Particle and Nuclear
  Physics, 66, 346

\bibitem[\protect\citeauthoryear{{Timmes} \& {Swesty}}{{Timmes} \&
  {Swesty}}{2000}]{timmes00a}
{Timmes} F.~X.,  {Swesty} F.~D.,  2000, ApJS, 126, 501

\bibitem[\protect\citeauthoryear{{Wanajo} \& {Janka}}{{Wanajo} \&
  {Janka}}{2012}]{wanajo12}
{Wanajo} S.,  {Janka} H.-T.,  2012, ApJ, 746, 180

\bibitem[\protect\citeauthoryear{{Wanajo}, {Sekiguchi}, {Nishimura}, {Kiuchi},
  {Kyutoku} \& {Shibata}}{{Wanajo} et~al.}{2014}]{wanajo14}
{Wanajo} S.,  {Sekiguchi} Y.,  {Nishimura} N.,  {Kiuchi} K.,  {Kyutoku} K.,
  {Shibata} M.,  2014, ApJL, 789, L39

\bibitem[\protect\citeauthoryear{Wang, Audi, Wapstra, Kondev, MacCormick, Xu \&
  Pfeiffer}{Wang et~al.}{2012}]{wang12}
Wang M.,  Audi G.,  Wapstra A.,  Kondev F.,  MacCormick M.,  Xu X.,    Pfeiffer
  B.,  2012, Chinese Physics C, 36, 1603

\bibitem[\protect\citeauthoryear{{W}endland}{{W}endland}{1995}]{wendland95}
{W}endland H.,  1995, Advances in Computational Mathematics, 4, 389

\bibitem[\protect\citeauthoryear{{Winteler}}{{Winteler}}{2012}]{winteler12}
{Winteler} C.,  2012, PhD thesis, University Basel, CH

\bibitem[\protect\citeauthoryear{{Winteler}, {K{\"a}ppeli}, {Perego},
  {Arcones}, {Vasset}, {Nishimura}, {Liebend{\"o}rfer} \&
  {Thielemann}}{{Winteler} et~al.}{2012}]{winteler12b}
{Winteler} C.,  {K{\"a}ppeli} R.,  {Perego} A.,  {Arcones} A.,  {Vasset} N.,
  {Nishimura} N.,  {Liebend{\"o}rfer} M.,    {Thielemann} F.-K.,  2012, ApJL,
  750, L22

\bibitem[\protect\citeauthoryear{{Wu}, {Fernandez}, {Martinez-Pinedo} \&
  {Metzger}}{{Wu} et~al.}{2016}]{wu16}
{Wu} M.-R.,  {Fernandez} R.,  {Martinez-Pinedo} G.,    {Metzger} B.~D.,  2016,
  MNRAS

\bibitem[\protect\citeauthoryear{{Yang}, {Jin}, {Li}, {Covino}, {Zheng},
  {Hotokezaka}, {Fan}, {Piran} \& {Wei}}{{Yang} et~al.}{2015}]{yang15}
{Yang} B.,  {Jin} Z.-P.,  {Li} X.,  {Covino} S.,  {Zheng} X.-Z.,  {Hotokezaka}
  K.,  {Fan} Y.-Z.,  {Piran} T.,    {Wei} D.-M.,  2015, Nature Communications,
  6, 7323

\bibitem[\protect\citeauthoryear{Zinner}{Zinner}{2007}]{zinner07}
Zinner N.~T.,  2007, PhD Thesis, University of Aarhus

\end{thebibliography}
\end{document}